\documentclass{mn2e}
\usepackage[dvips]{graphicx}
\usepackage{amssymb}

\begin{document}

  \title[Discs and spheroids in cosmological simulations]{Formation history,
structure and dynamics of discs and spheroids in simulated Milky Way mass galaxies}

\author[Scannapieco et al.]{Cecilia Scannapieco $^{1}$\thanks{E-mail: cscannapieco@aip.de}, 
Simon D.M White$^{2}$, Volker Springel$^{3,4}$ and  Patricia B. Tissera$^{5,6}$
\\
$^1$ Leibniz-Institute for Astrophysics Potsdam (AIP), An der Sternwarte 16, D-14482, Potsdam, Germany\\
$^2$ Max-Planck Institute for Astrophysics, Karl-Schwarszchild Str. 1, D85748, Garching, Germany\\
$^3$ Heidelberg Institute for Theoretical Studies, Schloss-Wolfsbrunnenweg 35, D-69118, Heidelberg, Germany\\
$^4$ Zentrum f\"{u}r Astronomie der Universit\"{a}t Heidelberg, Astronomisches Recheninstitut, 
     M\"{o}nchhofstr. 12-14, 69120 Heidelberg, Germany\\
$^5$ Instituto de Astronom\'{\i}a y F\'{\i}sica del Espacio, Casilla de Correos 67,
Suc. 28, 1428, Buenos Aires, Argentina\\
$^6$  Consejo Nacional de Investigaciones Cient\'{\i}ficas
y T\'ecnicas, CONICET, Argentina\\ 
}

   \maketitle

   \begin{abstract}
We study the stellar discs and spheroids
in eight  simulations of galaxy formation
within Milky Way-mass haloes in a $\Lambda$ Cold Dark Matter cosmology.
A first paper in this series concentrated on disc properties.
Here, we extend this analysis
to study how the formation history, structure and dynamics of
discs and spheroids relate to the assembly history and structure
of their haloes. We find that
discs are generally young, with stars spanning a wide range in stellar
age: the youngest stars define thin discs and have near-circular
orbits, while the oldest stars form thicker discs which rotate $\sim
2$ times slower than the thin components, and have $2-3$ times larger
velocity dispersions. Unlike the discs, spheroids form early
and on short time-scales, and are dominated by velocity dispersion.
We find  great variety in their structure. The inner regions are
bar- or bulge-like, while the extended outer haloes are rich in
complex non-equilibrium structures such as stellar streams, 
shells and clumps.  Our discs have very high {\it
  in-situ} fractions, i.e. most of their stars 
formed in the disc itself. Nevertheless, there is
a non-negligible contribution  ($\sim 15\%$) from 
satellites that are accreted  on nearly coplanar orbits. The
inner regions of spheroids  also have relatively high {\it in-situ}
fractions, but  $65-85\%$ of their outer stellar population
is accreted.  We analyse the
circular velocities, rotation velocities and velocity dispersions
of our discs and spheroids, both for gas and stars, showing that
the dynamical structure is complex
as a result of the non-trivial interplay between
cooling and SN heating.
   \end{abstract}


\section{Introduction}

Our Milky Way is the galaxy most accessible to us, and therefore the
one for which we have the most abundant and detailed data. 
Recently, significant information has been gathered on the dynamical
and chemical properties of its stellar components: the thin and thick
discs, the bulge and the stellar halo, and  on-going and
future surveys will provide exquisite new data, both for the Milky Way
and for nearby galaxies.  These observations provide important clues
to understand how the stellar components in galaxies formed, 
and to put constraints on their assembly histories. However,
a better theoretical understanding of galaxy evolution is
needed to interpret observations and to separate
the effects of the various underlying physical processes. 
For example, if radial mixing of stars in a pre-existing disc were significant (Sellwood \&
Binney 2002), it would be difficult to infer the birth places of
stars or to decipher the metallicity distribution of the gas at the time
they formed.

Furthermore, the Milky Way is a single galaxy and it is not
yet clear that it is ``typical'' in the context of the currently
accepted structure formation model (Boylan-Kolchin et al.  2010), the
$\Lambda$-Cold Dark Matter ($\Lambda$CDM) cosmology.  In order to understand galaxy
formation in more detail, it is necessary to explore the diversity
of galaxy properties inherent to $\Lambda$CDM through the variety of
possible formation and merger histories. Simulations
of galaxy formation in its proper cosmological
context have the advantage that they naturally capture important processes such as
accretion and mergers, as well as reproducing the early assembly of
dark matter haloes from smaller substructures.  However, due to
computational limitations, until recently such simulations  followed
one or a few haloes only (e.g. Abadi et al. 2003a,b; Brook et al.
2004; Governato et al. 2007; Scannapieco et al. 2008, among others),
preventing a statistical analysis of the distributions of galaxy
properties and merger histories.  In recent independent projects,
Brooks et al. (2009), Scannapieco et al. (2009) and Stinson et
al. (2010) have been able to simulate the evolution of larger samples
of haloes. This opens up the possibility of exploring a number of
important galaxy properties and their evolution, as a function of the
host halo's spin, concentration and assembly history.

In particular, Scannapieco et al. (2009, S09 hereafter) analysed eight
simulations of the formation of Milky Way-mass galaxies in a
$\Lambda$CDM universe, focusing on the formation of the disc
components.  These simulations are part of a campaign to investigate
in detail the formation of galaxies such as our Milky Way, as well as
to understand the diversity in galaxy properties expected in the
context of $\Lambda$CDM.  

This work is a continuation of that presented in S09 and a series
of companion papers. Scannapieco et al. (2010)
investigated how estimates of the disc-to-total ratio vary when
different techniques are used, Tissera et al. (2010) studied how
galaxy assembly affects the properties of the underlying dark matter
halo, and Tissera et al. (2011, in preparation) discusses the
chemical properties of the gas and stellar components of the simulated
galaxies.  In the present paper, we further explore formation
histories, structure and dynamical properties of our discs and spheroids.

This paper is organized as follows: Section~\ref{sims} summarizes the
main characteristics of the initial conditions, the simulation code and
the simulation setup, as well as recapping previous results relevant for this
work. In Section~\ref{sec_ages} we compare the formation time-scales
of discs and spheroids; Sections~\ref{sec_disk}
and~\ref{sec_spheroids} discuss the structure, {\it in-situ} fractions
and dynamical properties of simulated discs and bulges; and in
Section~\ref{sec_vcirc} we analyse circular velocity curves and gas dynamics.
Finally, in Section~\ref{conclu} we summarize our findings and
conclusions.

\section{Simulations and analysis}
\label{sims}

\subsection{Code and  initial conditions}

We use for this study the simulations already presented in S09, which
follow the formation of eight Milky Way-mass haloes in a $\Lambda$CDM
universe, including dark matter and baryonic physics. Full details can
be found in S09. Here we summarize the main characteristics of the
code and initial conditions.

The simulations are based on the initial conditions generated for the
Aquarius Project (Springel et al. 2008), adapted to include gas
particles.  Target haloes have been selected from a lower
resolution version of the Millenium-II Simulation
(MS-II; Boylan-Kolchin et al. 2009)\footnote{The MS-II simulation is
  an N-body cosmological simulation with $\sim 10$ billion
  dissipationless particles in a periodic box with a side of 137 Mpc.}, and required to have,
at $z=0$, similar mass to the Milky Way, and to be mildly isolated (no
neighbour exceeding half of the halo mass within $1.4$ Mpc).  The
simulations assume a $\Lambda$CDM cosmology with
$\Omega_\Lambda=0.75$, $\Omega_{\rm m}=0.25$, $\Omega_{\rm b}=0.04$,
$\sigma_8=0.9$ and $H_0=73$ km s$^{-1}$ Mpc$^{-1}$.  The eight
simulations have similar dark matter and gas particle masses, as
listed in Table~\ref{simulations_table}, and similar gravitational
softenings, either $0.7$ or $1.4$ kpc, which is assumed the same for
gas, stars and dark matter particles.

The simulations were run with an extended version of the Tree-PM SPH
code {\small GADGET-3} (last described by Springel 2005) which
includes star formation, chemical enrichment and supernova feedback
(from Type II and Type Ia), metal-dependent cooling (Sutherland \&
Dopita 1993) and an explicit multiphase model for the gas component
which allows coexistent, interpenetrating relative motion and exchange
of material between a cold, dense, possibly star-forming phase
and a hot diffuse phase. 
 This model
is described in detail in Scannapieco et al. (2005, 2006), and has
been already used to study the formation of disc galaxies in a
cosmological context (Scannapieco et al. 2008; S09; Scannapieco et
al. 2010) and the formation of dwarf galaxies (Sawala et al. 2010).
We note that the multiphase model and our implementation of star 
formation and feedback are
different from the one in Springel \& Hernquist (2003), although we do
use their treatment for a UV background, based on the formulation of
Haardt \& Madau (1996).  More details on the simulations, the initial
conditions and the input parameters can be found in S09.

At $z=0$, the simulated haloes have virial masses in the range $\sim
7-16\times 10^{11}$ M$_\odot$ and are represented by about $1$ million
particles within their virial radius.  In
Table~\ref{simulations_table}, we show their main $z=0$ properties:
virial radius ($r_{200}$) and virial mass\footnote{The virial radius
  is defined as the one enclosing an overdensity $200$ times the
  critical value.} ($M_{200}$), masses in stars and in gas (within
$r_{200}$ but excluding satellite objects), cold gas masses (within $r_{200}$,
again excluding satellites), optical radius ($r_{\rm
  opt}$, which encloses $83\%$ of the cold gas plus stellar mass),
baryon fraction (within $r_{200}$), and spin parameter (as defined in
Eq.~(5) of Bullock et al.~2001) at $r_{200}$.  We also show the dark
matter and (initial) gas particle masses, and the formation redshift
of the haloes (see below).

The eight resimulated haloes are representative of Milky Way-mass
haloes in $\Lambda$CDM. Their mass assembly histories spread
around the median relation found for $\sim 7500$ haloes formed in the
MS-II (Boylan-Kolchin et al., 2010), although most of them form
earlier than the median (Appendix~\ref{Ap_MAH}).  In
particular, Aq-A, Aq-C, and Aq-H form early, Aq-B and Aq-F form late, and
the rest more or less follow the median relation.  The formation
redshifts for the haloes, defined as the redshift when the total mass
reaches half its final value, are listed in
Table~\ref{simulations_table} (see also Fig.~\ref{MAH}).  Typical
formation redshifts are between $1$ and $2$, except for Aq-F which
forms very late as a result of a major merger occurring at $z\sim
0.6$.

\begin{table*} 
\begin{small}
\caption{Principal characteristics of the simulated haloes, at $z=0$:
  virial radius ($r_{200}$), virial mass ($M_{200}$), stellar, total
  gas and cold gas masses within the virial radius ($M_{\rm star}$,
  $M_{\rm gas}$ and $M_{\rm cold}$, respectively; excluding mass in
  satellites), optical radius ($r_{\rm opt}$), baryon fraction
  ($f_{\rm b}$, within $r_{200}$), and spin parameter ($\lambda^\prime$, as
  defined in Eq.~(5) of Bullock et al. (2001)).  We also show the dark
  matter and (initial) gas particle masses, as well as the formation
  redshift of each halo, defined as that for which half of the final
  mass is reached.}
\vspace{0.1cm}
\label{simulations_table}
\begin{center}
\begin{tabular}{lccccccccccc}
\hline
Halo  & $r_{200}$  & $M_{200}$ & $M_{\rm star}$&$M_{\rm gas}$&$M_{\rm cold}$ &   $r_{\rm opt}$& $f_{\rm b}$&$\lambda^\prime$ & $m_{\rm DM}$&  $m_{\rm gas,0}$ & $z_{\rm form}$\\
        & [kpc]& [$10^{11}$M$_\odot$] &   [$10^{10}$M$_\odot$] &  [$10^{10}$M$_\odot$] &  [$10^{9}$M$_\odot$]  &     [kpc] & & &  [$10^{6}$M$_\odot$] & [$10^{6}$M$_\odot$] &\\\hline

Aq-A-5   &   232   &   14.9 &    9.0 &  4.6 &  1.51 &      17.9 & 0.09 & 0.017 &2.6 & 0.56 & 1.9\\

Aq-B-5    &  181 &      7.1 &    4.0 &  1.7 &  0.33 &   17.7 & 0.08& 0.031 &  1.5 & 0.29 & 1.3 \\

Aq-C-5    &  237 &      16.1 &    10.8&    3.6& 1.18 & 16.0 & 0.09 &0.012 &  2.2 & 0.41 & 1.9 \\

Aq-D-5&      233 &     14.9 &     7.9&   3.2&  0.02 &    14.8 & 0.07 & 0.019 & 2.3 &0.22 & 1.4 \\

Aq-E-5   & 206 &      10.8&     8.4&   2.6&   0.44&        10.6 & 0.10 & 0.026  & 1.8 &0.33 & 1.6 \\

Aq-F-5 &   196  &   9.1    &    7.7 & 1.7   & 0.12      &  14.1 & 0.10 &  0.049  & 1.2 &0.23 & 0.6 \\

Aq-G-5     &    180   &   6.8&          4.4 & 1.53&  0.61&    14.1 & 0.09 &0.048  & 1.2 &0.23 & 1.2 \\

Aq-H-5     &   182&      7.4&        6.5 &  0.52  &   0.11 &    10.4 & 0.10 & 0.008  & 1.4 &0.25 & 1.5\\
\hline

\end{tabular}
\end{center}
\end{small}
\end{table*}

\subsection{Morphologies and disc/spheroid decompositions}\label{decomp}

\begin{figure*}
\includegraphics[width=180mm]{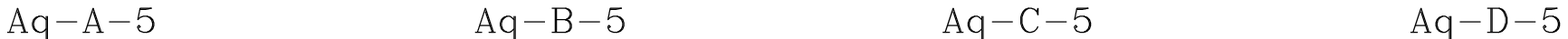}\vspace{-0.3cm}
{\includegraphics[width=45mm,angle=-90]{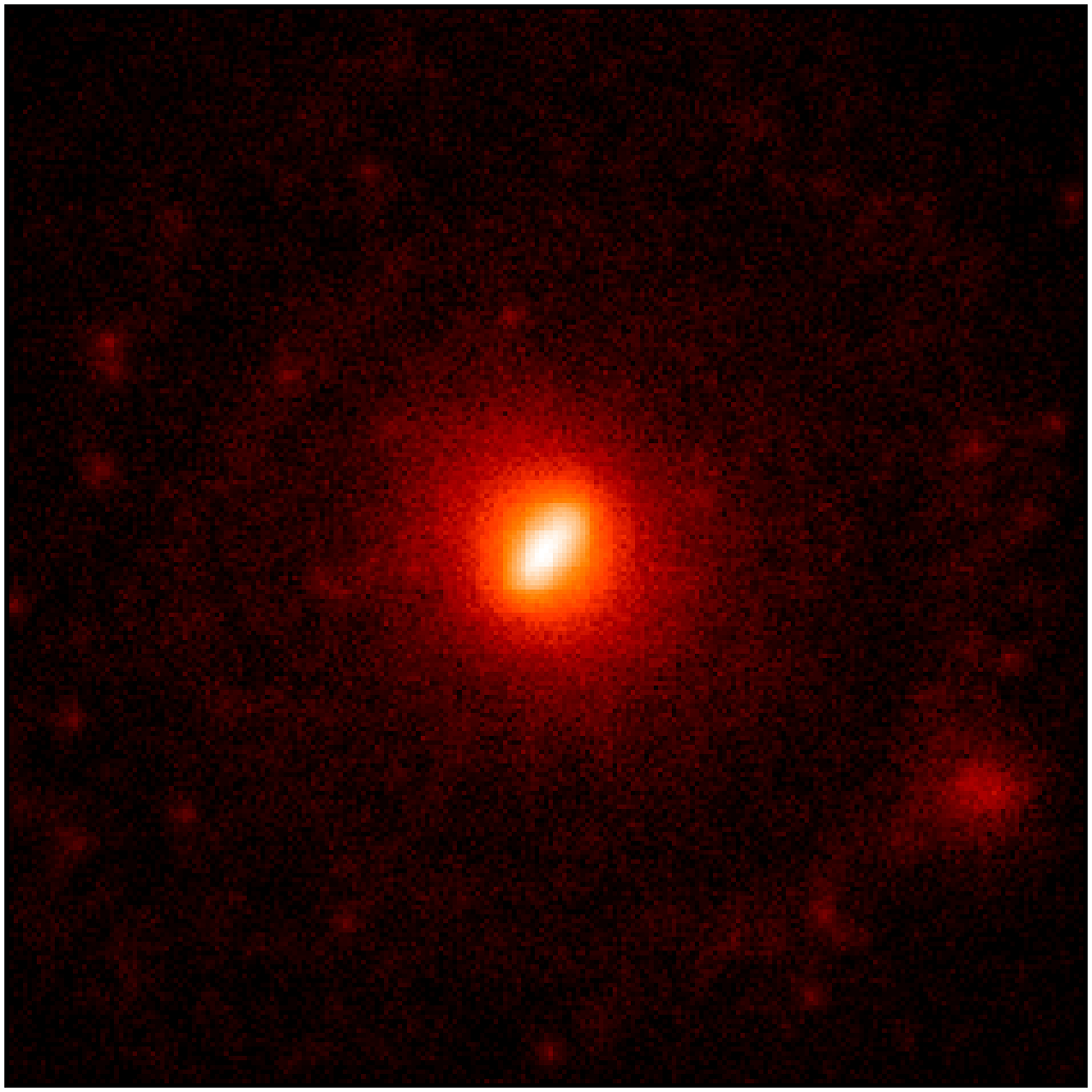}\includegraphics[width=45mm,angle=-90]{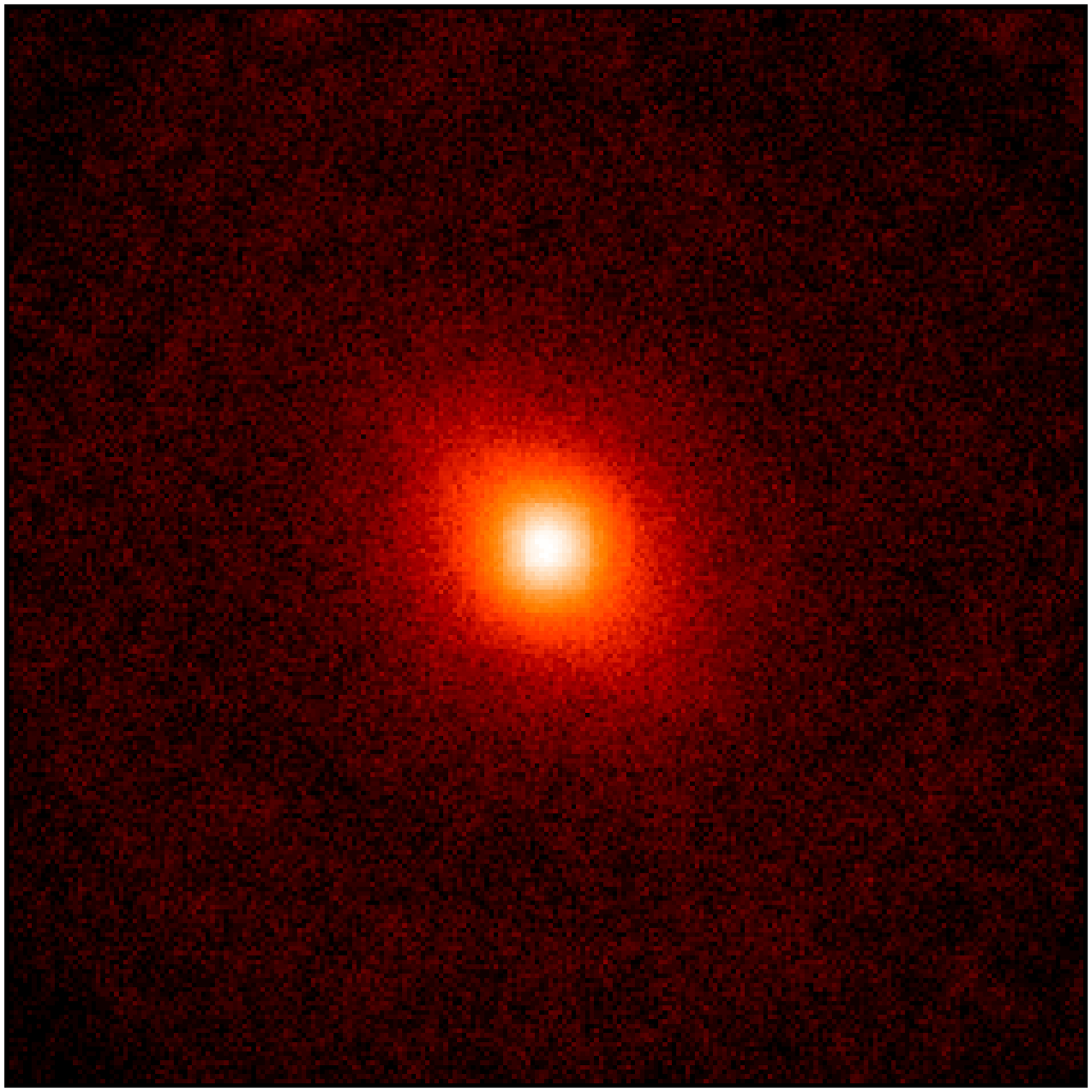}\includegraphics[width=45mm,angle=-90]{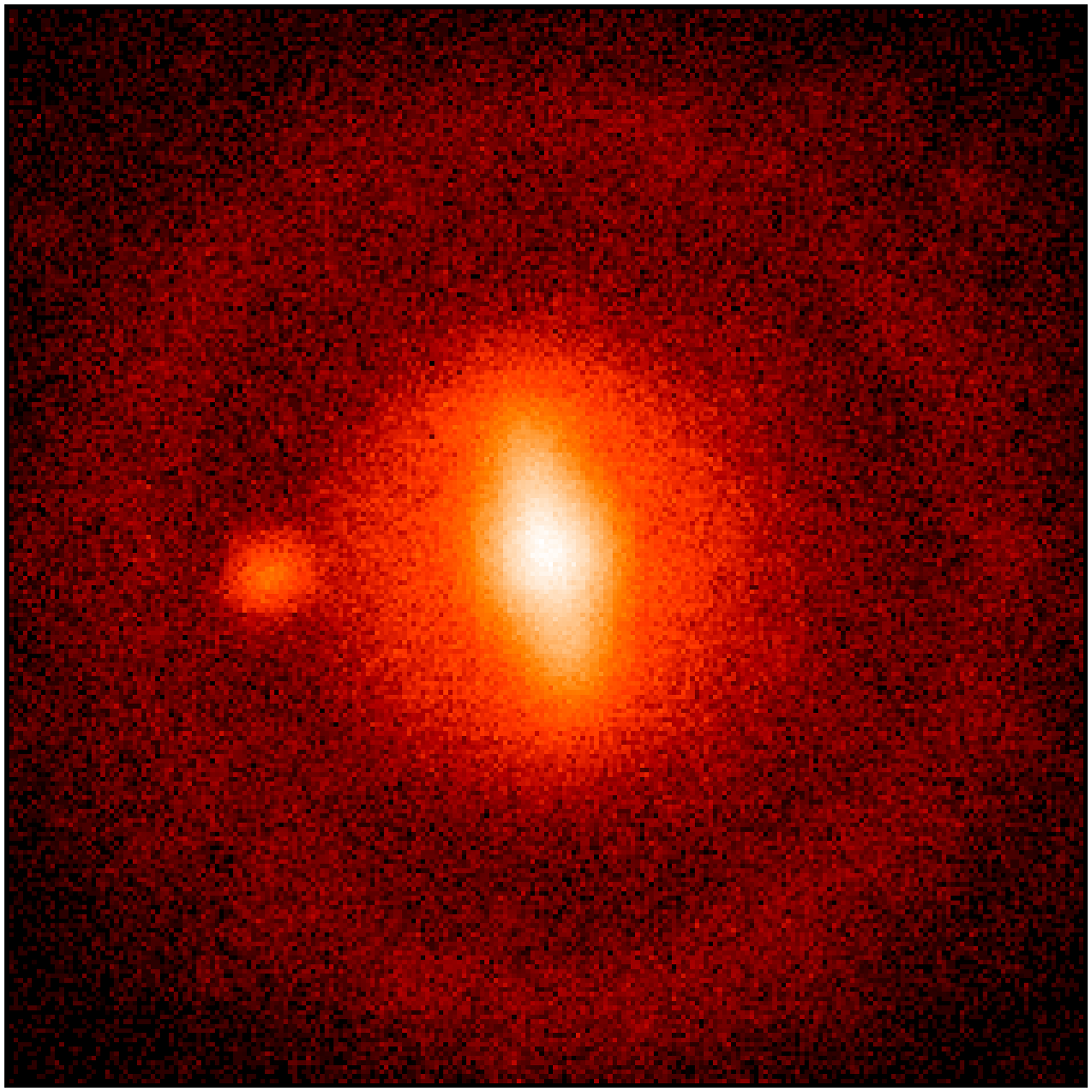}\includegraphics[width=45mm,angle=-90]{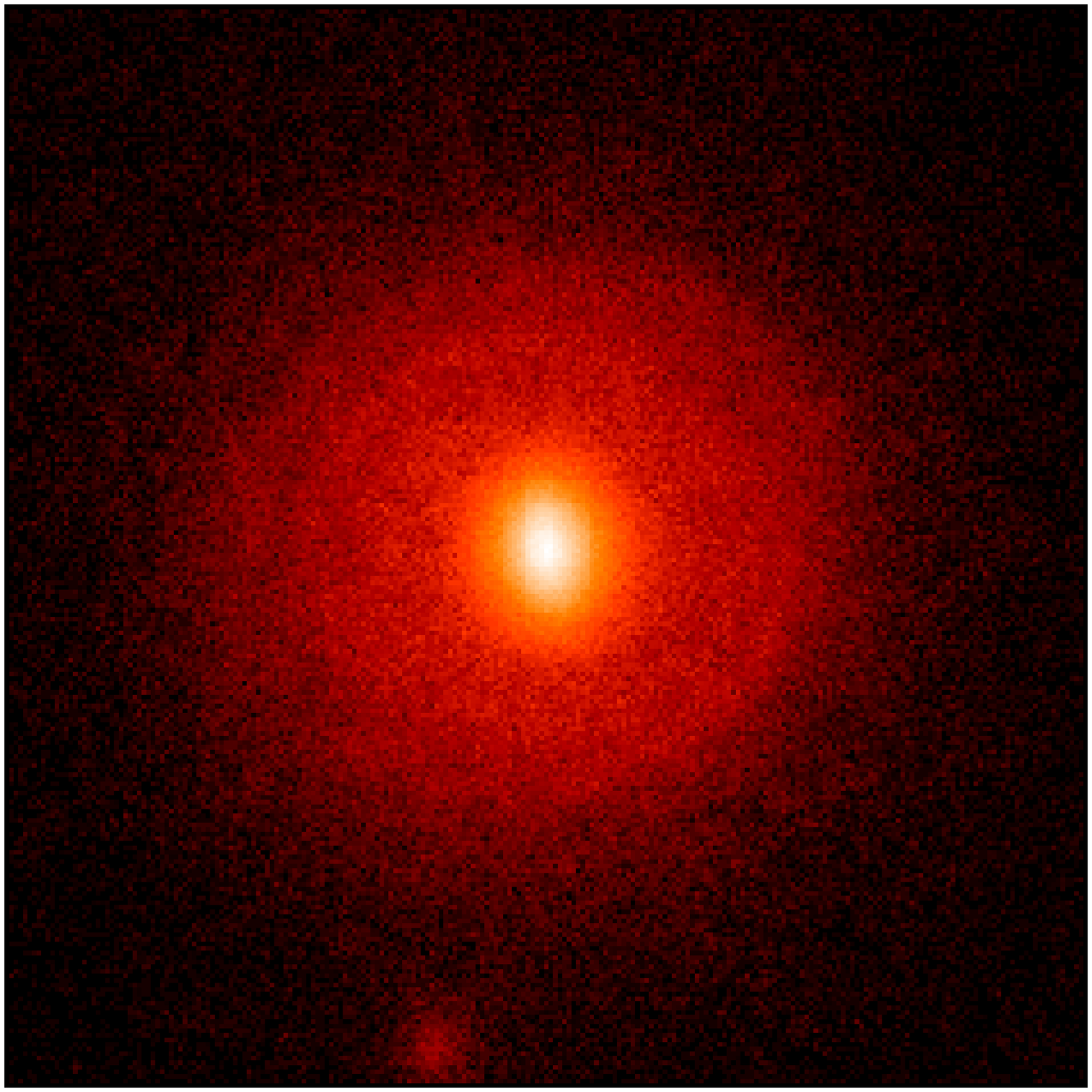}}
{\includegraphics[width=45mm]{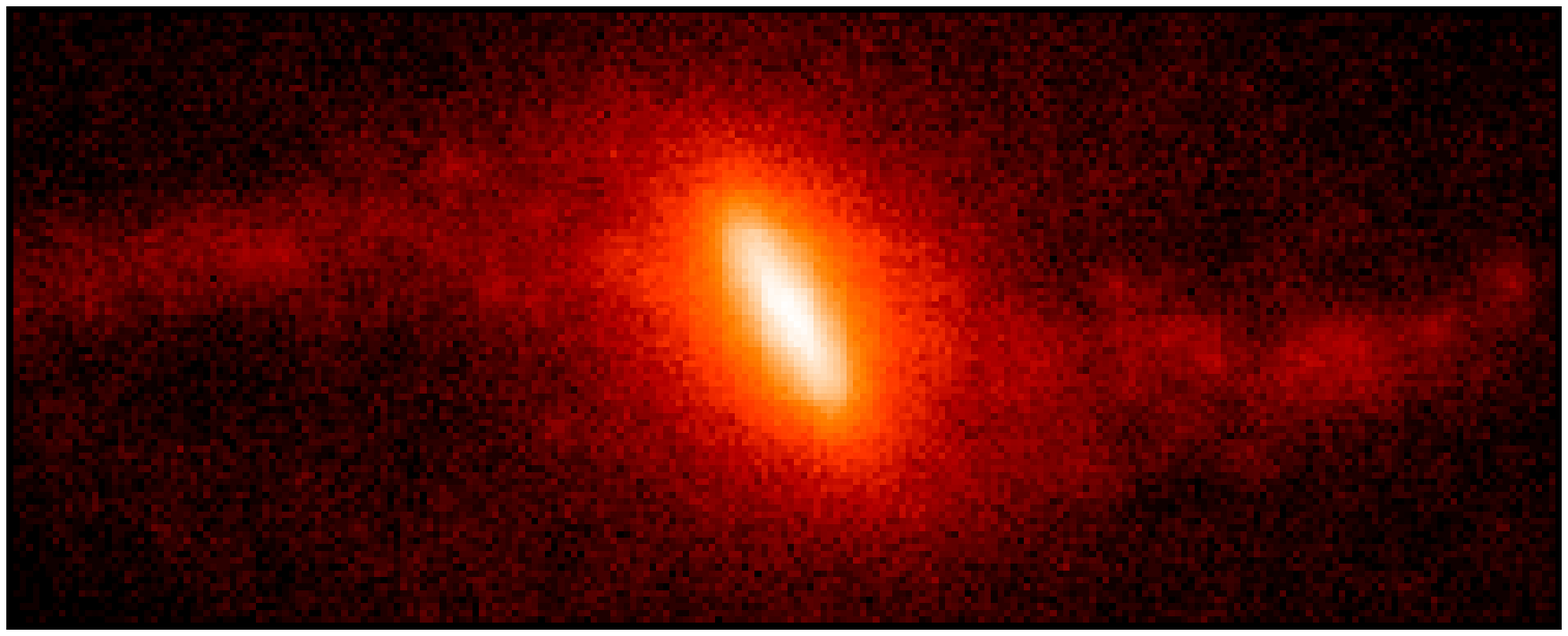}\includegraphics[width=45mm]{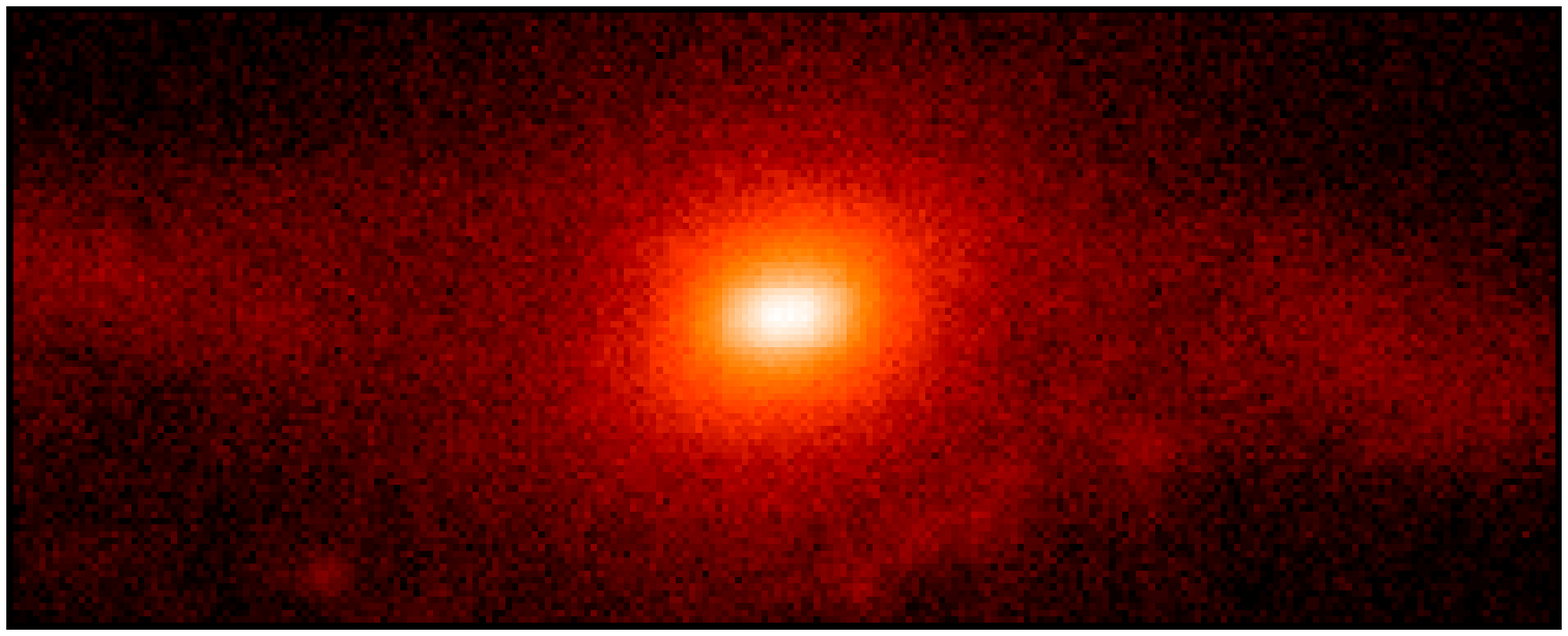}\includegraphics[width=45mm]{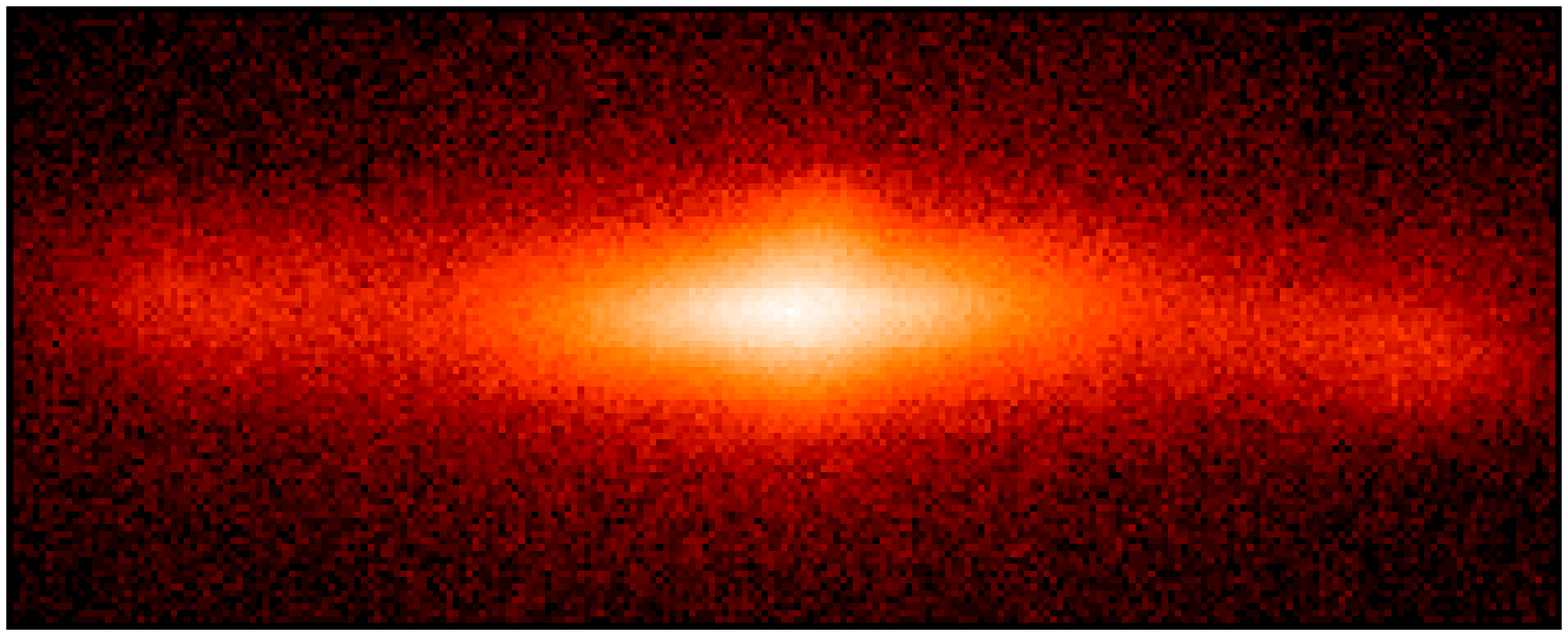}\includegraphics[width=45mm]{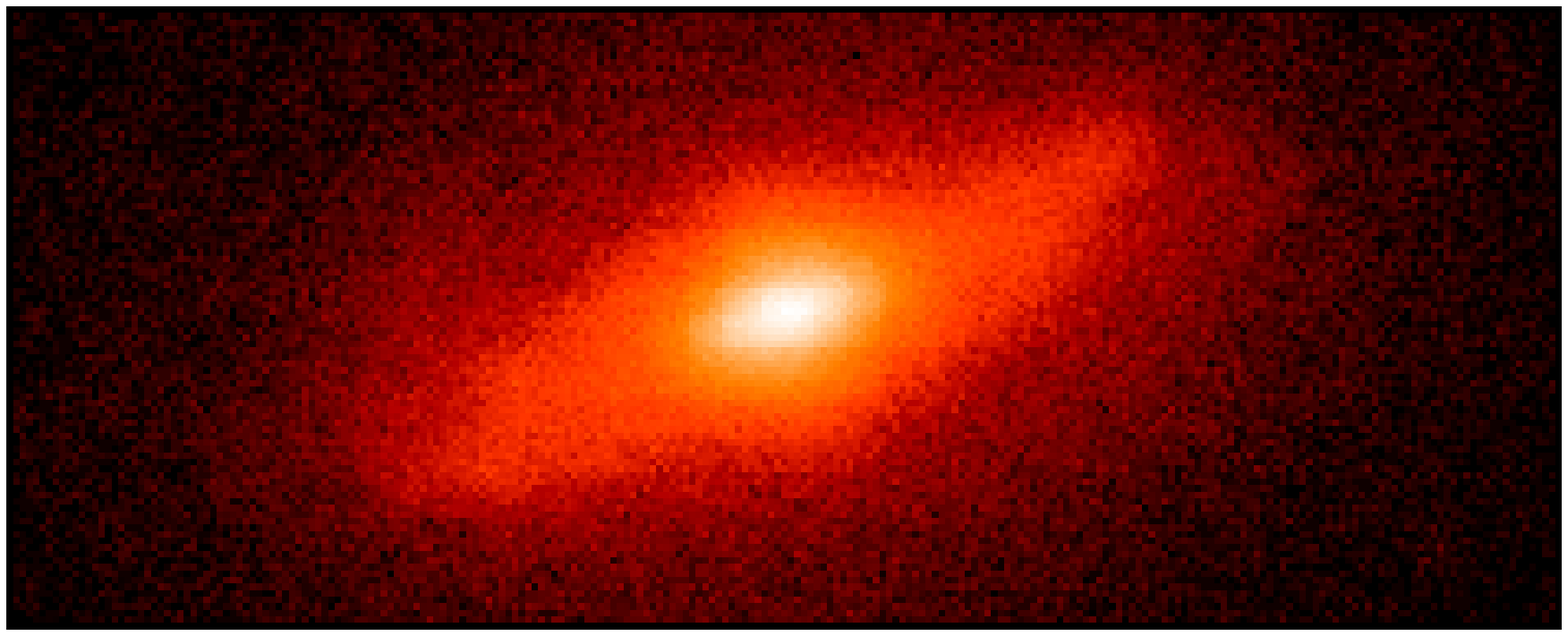}}

\includegraphics[width=180mm]{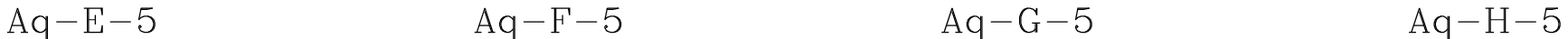}\vspace{-0.3cm}
{\includegraphics[width=45mm,angle=-90]{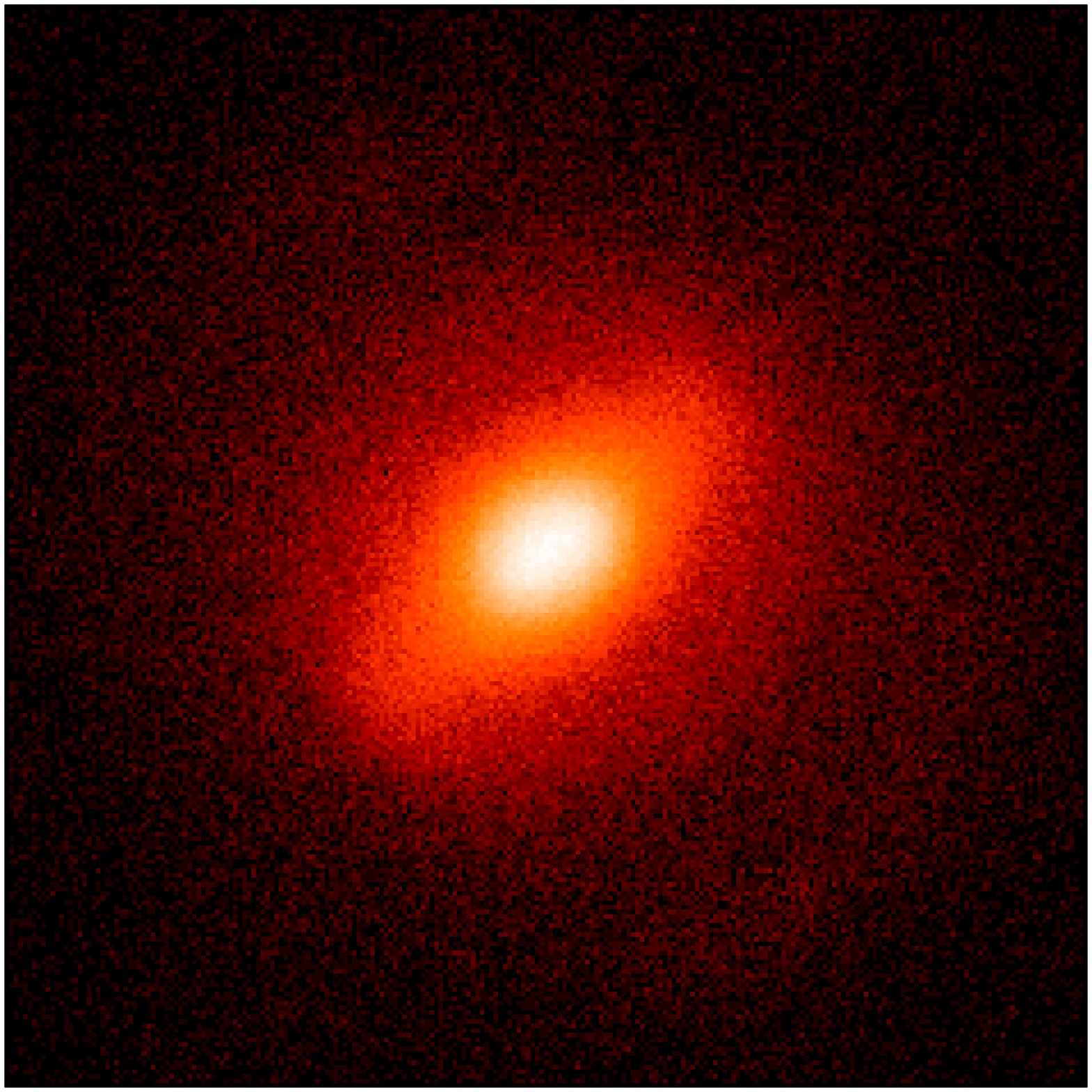}\includegraphics[width=45mm,angle=-90]{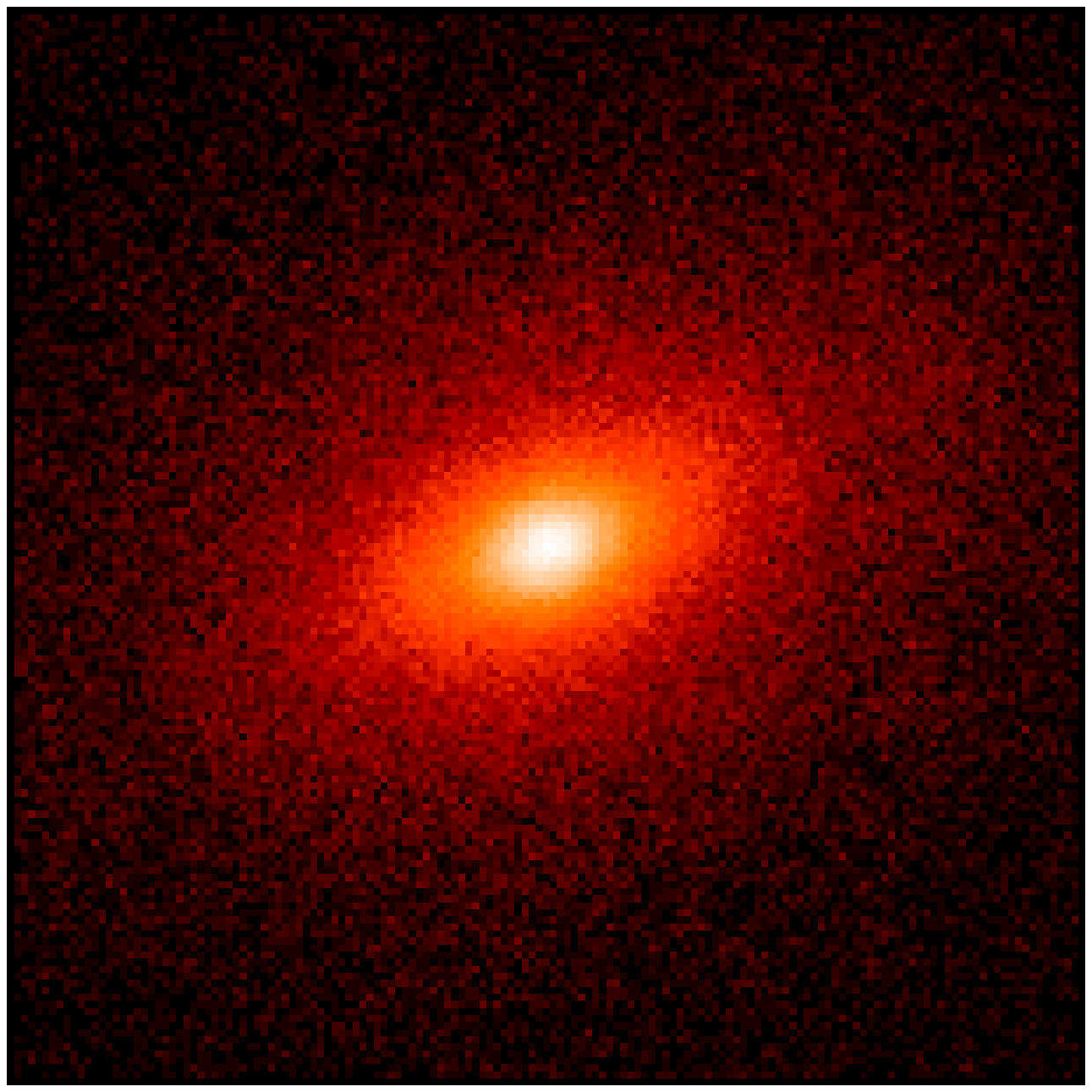}\includegraphics[width=45mm,angle=-90]{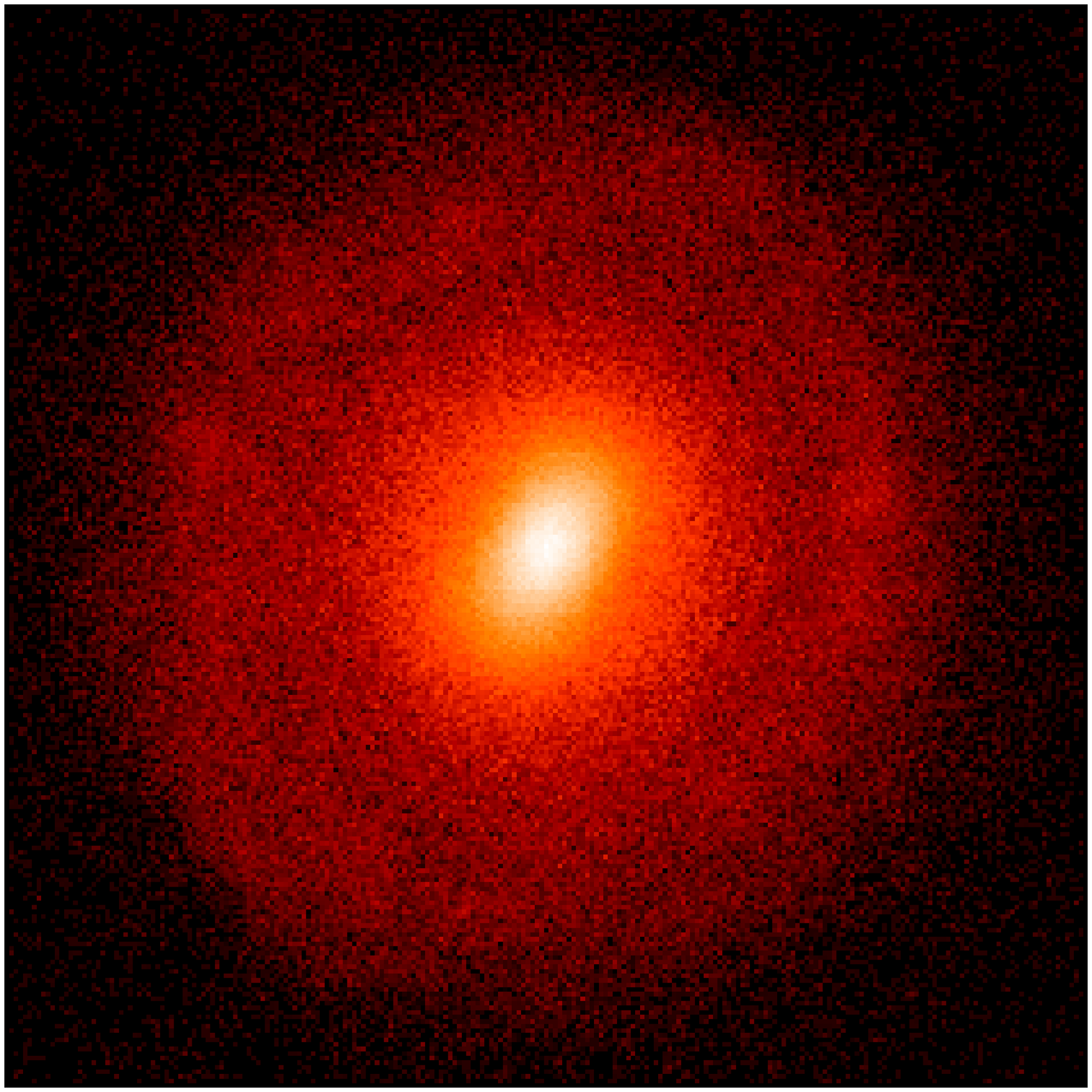}\includegraphics[width=45mm,angle=-90]{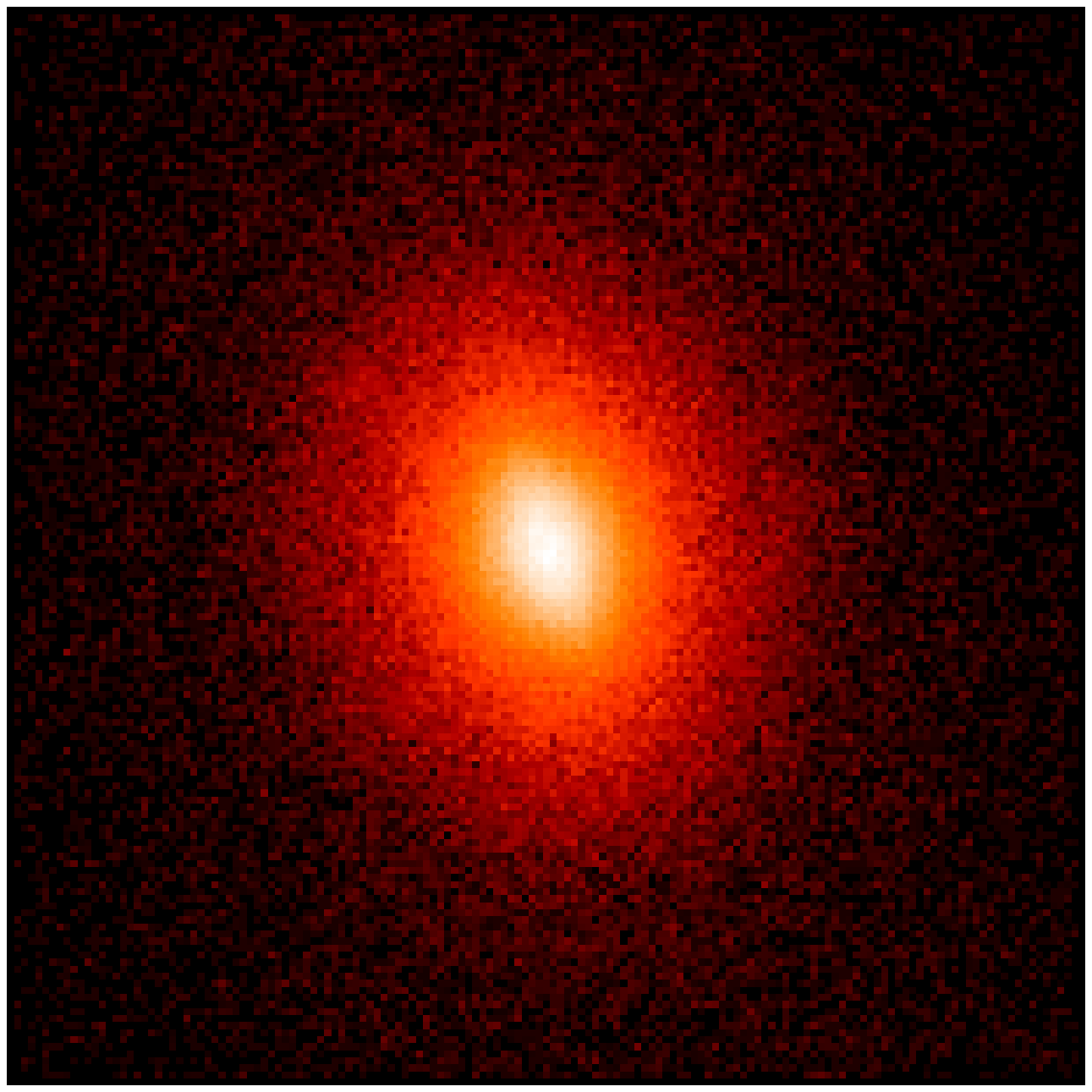}}
{\includegraphics[width=45mm]{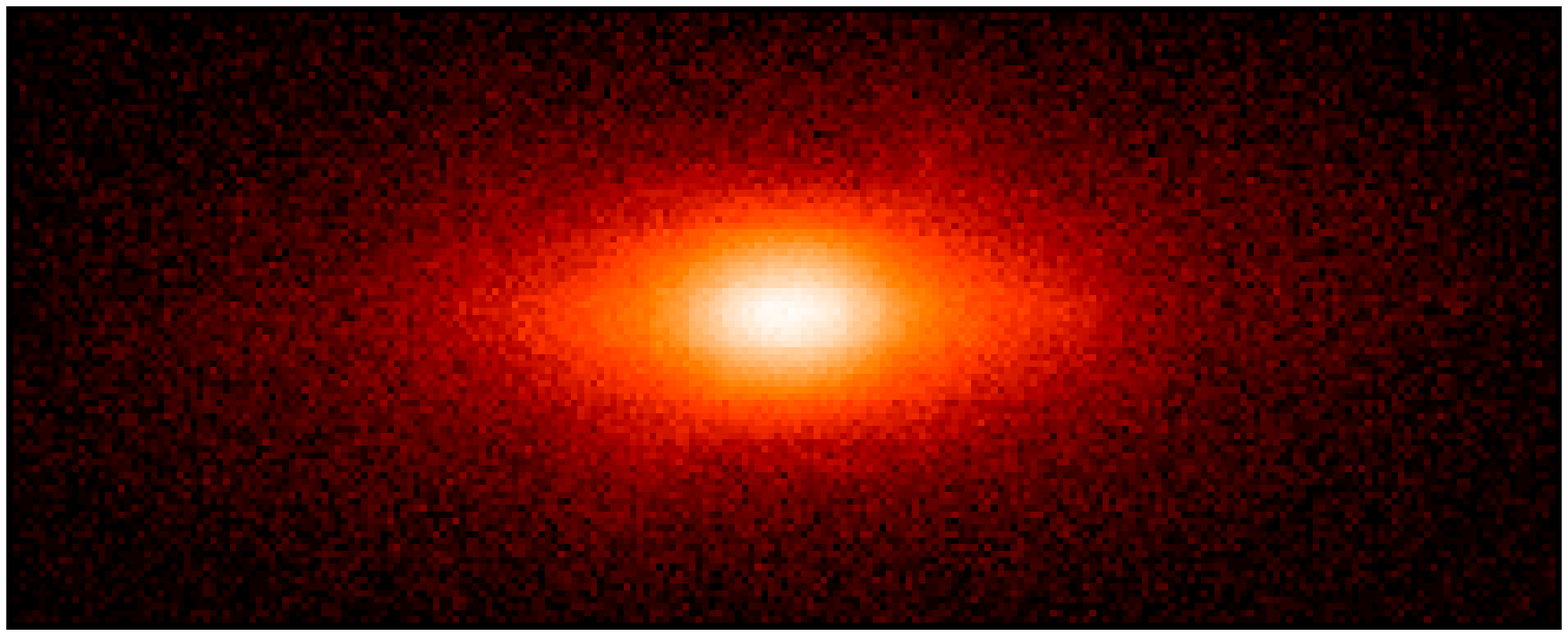}\includegraphics[width=45mm]{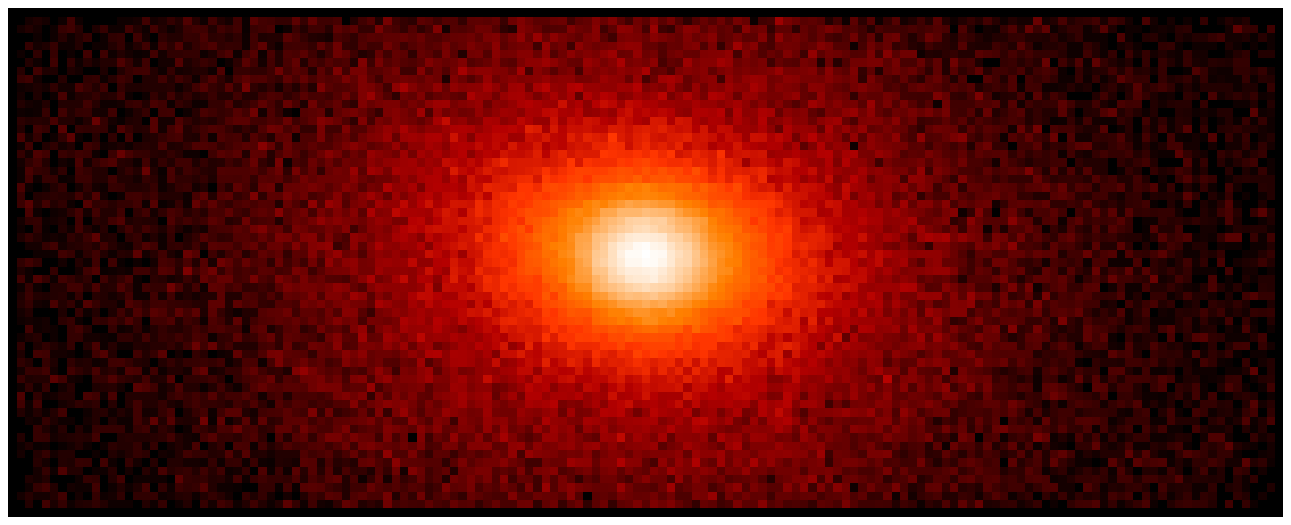}\includegraphics[width=45mm]{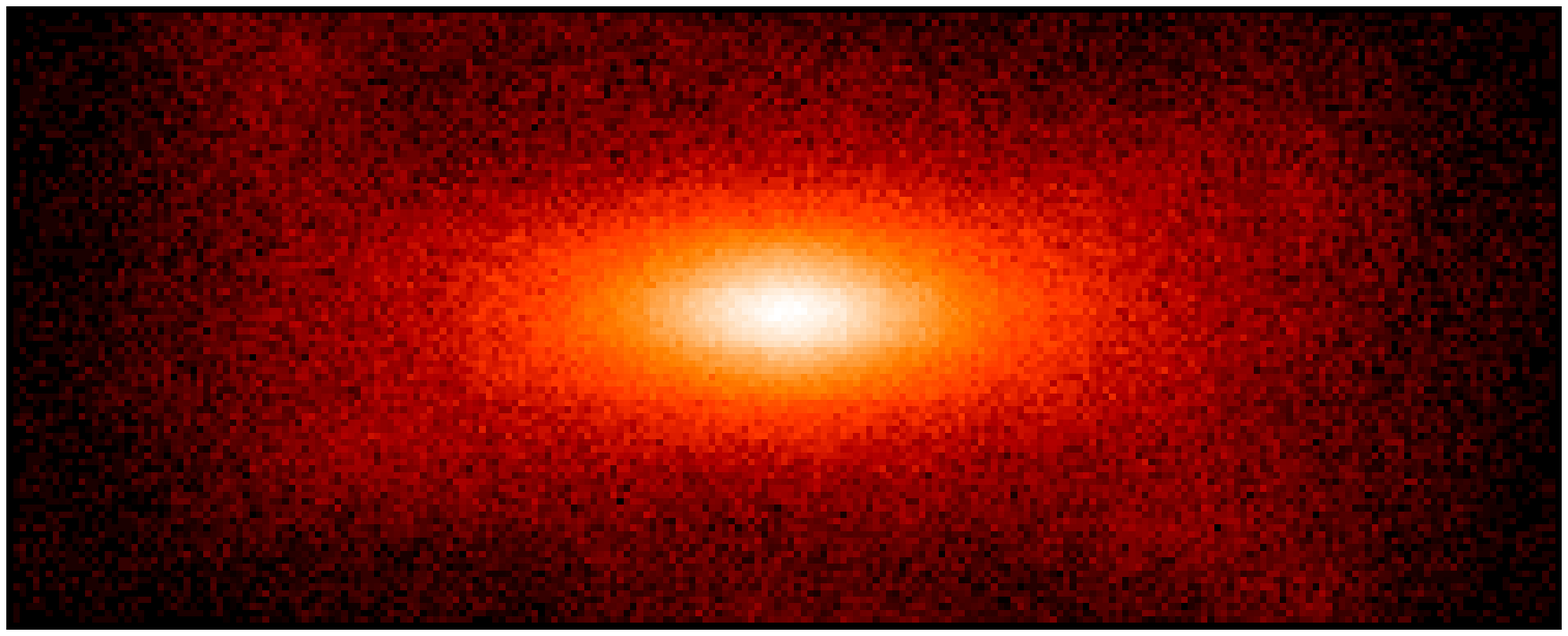}\includegraphics[width=45mm]{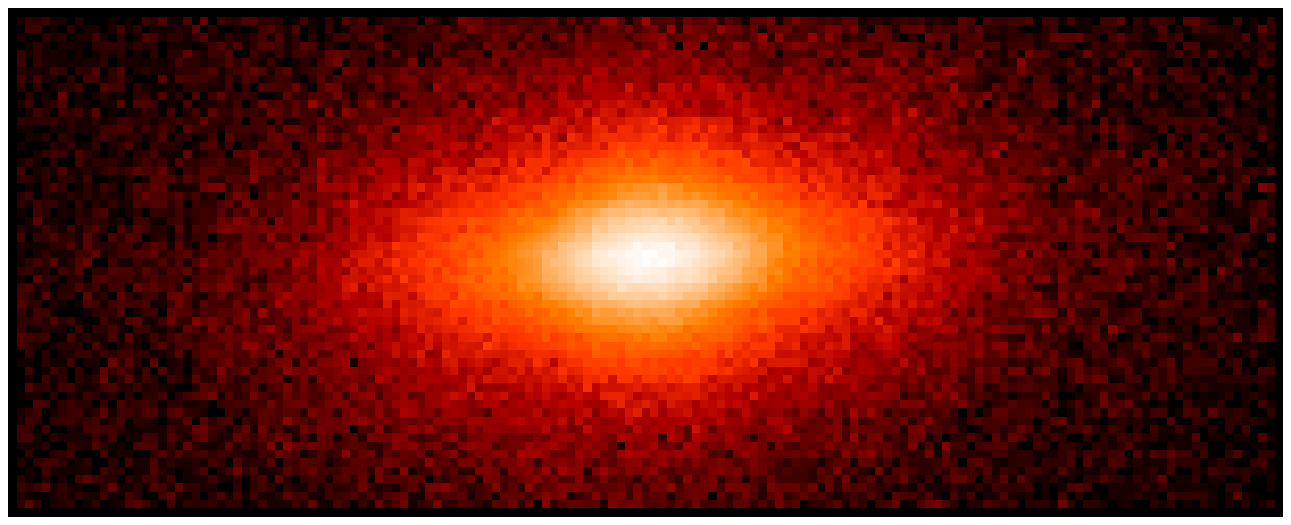}}

\caption{Face-on and edge-on maps of projected stellar luminosity
  ($i$-band) for our simulations, at $z=0$.  The images are $50$ kpc
  across, and the edge-on ones have a vertical height of $20$
  kpc. These plots correspond to the $XY$ and $YZ$ projections shown
  in S09, where projected mass density is shown instead of luminosity,
  and where the plotted scale is smaller than that used here.}
\label{maps_stars}
\end{figure*}

The simulated haloes host, at $z=0$, galaxies with a variety of 
morphologies (S09).  In Fig.~\ref{maps_stars}, we show face-on and
edge-on maps of projected stellar luminosity (in the $i$ band), for
the eight simulations.  In order to quantify the relative importance
of the disc and spheroidal components, we use the kinematic
disc/spheroid decomposition described in S09, which allows us to link
each simulated star to one of these components (we refer to S09 for
full details).  This method is based on the kinematic of stars, and
uses the parameter $\epsilon\equiv j_z/j_{\rm circ}$, where $j_z$ is
the angular momentum of each star in the $z$ direction (i.e. the
direction of the total baryonic angular momentum) and $j_{\rm circ}$
is the angular momentum corresponding to a circular orbit at the
position of the star. Values of $\epsilon$ similar to unity identify stars with
disc-like kinematics.  These are then tagged as disc stars, while the
remaining stars define the spheroidal component.

Using this decomposition technique, S09 divided simulated galaxies
into two groups according to the prominence of the disc with respect
to the spheroidal component: Aq-C-5, Aq-D-5, Aq-E-5 and Aq-G-5 have
significant disc components; while Aq-A-5, Aq-B-5, Aq-F-5 and Aq-H-5
have very small or no discs. Aq-F-5 is the only galaxy with pure
spheroid-like stellar kinematics and no sign of net stellar rotation.

\begin{figure*}
\includegraphics[width=170mm]{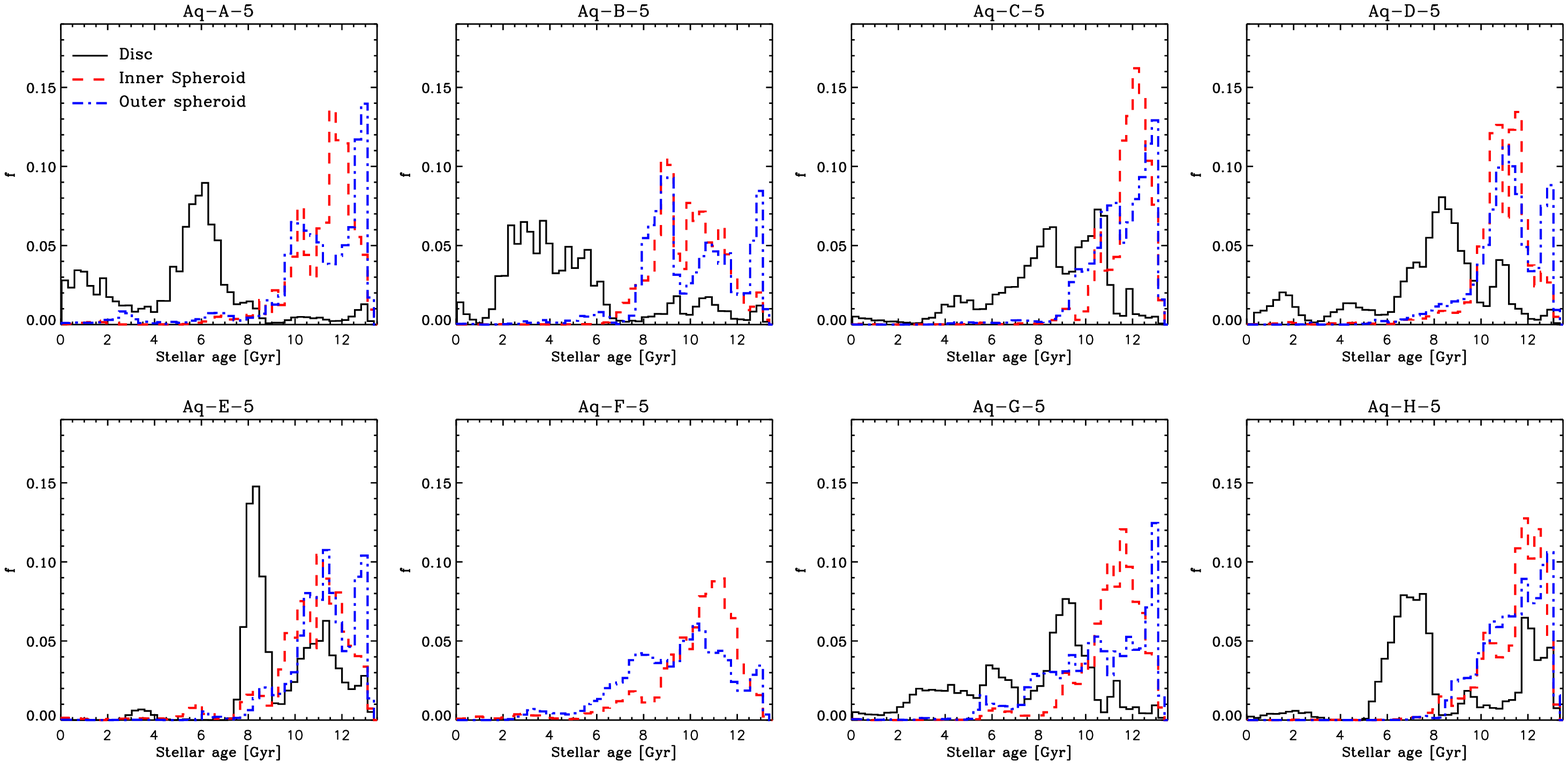}
\caption{Distribution of (mass-weighted) stellar ages for the disc (solid lines),
  inner spheroid (dashed lines) and outer spheroid (dotted-dashed lines) 
components in our simulations.
  For clarity, each component has been normalized to its total mass.}
\label{hist_stellarage}
\end{figure*}

An important advantage of the kinematic disc-spheroid decomposition is
that the disc component does not suffer from contamination by spheroid
stars (S09).  However, since spheroids are defined as those stars not
tagged as disc, their particles may be a superposition of distinct
dynamical structures. In particular, half of the simulated galaxies
have bars, which are seen not only by simple visual inspection, but
also when bar/disc/bulge decompositions are applied to synthetic
images (Scannapieco et al. 2010). As a result, bar particles are
classified as spheroidal particles by our decomposition.

We note that, as shown in S09, the (mass-weighted) disc-to-total
ratios obtained using the kinematic decomposition are of the order of
$0.2$, much lower than those of late-type spirals. However, when
applying techniques identical to those used in observations
(i.e. applying bulge/disc/bar decompositions to the 2D light
distribution), the disc-to-total ratios are significantly higher,
between $0.4$ and $0.7$ (Scannapieco et al. 2010). These are similar to
those estimated for early-type spirals but still substantially too
small to correspond to late types. Results for the disc-to-total
ratios obtained with the kinematic and photometric approaches are
given in Table~\ref{table_ages}.

Because spheroid stars extend all the way to the virial radius
(although the vast majority are in the central regions), we will often
subdivide the spheroids into ``inner'' ($r\le r_{\rm opt}$) and
``outer'' ($r> r_{\rm opt}$) components ($r_{\rm opt}$ is the radius
which encloses $83\%$ of the baryonic galaxy mass, see
Table~\ref{simulations_table}).  We adopt these names for two
reasons. First, because the ``inner'' region of the spheroid cannot,
in all cases, be thought as a pure bulge, since it may include a bar.
Secondly, the ``outer'' spheroid, although it can be thought as a
``stellar halo'', is not comparable to what is generally referred to
as the stellar halo in the Milky Way, since it contains stars which
lie well outside the solar radius (see the optical radii in
Table~\ref{simulations_table}).  The use of the optical radius to
separate the inner and outer spheroids is somewhat arbitrary; but 
since all properties (e.g. stellar age distributions, density
profiles) vary with radius in a smooth way such arbitrariness is unavoidable.
  In any case, none of our
conclusions is affected by this choice.

\subsection{Resolution effects}\label{resolution}

In Table~\ref{resolution_global} we compare the global properties of simulated
galaxies with lower resolution versions of Aq-C-5 (Aq-C-6)  and Aq-E-5 (Aq-E-6 and Aq-E-6b).
In Aq-C-6  and Aq-E-6, the mass  resolution is a factor of $\sim 8$ lower 
than in their level 5 resolution
counterparts, while  Aq-E-6b has a  $\sim 4$ times lower resolution than 
Aq-E-5. 
The Table shows the virial radii and virial masses, as well as 
the final gaseous and stellar masses within the virial
radius.

We find very good agreement for the virial radii and virial masses in all cases. 
The final stellar masses show larger variations, specially for the lowest resolution
simulations (Aq-C-6 and Aq-E-6), where
there is up to a $40$ percent change with respect to the level 5  resolution cases. 
In contrast, the stellar masses of Aq-E-6b and Aq-E-5 are very similar, with a
difference of the order of $10$ percent. 
For the gas masses we find large differences between Aq-C-6 and Aq-C-5 ($>60\%$), and 
small differences among the three Aq-E simulations ($<5\%$). 
Throughout the paper, we will use these additional simulations to investigate the effects
of resolution on the properties of simulated galaxies.
We shall see that differences are relatively small for the formation
time-scales and structural properties of discs and spheroids, while disc
 dynamical properties
are more strongly affected.

\begin{table} 
\begin{small}
\caption{Global  properties of the low resolution simulations 
at $z=0$. We show
  virial radius ($r_{200}$) and virial mass ($M_{200}$), 
as well as the stellar and
gaseous masses within the virial radius ($M_{\rm star}$ and $M_{\rm gas}$ respectively). }
\vspace{0.1cm}
\label{resolution_global}
\begin{center}
\begin{tabular}{lcccc}
\hline

Halo  & $r_{200}$  & $M_{200}$ & $M_{\rm star}$&$M_{\rm gas}$\\
    & [kpc]& [$10^{11}$M$_\odot$] &   [$10^{10}$M$_\odot$] &  [$10^{10}$M$_\odot$] \\\hline

Aq-C-5    &  237 &      16.1 &    10.8&   3.6 \\
Aq-C-6    &  230 &      15.1 &    6.6&    1.2 \\\\

Aq-E-5   & 206 &      10.8&     8.4&   2.6  \\
Aq-E-6b   & 206 &      10.7&     7.4&   2.7 \\
Aq-E-6   & 200 &      10.0&     4.7&   2.7  \\

\hline

\end{tabular}
\end{center}
\end{small}
\end{table}

In order to test the effects of
resolution on disc and spheroid properties, we
applied  our disc/spheroid decomposition technique to the low resolution
simulations. As found in other simulation studies, discs become more 
significant
as resolution is increased. 
 We find that the disc-to-total ratios for Aq-C-6 and Aq-E-6 are  $\sim 30\%$ lower
than for Aq-C-5 and Aq-E-5.
For Aq-E-6b, 
the D/T ratio is only $7\%$ smaller than in Aq-E-5 (see Table~\ref{table_ages}). 
We conclude that Aq-E-6b and Aq-E-5 have
converged reasonably well, also in terms of the disc/spheroid prominence, while Aq-C-6 and Aq-E-6
might still be affected by resolution effects.

\section{Formation time-scales of discs and spheroids}\label{sec_ages}

The various dynamical components of the Milky Way and external
galaxies are believed to have formed in different ways and on
different time-scales. In the context of $\Lambda$CDM, where mergers
are frequent with a rate that decreases with time, bulges are believed
to originate during the early epochs of galaxy assembly, while discs
are thought to form later during more quiescent periods, when galaxies
grow by smooth accretion rather than by mergers. Small bulges may also
form from bars through secular processes.

\begin{table*} 
\begin{small}
\caption{Mass-weighted ages of discs ($\tau_{\rm disc}$), inner
  spheroids ($\tau_{\rm inner\ sph}$) and outer spheroids ($\tau_{\rm
    outer\ sph}$).  Luminosity-weighted (SDSS $i$-band) ages for the
  three components are shown in parentheses.  We also show the typical
  formation time-scales for the three components, estimated as the
  standard deviation of the corresponding (mass-weighted) stellar age
  distributions.  All quantities are expressed in Gyr. The last two
  columns show the (mass-weighted) disc-to-total ratios calculated
  using the kinematic decomposition (D/T$^{\rm kin}$, from S09) and
  the (luminosity-weighted) D/T ratios obtained using an
  observationally based photometric approach (D/T$^{\rm phot}$, from
  Scannapieco et al. 2010). For Aq-F-5 and Aq-H-5, a reliable
  photometric decomposition was not possible (Scannapieco et
  al. 2010); these results should be interpreted with caution. }
\vspace{0.1cm}
\label{table_ages}
\begin{center}
\begin{tabular}{lcccccccc}
\hline
Galaxy  &  $\tau_{\rm disc}$  & $\tau_{\rm inner\  sph}$ & $\tau_{\rm outer\  sph}$ &  $\Delta\tau_{\rm disc}$  & $\Delta\tau_{\rm inner\  sph}$ & $\Delta\tau_{\rm outer\  sph}$ & D/T$^{\rm kin}$ & D/T$^{\rm phot}$\\\hline

Aq-A-5     &  5.7 (2.4)& 11.7 (11.7) & 11.4 (10.8) & 2.9 & 1.7 & 2.4 & 0.06 & 0.32\\

Aq-B-5    &   4.2 (3.3) & 10.0 \ (9.5)   &  9.8 (9.3) & 3.0 & 1.5 & 2.0 & 0.09 & 0.42\\

Aq-C-5   &  8.8 (7.7)&12.2 (12.2) &  11.9 (11.6) & 2.4 & 1.2 & 1.4 & 0.21 & 0.49\\

Aq-D-5&  8.4 (6.8) &11.1 (10.8) &  11.3 (11.2) & 2.8 & 1.5 & 1.5 & 0.20 & 0.68 \\

Aq-E-5 & 9.0 (8.7) &11.2 (10.6) &  11.5 (11.2) & 1.9 & 1.9 & 1.4 & 0.14 & 0.40 \\

Aq-F-5 &  -&10.8 (10.7) &  9.8 (9.2)\  & - & 2.1 & 2.2 & - & (0.44) \\

Aq-G-5  & 8.6 (6.5) &11.4 (11.3) &  10.8 (10.5) & 2.8 & 1.5 & 2.2 & 0.23 & 0.60\\

Aq-H-5&   7.6 (6.9) &11.9 (11.5) &   11.7 (11.2) & 2.9 & 1.3 & 1.3 & 0.04 & (0.05)\\

\\

Aq-C-6 & 9.3 & 12.3 & 12.4 & 2.6 & 1.0 & 1.0  & 0.15 & -\\

Aq-E-6b & 9.0 & 10.7 & 11.2 & 2.3 & 2.6 & 2.0 & 0.15 & - \\

Aq-E-6 & 10.8 & 11.0 & 11.7 & 1.7 & 2.4 & 1.7 & 0.10 & - \\

\hline
\end{tabular}
\end{center}
\end{small}
\end{table*}

Here we study the typical ages and formation time-scales of the discs,
inner and outer spheroids defined in section~\ref{decomp}.  Histograms
of stellar ages for these components are shown in
Fig.~\ref{hist_stellarage}.  As expected, we find that discs are
systematically younger than spheroids, and their typical formation
time-scales span several Gyrs.  On the other hand, both the inner and
outer spheroids show relatively narrow age distributions; in all
simulations the spheroidal components form very early and on very
short time-scales.  Moreover, we detect no systematic difference
between inner and outer spheroids, in terms of their stellar ages.
Interestingly, we find that, in the case of the discs, the stellar age
distributions are in some cases better described by a combination of
two or more populations of different ages -- this might be indicative
of two different components, such as a thick and a thin disc (see the next
section).

In Table~\ref{table_ages} we show, for the discs and the inner/outer
spheroids, the ``ages'' defined by the median of the corresponding
distributions.  Typical ages are in the range [$9$-$12$] Gyr for
spheroids and [$4$-$9$] Gyr for discs. In Table~\ref{table_ages}, we
also show the values of the standard deviation of the corresponding
stellar age distributions, which can be taken as an estimate of the
spread in formation times. As is evident from
Fig.~\ref{hist_stellarage}, discs form over larger time-scales, with
$\Delta\tau$ values between $2$ and $3$ Gyr.  In contrast, the
formation time-scales of spheroids are small, typically in the range
[$1-2$] Gyr.  Comparing the time-scales for inner and outer spheroids,
we find that the latter have systematically longer formation time
spreads. As we shall see later, outer spheroids have low {\it in-situ}
fractions which may explain their broader stellar age distributions.

Finally, we compare the ages obtained for our discs and spheroids
with observational results for the discs and bulges of spiral
galaxies. To this end, we calculate luminosity-weighted mean ages
(SDSS $i$-band) for our simulated discs, inner and outer spheroids,
using the Bruzual \& Charlot (2003) population synthesis models (for a
Salpeter initial mass function (IMF), as assumed in the
simulations). Results are listed in Table~\ref{table_ages}, in
parentheses. These luminosity-weighted estimates do not differ
significantly from the mass-weighted ones for the inner and outer
spheroids but, in the case of the discs, they can
be significantly lower than the mass-weighted ones, due to the younger
stars.  The mean ages of our discs and spheroids are in relatively
good agreement with observational results. Typical mean ages of discs
and bulges are [$4-12$] Gyr and $>8$ Gyr, respectively (MacArthur et
al. 2004, 2009).  However, we note that observational estimates can
vary significantly depending on the assumed star formation histories
and the modelling of dust. In addition, mean ages are found to depend on
Hubble type, such that earlier-type, faster-rotating and more luminous
galaxies are older (MacArthur et al. 2004).

The results found for our level 5 simulations are recovered in the
lower resolution runs. Discs are in all cases younger than spheroids, and we detect
no significant differences in the ages of inner and outer spheroids. 
The mean ages for disc, inner spheroids and outer spheroids  converge
very well (see Table~\ref{table_ages}), with
differences of the order of $5\%$  with respect to the higher resolution
simulations (except for Aq-C-6, where differences reach $20\%$).
Larger differences are detected for the time-scales, ranging from $10\%$ 
to $40\%$.

\subsection{Disc mass and disc age}

As discussed above, the ages of simulated discs are in relatively good
agreement with observational results.  However, observations reveal a
great variety of disc-bulge systems and the estimation of their ages
still suffers from large uncertainties. As a result, the scatter in
disc and bulge ages obtained observationally is large, and ages cover
almost the whole possible range between 2 Gyr and the Hubble
time. Theoretically, it is however expected that discs grow
significantly at low redshifts, because they are easily disrupted and they
have higher survival probability if they form late, during more
quiescent evolutionary periods.

In this context, our simulated discs are perhaps older than expected
-- or, alternatively, not massive enough, while bulges are too
massive.  As shown in Table~\ref{table_ages}, the disc-to-total ratios
are of the order of $0.2$ (kinematically-defined) or $0.4-0.7$
(photometrically-defined).  Recent estimations for the mass of the
bulge and the disc of our Milky Way are $\sim 2\times 10^{10}$
M$_\odot$ and $\sim 6\times 10^{10}$ M$_\odot$, respectively (Sofue,
Honma \& Omodaka 2009).  This translates into a disc-to-total mass
ratio of $0.75$ (ignoring the mass in the stellar halo which anyway
contributes very little to the total mass).  In simulations, the
formation of overly massive bulges and less massive discs might be due to
a number of causes that have been already discussed in the
literature: inappropriate modelling of the involved physical processes
(e.g. Piontek \& Steinmetz 2010; Agertz, Teyssier \& Moore 2010),
insufficient resolution (e.g. Governato et al. 2007), or even a
failure in the cosmological model (e.g.  Sommer-Larsen \& Dolgov 2001;
Mayer, Governato \& Kaufmann 2008).  In particular, Agertz et
al. (2010) claim that the inability to form massive discs is due to
the adoption of strong feedback combined with high star formation
efficiencies.
Note however that although their simulations produce massive
discs, they convert baryons into stars with an overall efficiency
which is much too high to be consistent with a $\Lambda$CDM cosmology 
(e.g. Guo et al. 2010).

\begin{figure*}
\includegraphics[width=170mm]{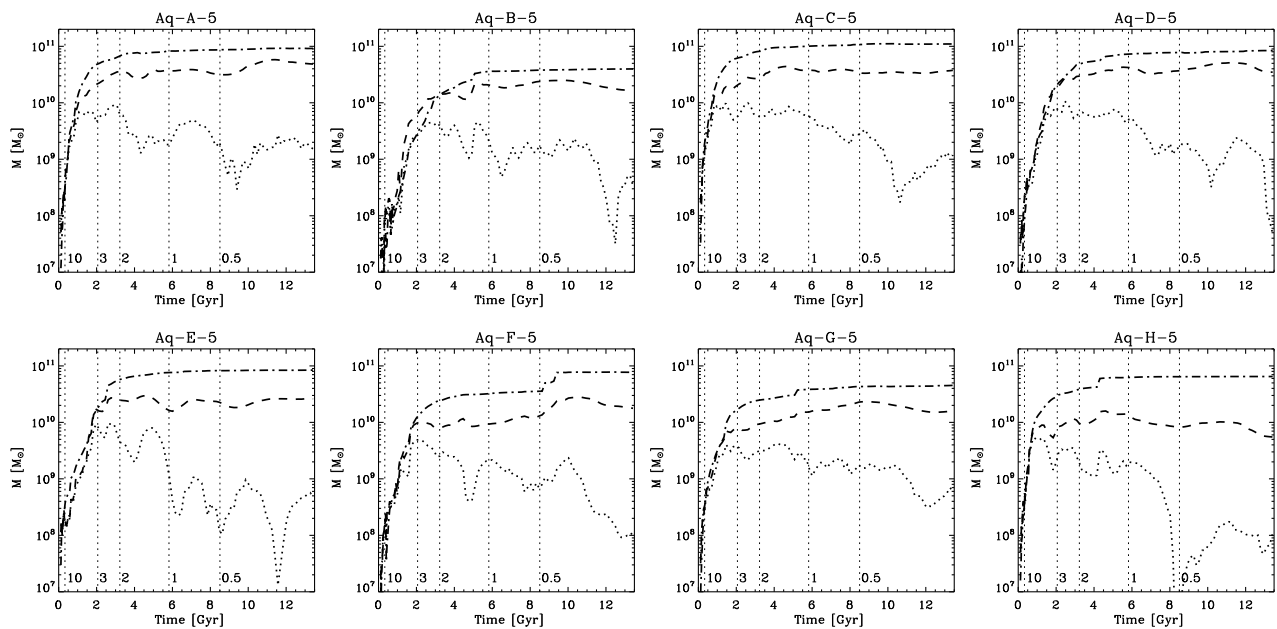}
\caption{Evolution of the baryonic mass within $r_{200}$ in the
  simulated haloes. Dot-dashed, dashed and dotted lines correspond to
  stars, hot (non-star forming) gas, and cold (star-forming) gas,
  respectively. The vertical dotted lines indicate a number of
  reference redshifts, from $z=10$ to $z=0.5$.}
\label{baryonic_mass_evol}
\end{figure*}

In models where supernova feedback is the main regulating mechanism of
star formation, the early formation of very massive bulges may be responsible
of the inability to form  massive discs later on. This
is because feedback at high redshift drives a wind from
the inner regions where star formation is taking place, leaving less
gas available for late star formation.  This occurs in our model, and
is illustrated in Fig.~\ref{baryonic_mass_evol}: the stellar masses
(dotted-dashed lines) grow rapidly at early times, leaving little cold
gas (dotted lines) at late times.  Most of the gas (gas masses are
typically $1.5-4$ times lower than the stellar masses) is hot (dashed
lines), and forms gaseous haloes with typical temperatures of $10^7$
K. Much gas is also expelled from our haloes, as
reflected in the baryon fractions we find, 
which are in the range $0.07$ and $0.10$, substantially smaller than the cosmic
baryon fraction assumed (Table~\ref{simulations_table}).

Finally, another possible reason for the generally low SFRs at low
redshifts is that our model does not include the effects of gas return
from low- and intermediate-mass stars 
(Tinsley 1974).  The gas return fraction of a stellar population of
given age over the Hubble time can be as high as $40-50\%$, depending
on the IMF and, less strongly, on the metallicity (Jungwiert, Combes
\& Palou\v{s} 2001).  This gas can in principle be added to the disc
and form new stars. 
Thus, up to about half of the stellar mass forming
at redshift $z\sim 2$ may be returned in the form of gas by
$z=0$ and be available to make new disc stars
(Martig \& Bournaud 2009).  However, if gas return comes from
stars in the bulge, this effect cannot build extended 
discs, since such stars have low angular momentum.

\section{The discs}\label{sec_disk}

\subsection{Structure}\label{structure_disks}

\begin{figure*}
\begin{center}

\includegraphics[width=85mm]{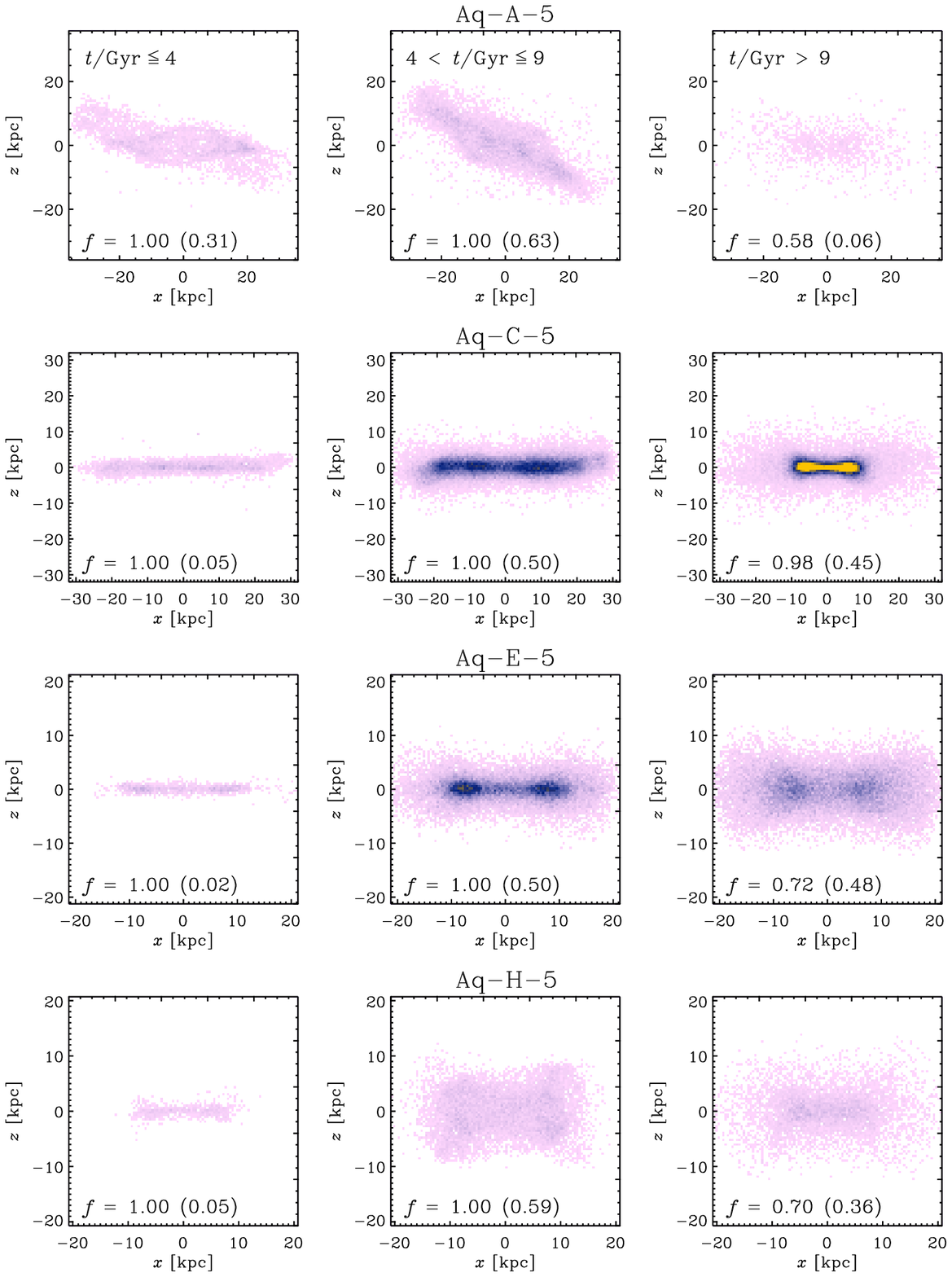}\hspace{0.5cm}\includegraphics[width=85mm]{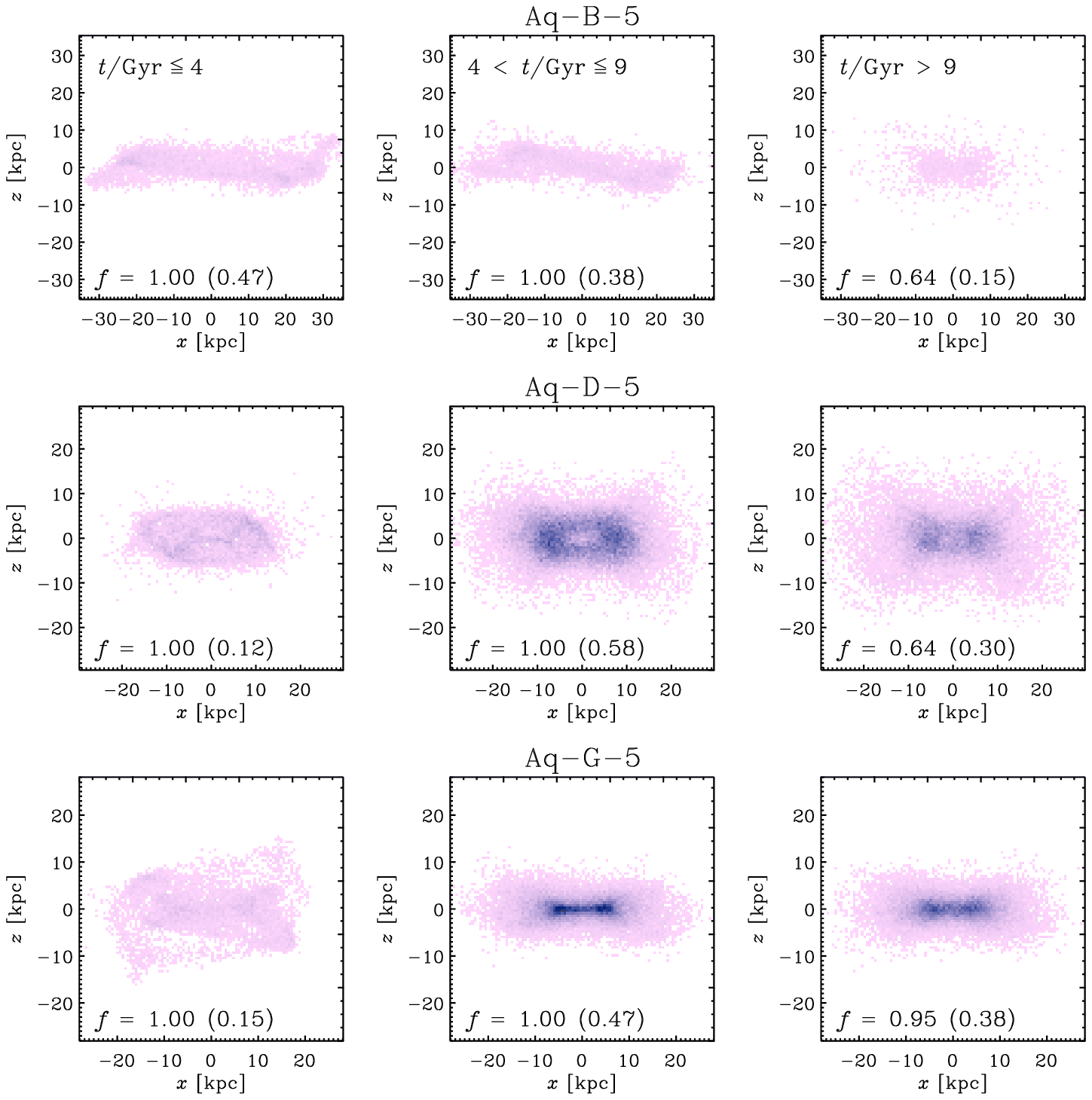}
\end{center}
\caption{Spatial distribution of (kinematically defined) disc
  particles (up to twice the corresponding optical radius) for the
  simulations with a disc component.  For each simulation, we show
  separately the distribution of disc stars for three stellar age
  bins: young stars ($t\le 4$ Gyr), intermediate age stars ($4$ Gyr $<
  t \le 9$ Gyr), and old stars ($t > 4$ Gyr).  Colours represent
  surface mass density, and cover $4$ orders of magnitude ($10^4-10^8$
  M$_\odot$ kpc$^{-2}$).  We also show the {\it in-situ} fraction of
  disc stars for the three stellar age bins, together with the
  corresponding disc mass fraction (in parentheses).}
\label{disk_structure}
\end{figure*}

As already shown in Fig.~\ref{maps_stars}, our eight simulated
galaxies have a variety of morphologies and, in particular, their discs
show a great variety of shapes and masses. Moreover, the distributions
of disc stellar ages (Fig.~\ref{hist_stellarage}) are in general
characterized by a superposition of bursts of different
age.  In order to investigate in more detail the structure of discs
at $z=0$, we show in Fig.~\ref{disk_structure} the distribution of
disc particles, in an edge-on view (note
that Aq-F-5 has no disc), up to twice the corresponding optical
radius.  We plot separately stars formed in three different age
bins\footnote{For simplicity, we adopt the same age bins for all
  galaxies, although the valleys in SFR are slightly different
  in each case. Using individual age bins for the different galaxies,
  more consistent with their particular distributions, does not alter
  our results in any significant way.}: $t/{\rm Gyr}\le 4$, $4 <
t/{\rm Gyr} \le 9$ and $t/{\rm Gyr}>9$.  The plots are color-coded
according to the surface mass density at each point; covering 4 orders
of magnitude ($10^4-10^8$ M$_\odot$kpc$^{-2}$).

The diversity of discs is evident: we find not only very thin discs
(Aq-C-5, Aq-E-5), but also a variety of shapes: ``boxy'' or
``X''-shaped discs (Aq-D-5, Aq-G-5, Aq-H-5), warps (Aq-B-5), and a
case where two misaligned discs of different age are present
(Aq-A-5). This diversity arises naturally in the context of
$\Lambda$CDM due to the different formation, merger and accretion
histories of galaxies.

\begin{figure*}
\begin{center}
\includegraphics[width=170mm]{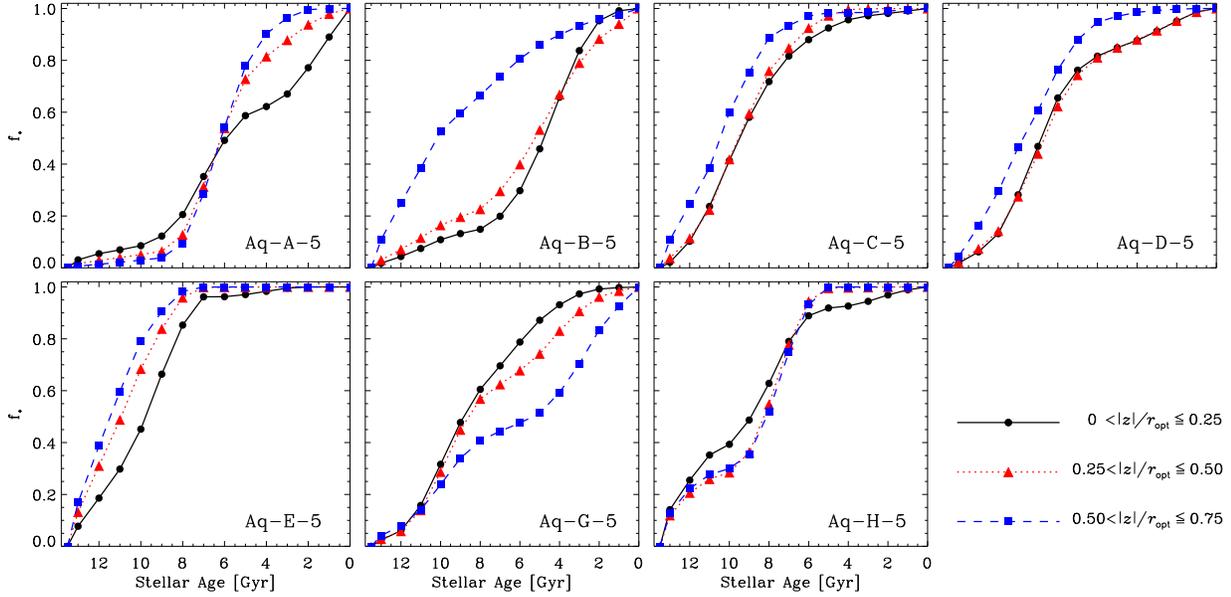}
\end{center}
\caption{Cumulative fraction of stellar mass formed as a function of
  stellar age for the kinematically-defined discs (note that Aq-F-5
  has no disc component), and for three bins in height above the disc
  plane, as indicated.}
\label{mean_z}
\end{figure*}

From Fig.~\ref{disk_structure} we can also observe that stars in the
youngest age bin (left-hand panels) tend to define thinner discs than 
older populations. In Fig.~\ref{mean_z} we show the
cumulative fraction of stellar mass formed $f_*$ as a function of
stellar age for stars located in three different bins of (absolute
value of) height above the disc plane: $|z|= [0-0.25]$, $(0.25-0.5]$
  and $(0.5-0.75]$ times the corresponding optical radius. In general,
    we find that stars in the lower $|z|$ bins, i.e. closer to the
    disc plane, are younger than stars at larger vertical distances.
    This behaviour is more clear in those galaxies with significant
    disc components (Aq-C-5, Aq-D-5, Aq-E-5). Aq-G-5 has also an
    important disc component, but we detect the opposite trend:
    further above the disc plane we find younger stars.  It is
    interesting to note that this galaxy has a strong bar and mixing
    might play an important role in the distribution of young stars.

Galaxies with less prominent discs also show a trend of younger stars
being closer to the disc plane, however, the behaviour is more
diverse. In Aq-A-5 and Aq-H-5, the dependence of vertical extent on
 stellar age is not monotonic.  On the contrary, in Aq-B-5 we detect a very well
defined relation between stellar age and thickness. 
We also note that in Aq-A-5 the results might be influenced by
the fact that the young stars define a second stellar disc, misaligned
with the older disc component.  The projection we adopt is such that
the older disc is contained in the $xy$ plane, while the younger disc
is not.  In fact, if projected with respect to the latter, the young
disc appears very thin.

\begin{figure*}
\begin{center}
\includegraphics[width=170mm]{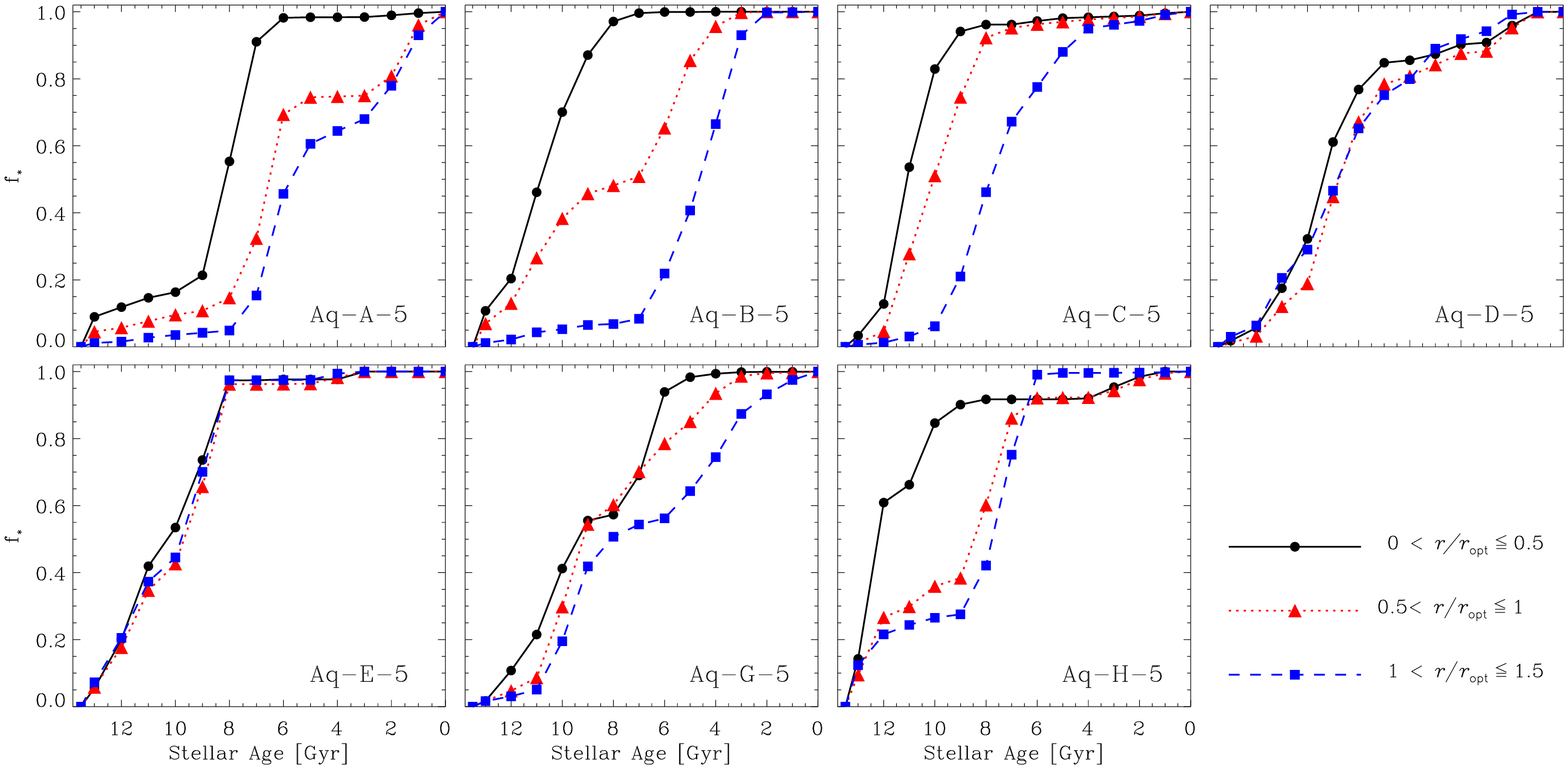}
\end{center}
\caption{Cumulative fraction of stellar mass formed as a function of
  stellar age for the kinematically-defined discs (note that Aq-F-5
  has no disc component), and for three bins in cylindrical radius as
  indicated. For radii larger than $1.5\times r_{\rm opt}$, the relation
  reverses.}
\label{cum_Mstar}
\end{figure*}

Furthermore, we find that the youngest stars define more extended
discs. In Fig.~\ref{cum_Mstar} we show the cumulative fraction of stellar mass
$f_*$ as a function of stellar age, now for disc stars in three
different radial bins according to their present-day positions: $r\le
0.5\,r_{\rm opt}$, $0.5\,r_{\rm opt}<r\le r_{\rm opt}$ and $r_{\rm
  opt}<r\le 1.5\,r_{\rm opt}$.  The behaviour of the cumulative
stellar mass fraction demonstrates an inside-out disc formation
pattern for most simulated discs.  While the innermost regions are
populated with stars  older than $8$ Gyr, disc stars in
the outermost regions are significantly younger.  There are two cases,
Aq-D-5 and Aq-E-5, for which we detect no significant difference
between the three radial bins.

Our findings clearly show that discs in our simulations have
preferentially formed from the inside-out. This is reminiscent of
recent observational results, which also show evidence for inside-out disc
formation in M33 (Barker et al. 2007; Williams et al. 2009) and in
local HI-rich galaxies (Wang et al. 2010).  Finally, we note that, as
already shown in S09 and Scannapieco et al. (2010), simulated discs
have half-mass radii that are consistent with observational results 
(Table~\ref{disk_bulge}).

The results presented thus far evidently show that simulated discs are
better described by a combination of two (or perhaps more) components,
reminiscent of thin and thick discs. We shall see later that these two
components also have different dynamical properties.  Assuming that we
can divide disc stars into thin and thick components depending on
their age (we use $t\le 9$ Gyr and $t> 9$ Gyr, respectively), we can
estimate the relative masses of the simulated thin and thick discs.
We adopt this criterion because it is easy to apply to simulated
stars, and follows the observation that, in our Milky Way, the bulk of
stars in the thick disc are older than $10-12$ Gyr while the thin disc
has a low contribution of stars older than $8$ Gyr (Norstr\"om et
al. 04; Bensby et al. 2007).  Using this definition, we find that
thick discs contribute between $6\%$ and $45\%$ to the total disc mass
(as can be inferred from the mass fractions 
shown in parentheses in Fig.~\ref{disk_structure}).  These
results are in relatively good agreement with observational results:
in our Milky Way, the contribution of the thick component to the total
disc mass is found to be $\sim 23\%$ (Juri\'c et al. 2008), while for
thick discs in external galaxies values of the order of $20-40\%$ are
found (Yoachim \& Dalcanton 2006), depending on the galaxy stellar
mass. As emphasized by these authors, thick discs of external galaxies
are very diverse.

The results discussed in this section change little in our lower
 resolution simulations. We again find that disc populations with
different stellar ages have distinct properties, and the trends with stellar
age found in the level 5  simulations are recovered. However, we note that
the youngest disc components are poorly resolved in the low
resolution simulations. In particular, the number of stars younger
than $4$ Gyr is significantly ($\sim 60-90\%$) smaller, 
and it is thus not possible to estimate  reliably the properties of the
youngest 
subcomponents. However, we find that the relative fraction of disc 
stars in the different age bins is very similar, suggesting that resolution does not
strongly affect our results.

\begin{table} 
\begin{small}
\caption{{\it In-situ} fractions ($f$) for simulated discs, inner spheroids and outer spheroids.
}
\vspace{0.1cm}
\label{table_insitu}
\begin{center}
\begin{tabular}{lccc}
\hline
Galaxy  &   $f_{\rm disc}$ & $f_{\rm inner\ sph}$&$f_{\rm outer\ sph}$   \\\hline

Aq-A-5    &  0.98 &  0.93 &  0.35 \\

Aq-B-5    & 0.95 & 0.85 &  0.19\\

Aq-C-5   &  0.99  & 0.90 &  0.22\\

Aq-D-5   &  0.89   & 0.87 &  0.22\\

Aq-E-5   &  0.86 &  0.73 &   0.28\\

Aq-F-5   & -     & 0.54  & 0.14 \\

Aq-G-5  &  0.98 &  0.78 & 0.31\\

Aq-H-5   & 0.88 & 0.79 &  0.15\\

\\

Aq-C-6 & 0.94  & 0.91 & 0.13 \\

Aq-E-6b & 0.90 & 0.74 & 0.31 \\

Aq-E-6 & 0.75 & 0.76 & 0.27 \\

\hline
\end{tabular}
\end{center}
\end{small}
\end{table}

\subsection{In-situ fractions}\label{in_situ_disks}

\begin{figure*}
\begin{center}
{\includegraphics[width=90mm]{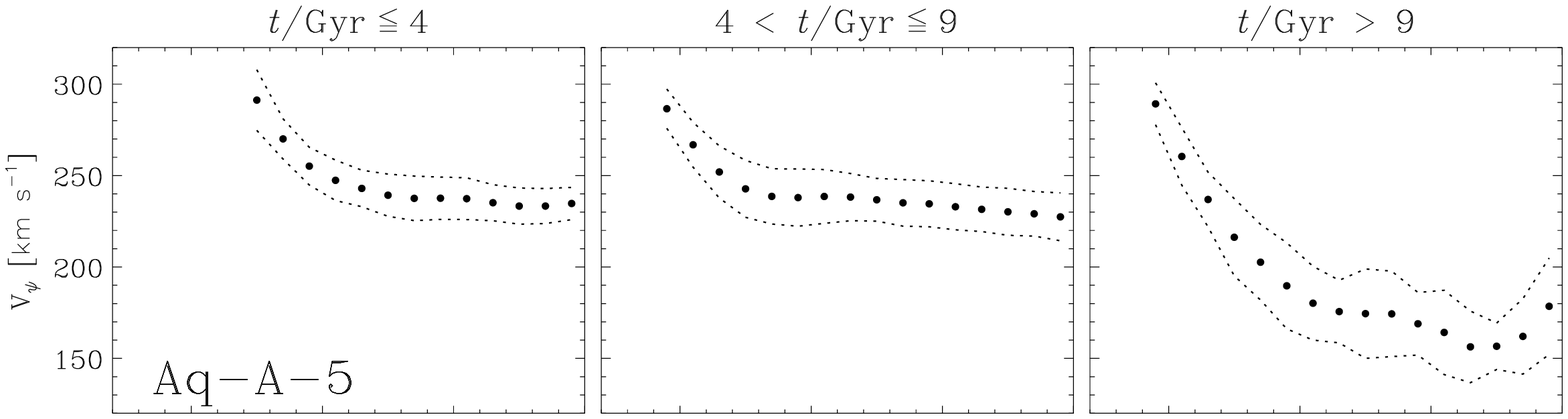}\includegraphics[width=90mm]{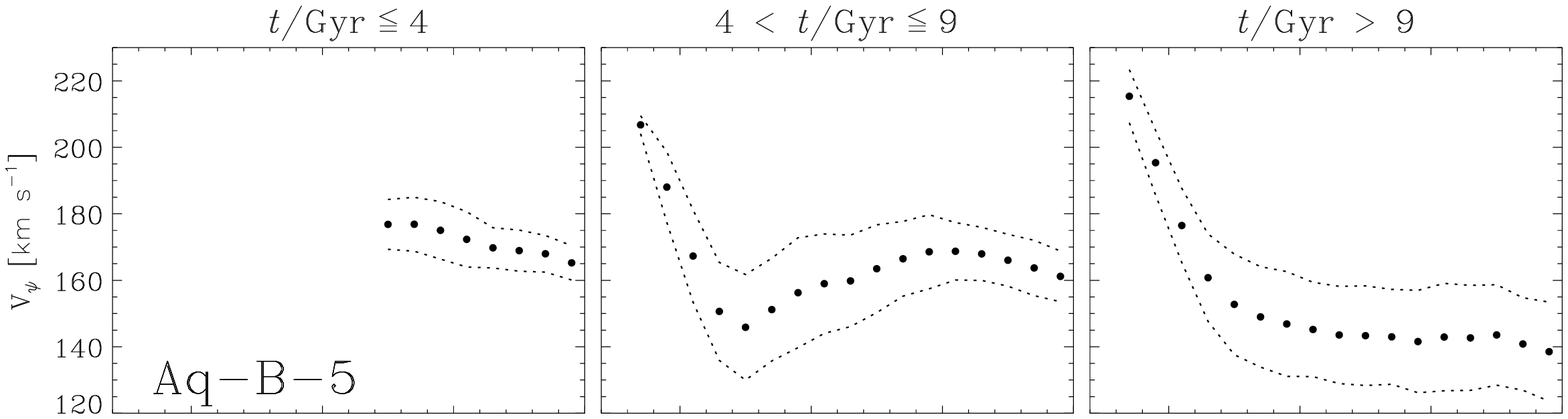}}

\vspace{-0.4cm}
{\includegraphics[width=90mm]{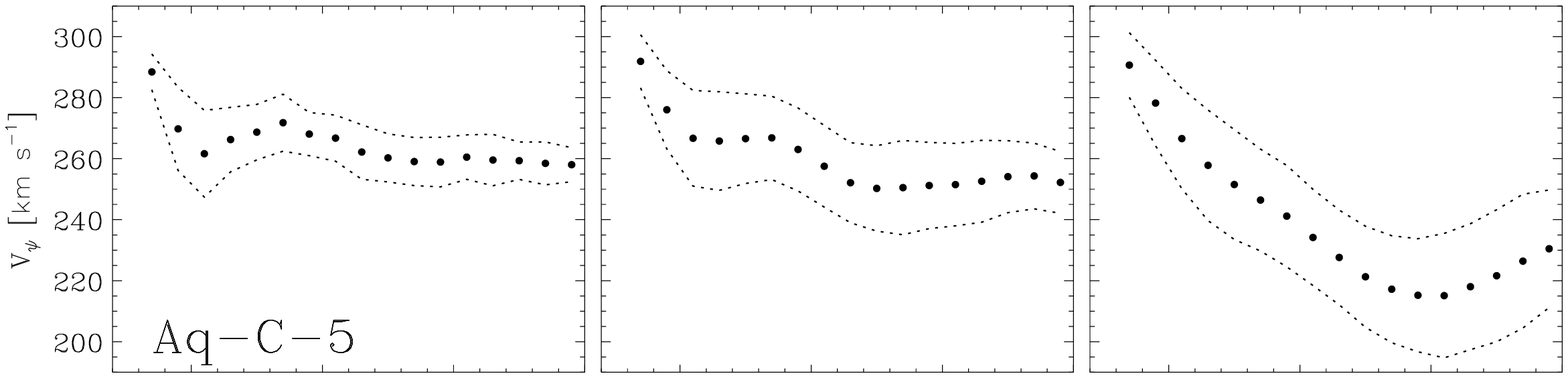}\includegraphics[width=90mm]{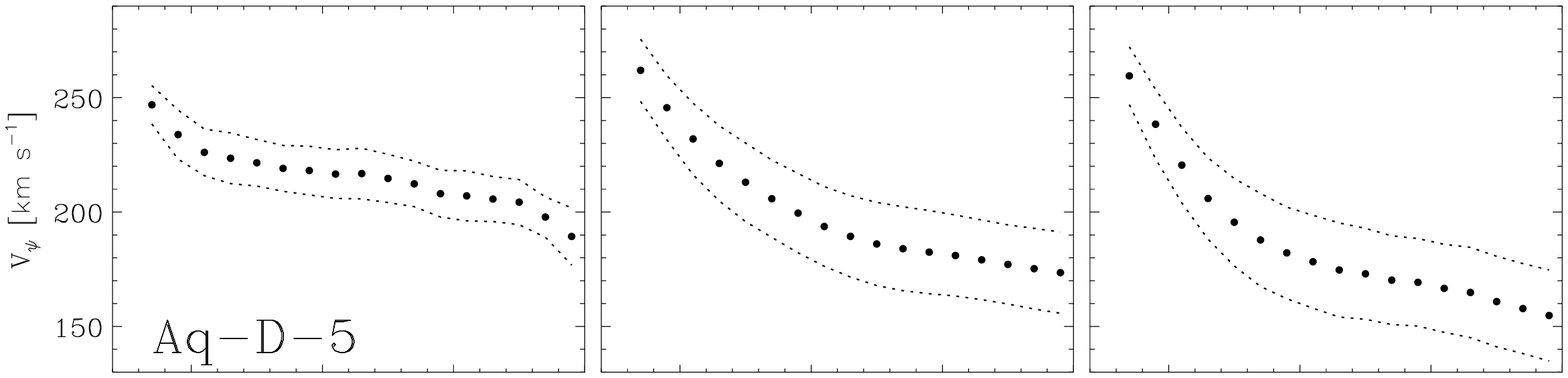}}

\vspace{-0.4cm}
{\includegraphics[width=90mm]{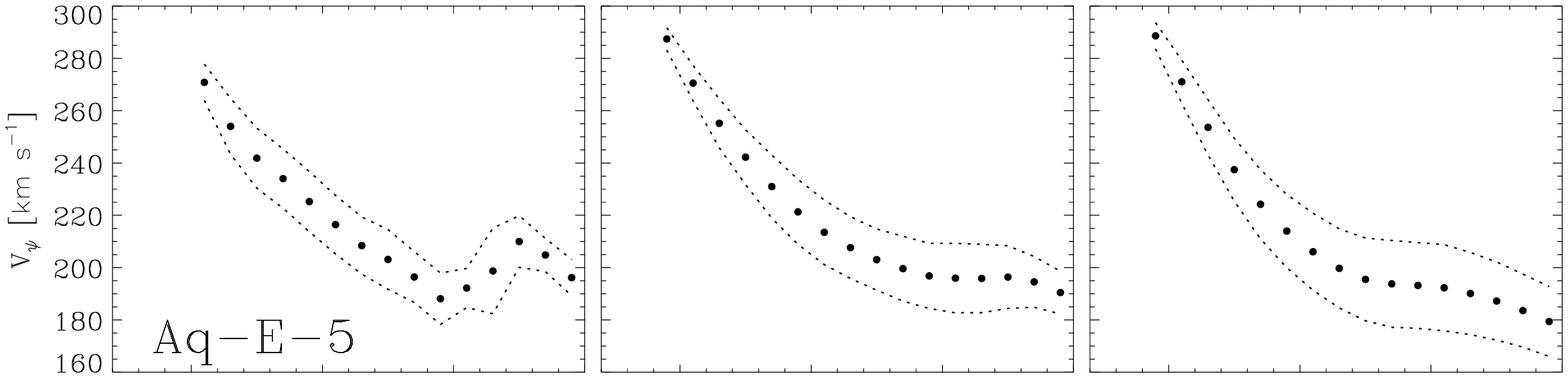}\includegraphics[width=90mm]{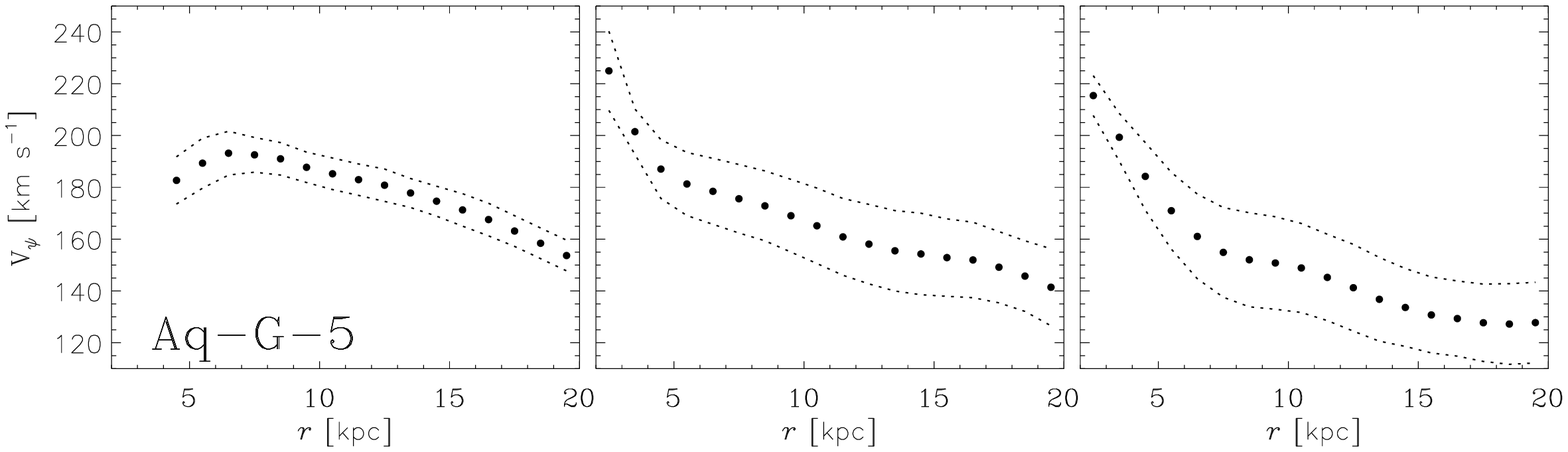}}

\vspace{-0.4cm}
\hspace{-87mm}\includegraphics[width=90mm]{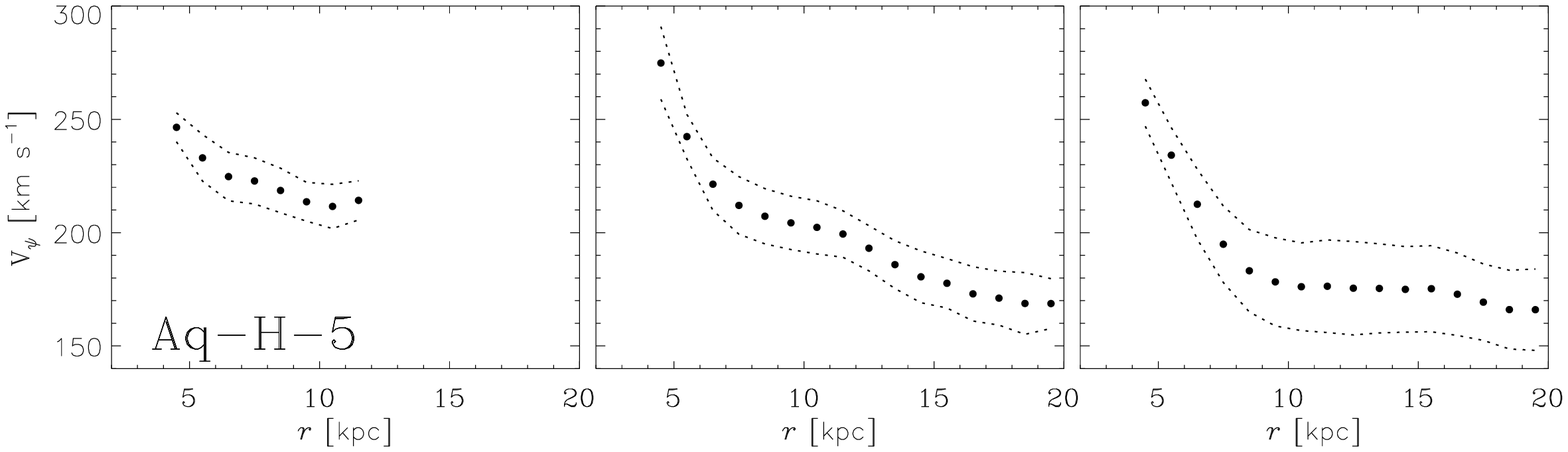}

\end{center}
\caption{Mean tangential velocity 
of disc stars as a function of projected radius in the disc plane (filled circles). 
The width between the dotted lines is the tangential velocity dispersion.}  
\label{vtita_vs_r}
\end{figure*}

In order to investigate the formation sites of the stars that end up
in each stellar component, we calculated the {\it in-situ} fractions
($f$) defined as the fraction of stellar mass that was formed in the
main progenitor of the final galaxy. These can be calculated
separately for the different components: the disc, and the inner and
outer spheroids. For the discs we also estimate $f$ separately for the
three stellar age bins defined above.  We start this calculation at
$z=10$ (at higher redshift the main progenitor can be ill-defined),
and assume that older stars formed in systems other than the main
progenitor.  We checked that only a small fraction (less than $4$
percent) of the stellar mass is formed before $z=10$ for all
simulations, and thus our results are not affected by this choice.
Note that stars that are not formed within the main progenitor are
thus ``accreted" stars (formed in systems other than the main
progenitor); consequently, the fraction of accreted stars is given by
$1-f$.

Labels at the bottom of the various panels of
Fig.~\ref{disk_structure} show the {\it in-situ} fractions for discs
in each age bin, as well as the the fraction of disc stellar mass in
that age bin (quantities in parenthesis).  We find that {\it all} disc
stars younger than $9$ Gyr formed {\it in-situ}, and that even the
oldest disc stars (representing between $30$ and $50\%$ of the total
disc mass) have relatively high {\it in-situ} fractions. The overall
{\it in-situ} fractions for simulated discs are shown in
Table~\ref{table_insitu}.  The {\it in-situ} fractions for discs are
$f> 0.85$ for all simulated galaxies.  We note that although the bulk
of disc stars are formed {\it in-situ}, a non-negligible fraction of
stars ($\le 15\%$) can be contributed by disrupted satellites that
come in on nearly coplanar orbits, as is the case for Aq-E-5, but in
general these bring relatively old stars to the final disc.  This
latter result is very similar to that obtained by Abadi et
al. (2003b), who also found that debris from disrupted satellites can be
found in $z=0$ thin/thick disc components.

\begin{figure*}
\begin{center}
{\includegraphics[width=90mm]{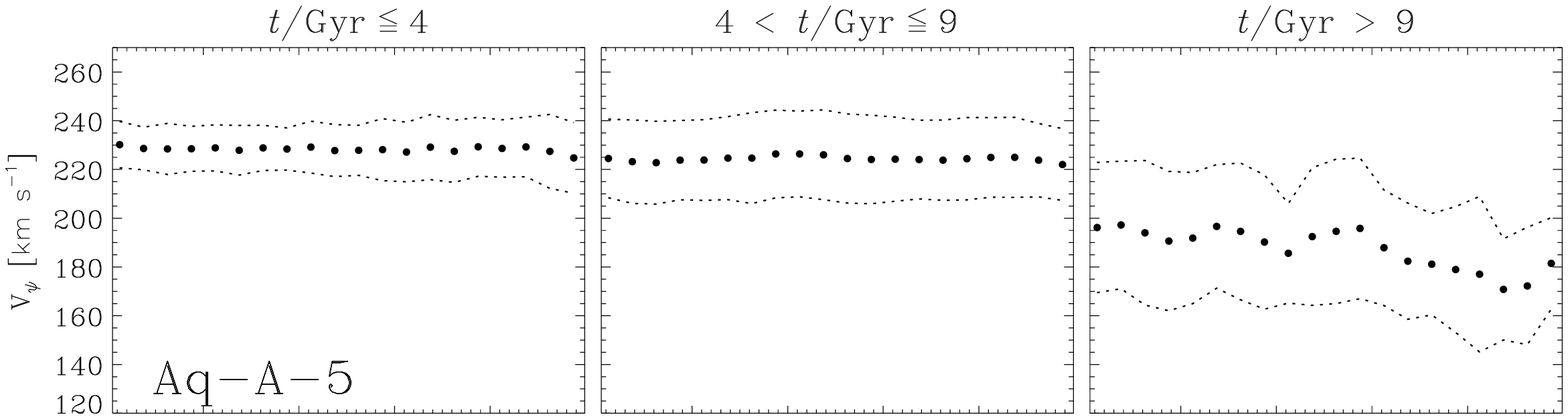}\includegraphics[width=90mm]{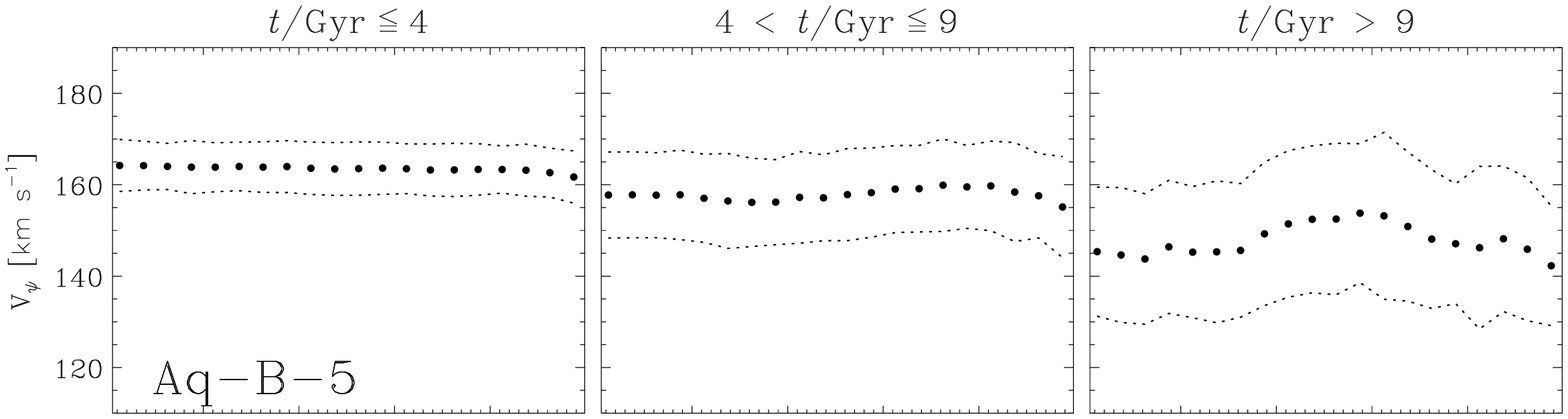}}

\vspace{-0.4cm}
{\includegraphics[width=90mm]{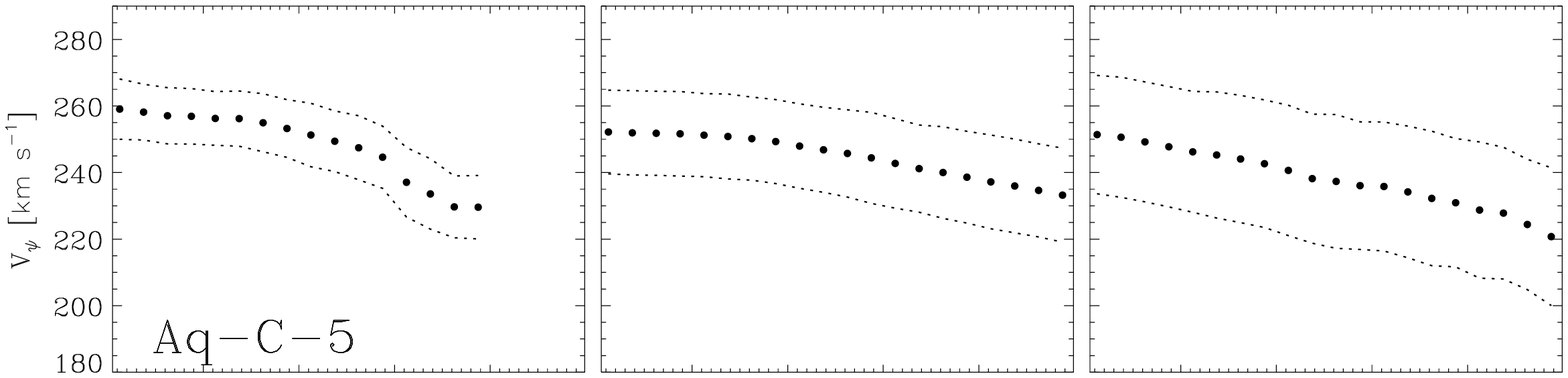}\includegraphics[width=90mm]{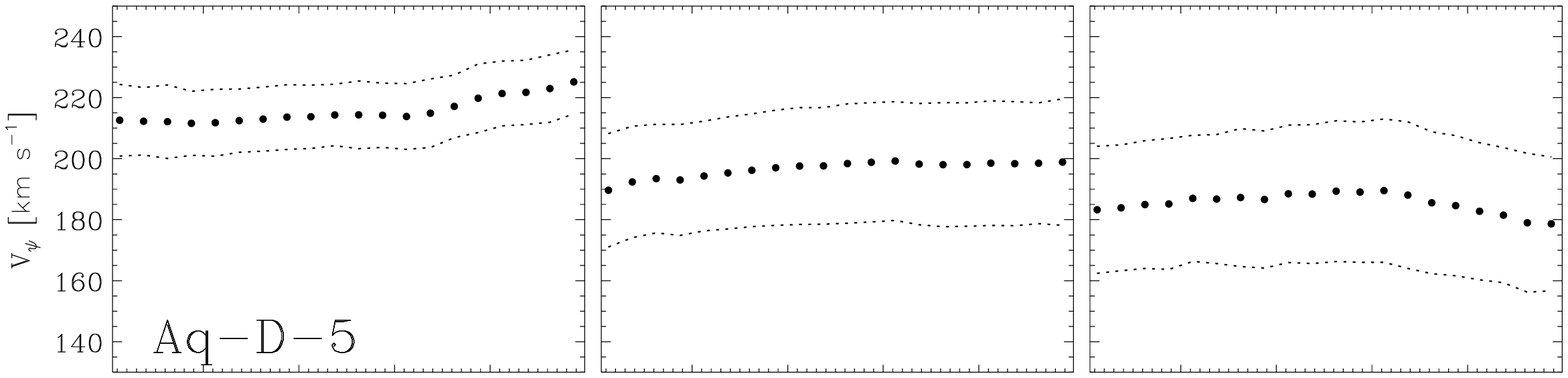}}

\vspace{-0.4cm}
{\includegraphics[width=90mm]{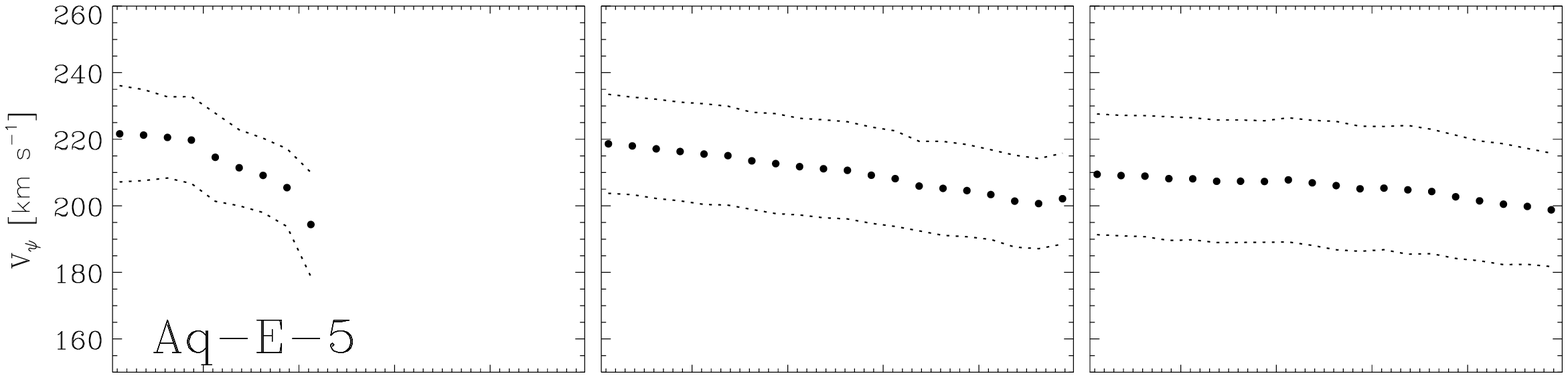}\includegraphics[width=90mm]{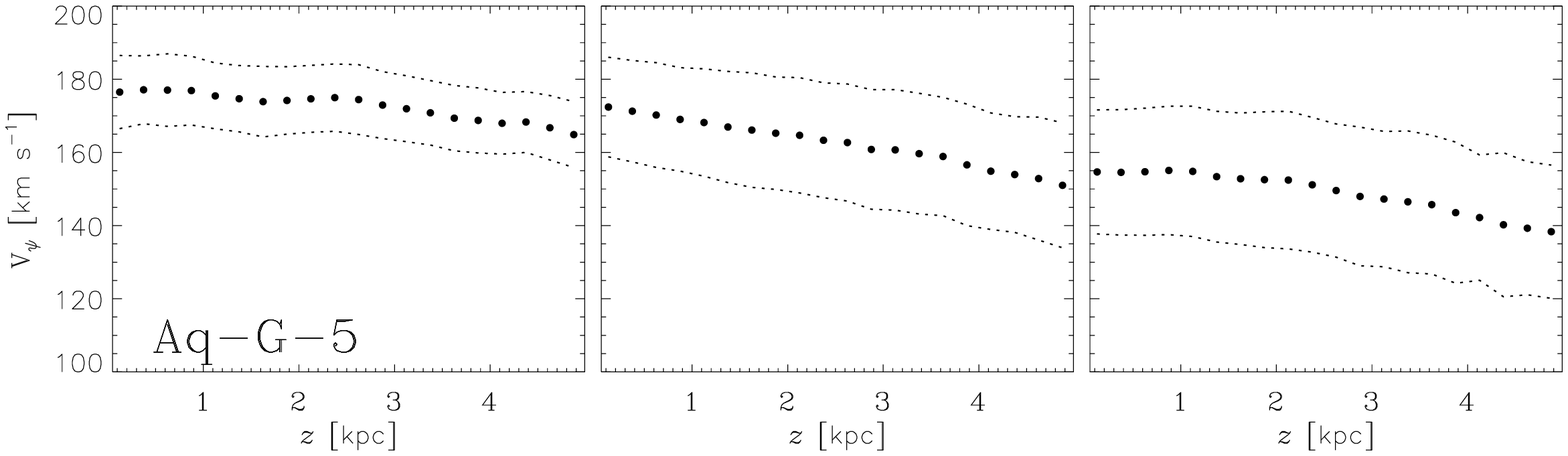}}

\vspace{-0.4cm}
\hspace{-87mm}\includegraphics[width=90mm]{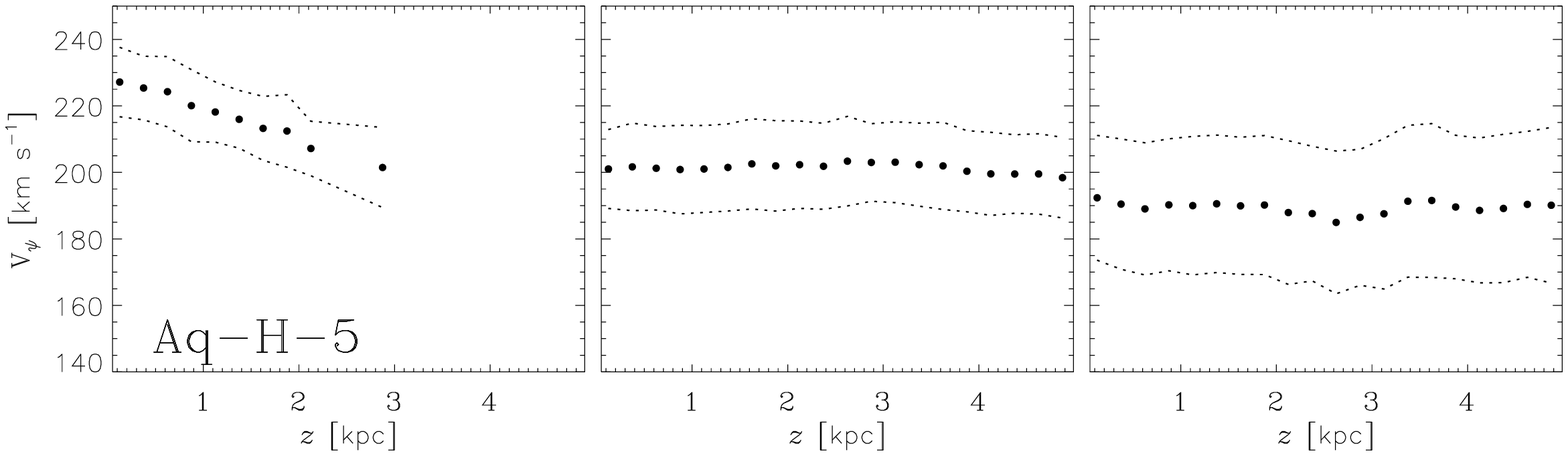}
\end{center}

\caption{Tangential velocity 
of disc stars as a function of height over the disc plane (filled circles). 
The width between the dotted lines is the tangential velocity dispersion.}  
\label{vtita_vs_z}
\end{figure*}

We find very good agreement of the overall {\it in-situ} fractions for discs in Aq-C-6 and Aq-E-6b
compared to Aq-C-5 and Aq-E-5, with changes of $5\%$ and $3\%$, respectively (Table~\ref{table_insitu}). 
In the case of Aq-E-6, the disc {\it in-situ} fraction is $17\%$ lower 
than in Aq-E-5. (Note that 
the disc/spheroid decomposition can also introduce some differences since,
as explained in S09, Aq-E
 has a rotating bulge and the decomposition is difficult.) 
The {\it in-situ} fractions for stars of given age
(in particular for the three stellar age bins used above)
 show small differences
 with varying resolution. The largest differences ($\lesssim 10\%$) are found for
the oldest stellar populations. For stars younger than $9$
 Gyr, we recover the result from
the level 5 simulations: all such stars formed {\it in-situ}.

\subsection{Dynamical properties}\label{dynamics_disks}

The motion of stars in the simulated discs are dominated by the tangential velocity
component.  In Fig.~\ref{vtita_vs_r} we show the mean tangential
velocity $V_\phi$ as a function of radius (filled circles), as well as
$V_\phi\pm \sigma_\phi/2$, where $\sigma_\phi$ is the dispersion in
$V_\phi$ in the corresponding radial bin.  For each simulation, we
divided stars in the three age bins we used to analyse the disc
structure.  Young stars ($t\le 9$ Gyr) usually
have higher tangential velocities than older
stars, particularly at large radii, 
and lower tangential velocity dispersions\footnote{We note that the separation
at $4$ Gyr of the two youngest stellar age bins is not in all cases separating two different
populations, as it can be inferred from Fig.~\ref{hist_stellarage}.}.  The
tangential velocity decreases with increasing radius, but the decrease
for old stars is stronger than for young stars: the difference in
$V_\phi$ between inner and outer regions is $60-110$ km s$^{-1}$ for
the former ($> 90$ km s$^{-1}$ for $5$ of the $7$ galaxies) and
$15-70$ km s$^{-1}$ for the latter ($< 35$ km s$^{-1}$ for $4$ of the
$7$ galaxies).

Typical values for the tangential velocity dispersions are $10-20$ km
s$^{-1}$ for disc stars younger than $4$ Gyr, $20-30\,{\rm km\,
  s}^{-1}$ for stars with $4<t/{\rm Gyr}\le 9$ and $30-40\,{\rm km\,
  s}^{-1}$ for stars older than $9$ Gyr.  These results are consistent
with discs having more than a single stellar population: older stars
populate discs that are thicker, with lower tangential velocities and
higher velocity dispersions, while younger stars populate thinner
discs that rotate faster and have lower tangential velocity
dispersions.  These results are also consistent with our previous
findings on the {\it in-situ} fractions: while young disc stars were
almost entirely  formed {\it in-situ}, the oldest component has a
higher contribution of accreted stars. Thus, it is expected that these
have higher velocity dispersions.

\begin{figure*}

\hspace{-1cm}\includegraphics[width=160mm]{sim_names_A-D.ps}\vspace{-0.1cm}

\vspace{0.2cm}

{\includegraphics[width=40mm]{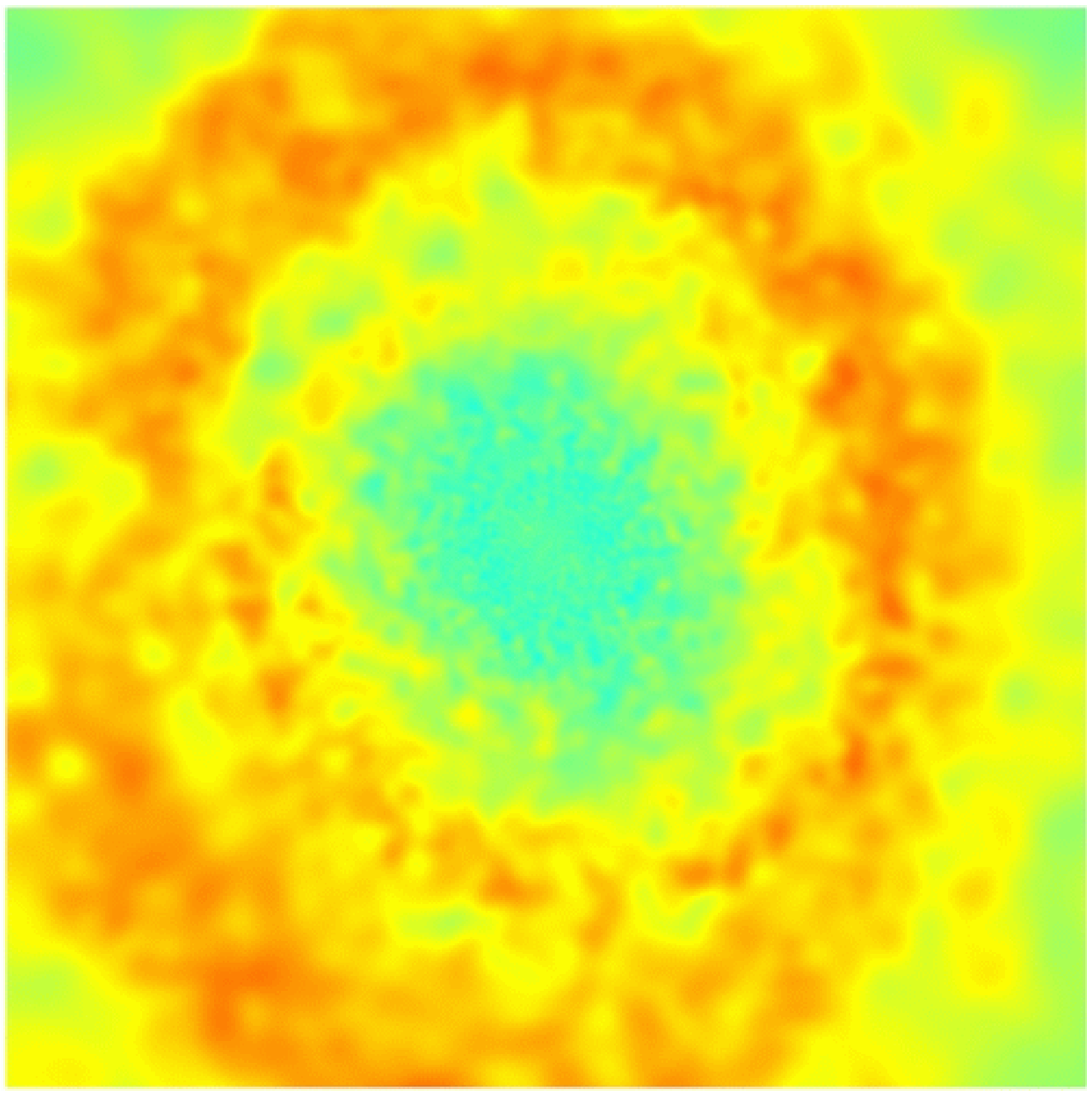}\includegraphics[width=40mm]{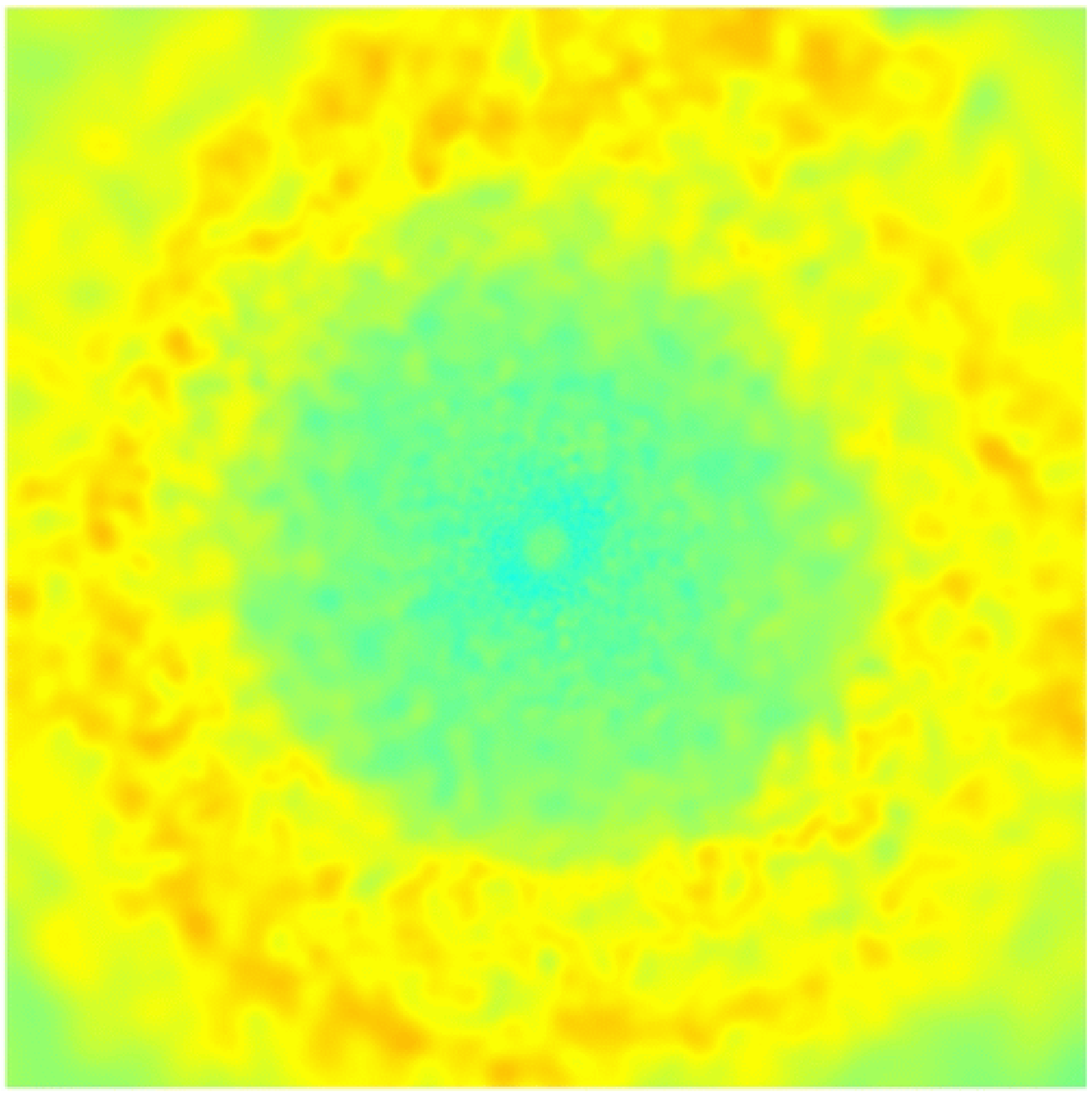}\includegraphics[width=40mm]{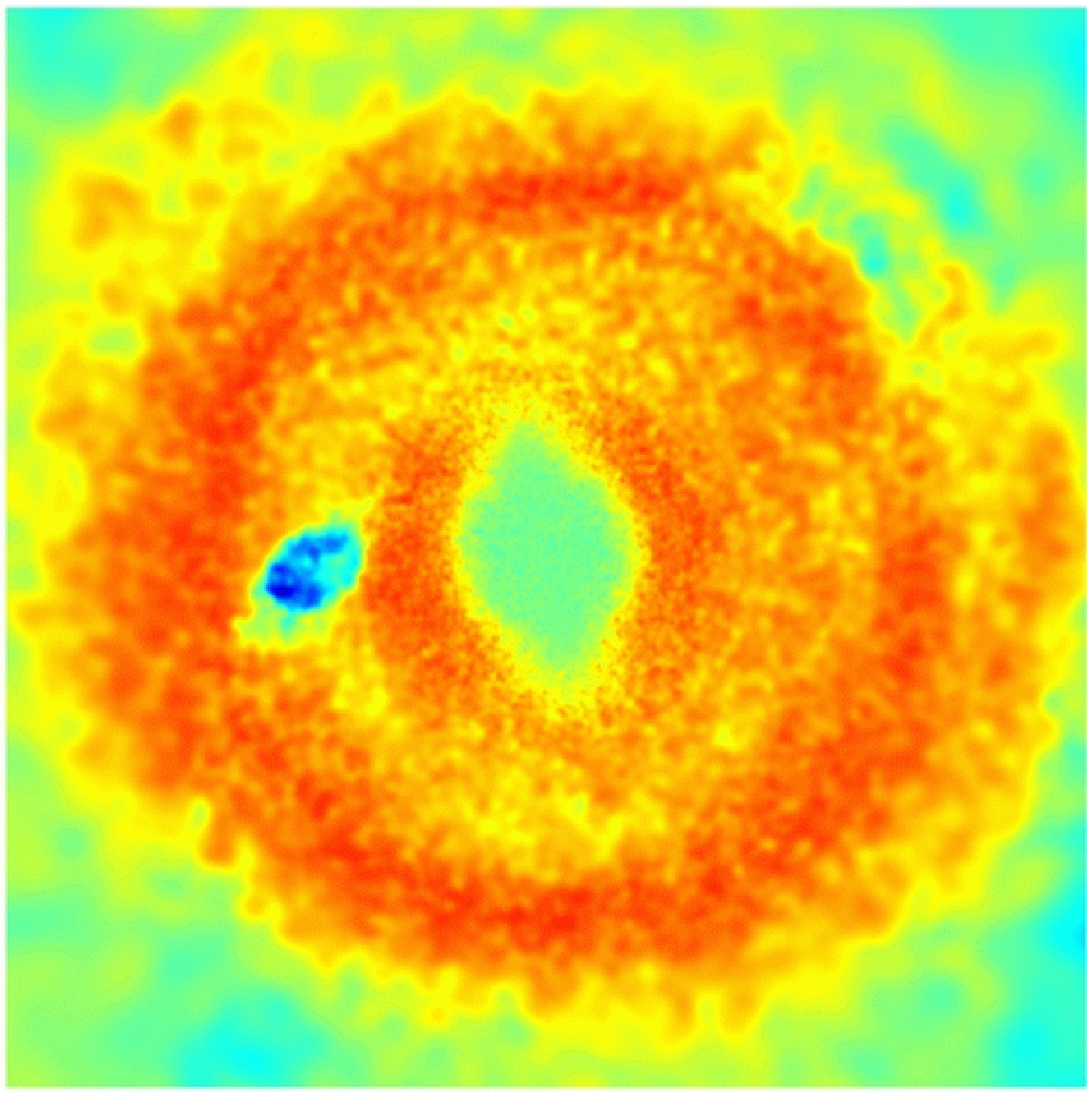}\includegraphics[width=40mm]{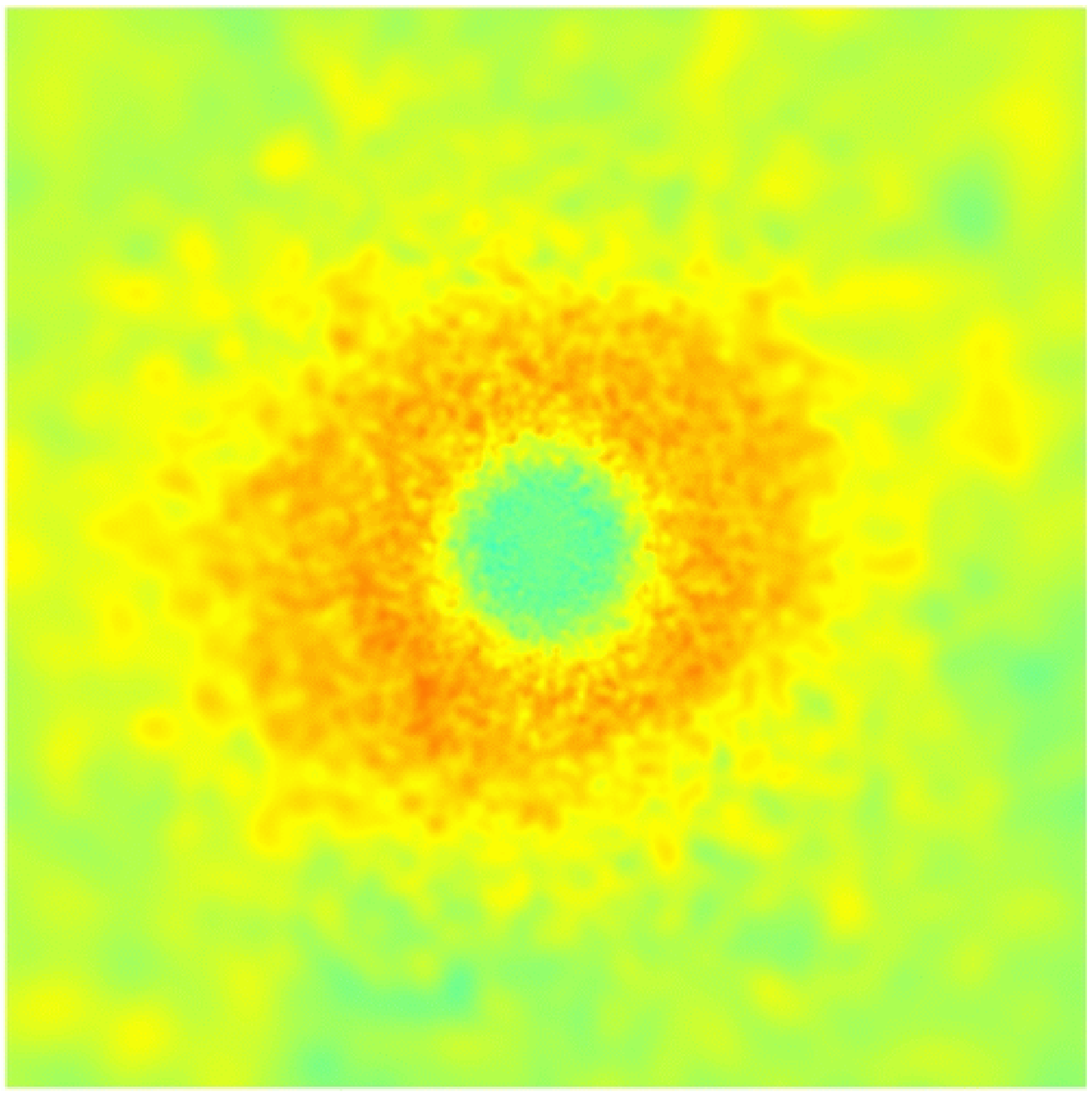}\includegraphics[width=14mm]{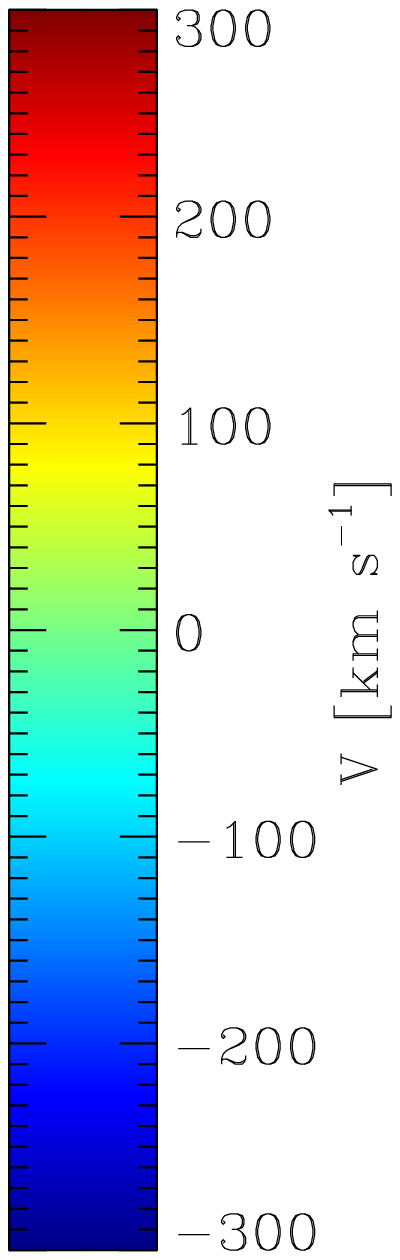}}

\hspace{-1cm}\includegraphics[width=160mm]{sim_names_E-H.ps}\vspace{-0.1cm}

\vspace{0.2cm}

{\includegraphics[width=40mm]{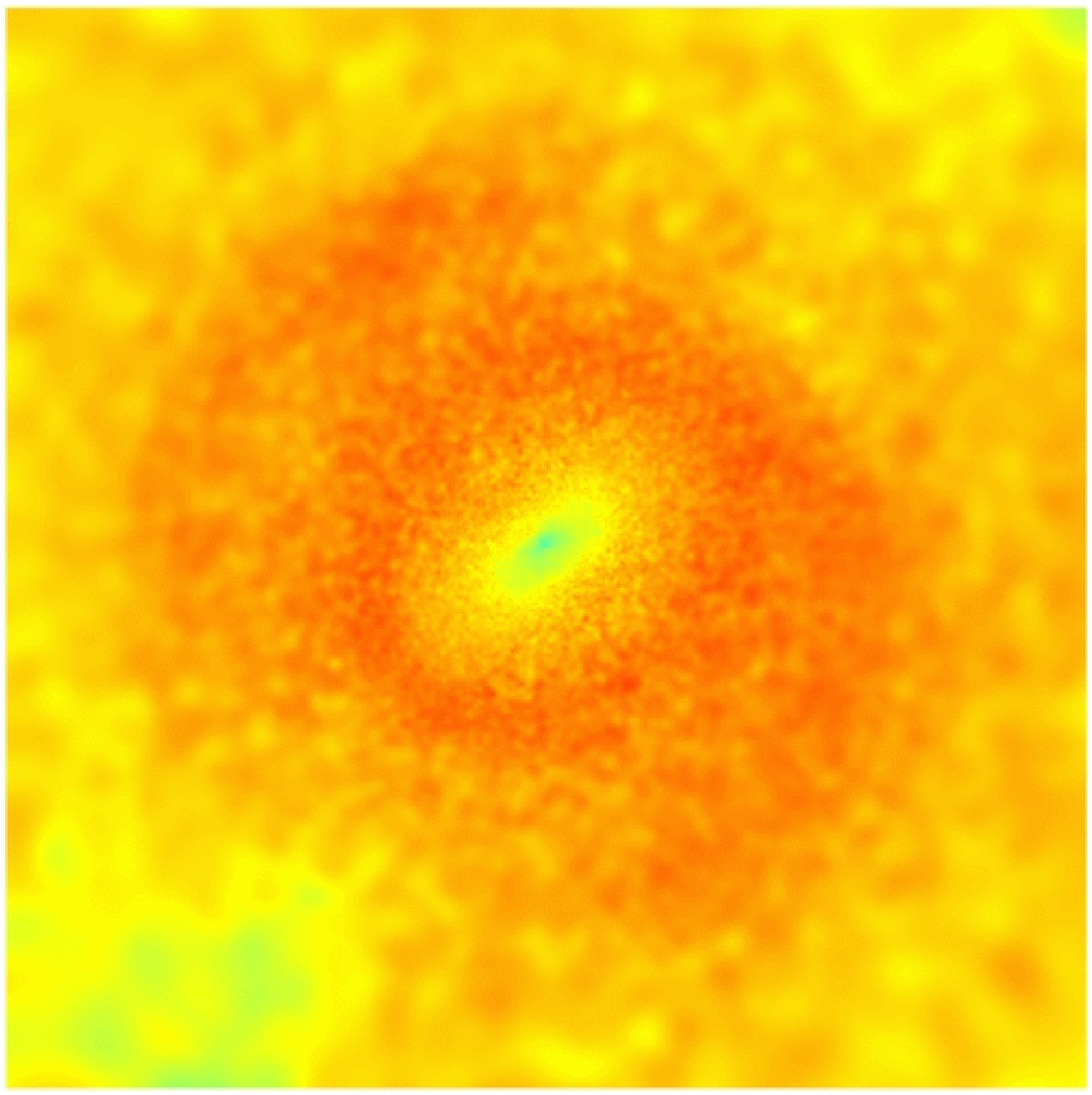}\includegraphics[width=40mm]{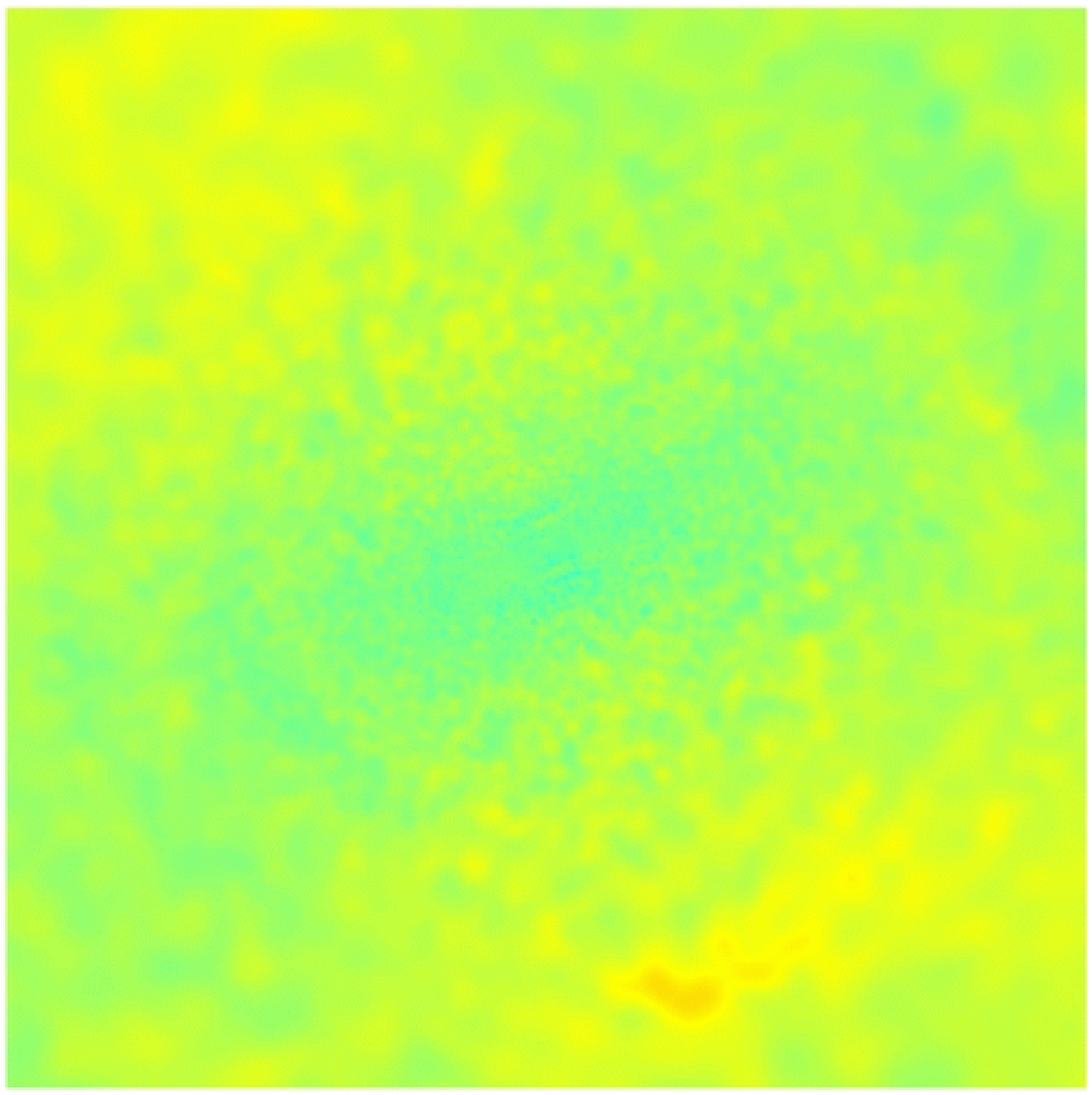}\includegraphics[width=40mm]{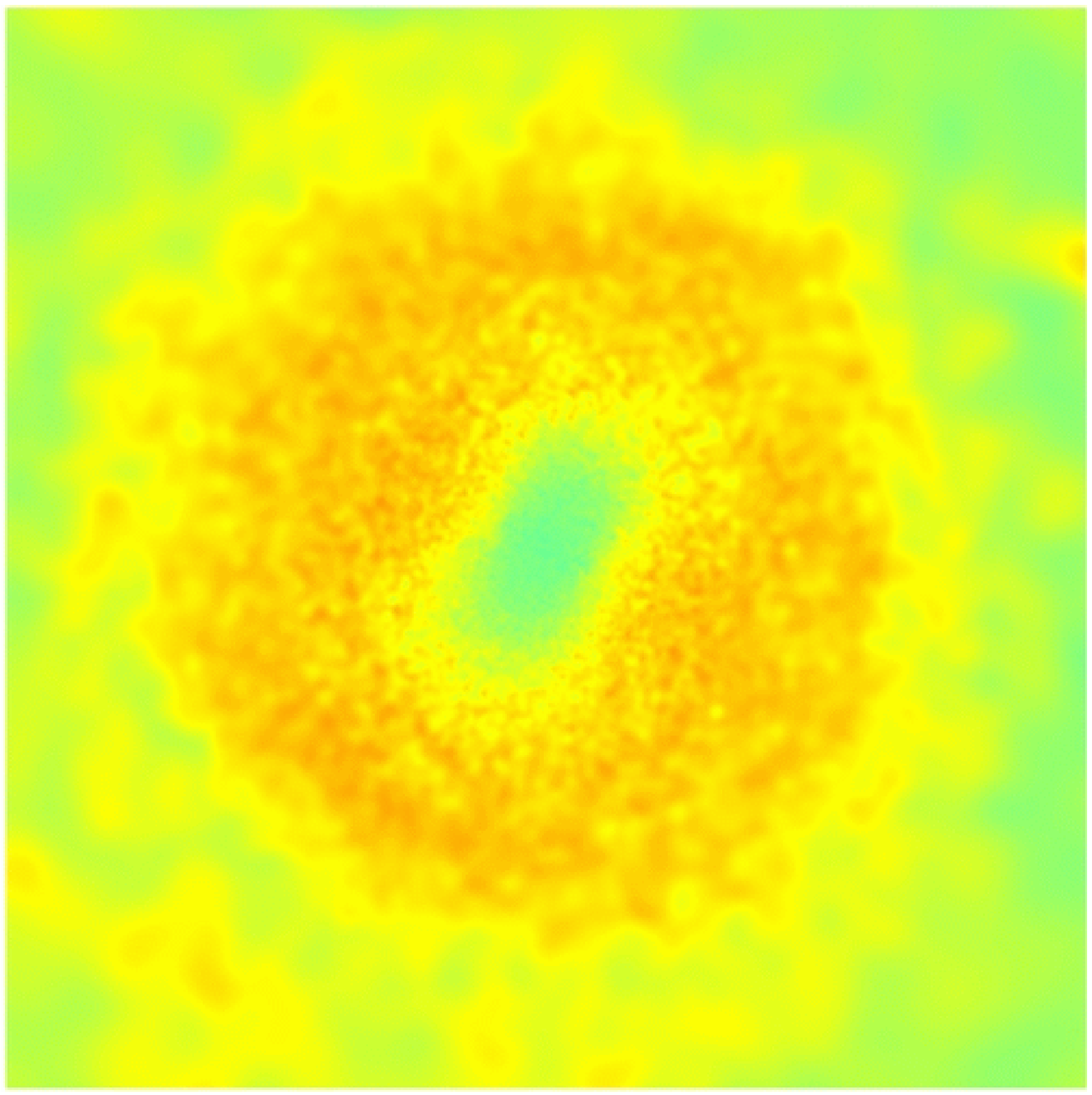}\includegraphics[width=40mm]{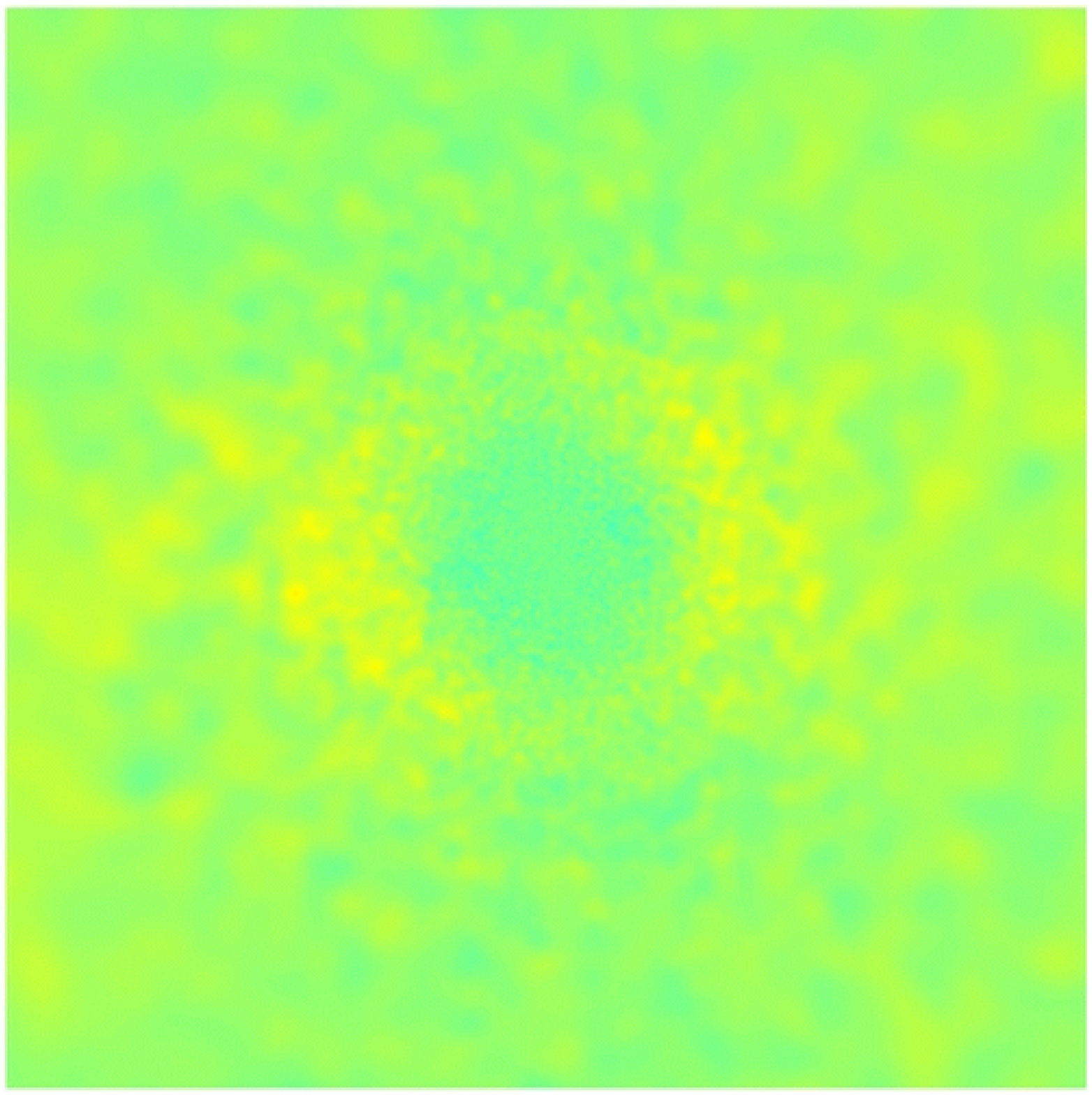}\includegraphics[width=14mm]{legend_vel_vert.eps}}

\caption{Maps of tangential velocity for the stellar components of our
 simulated galaxies. In this projection, the stellar discs are face-on, and colors represent the mean tangential
velocity in the corresponding bin, as indicated in the velocity scale. The
plots are $30$ kpc across.}
\label{vtita_maps_stars}
\end{figure*}

Old and young disc stars also show differences in their vertical
velocity structure, as shown in Fig.~\ref{vtita_vs_z}.  The figure
shows the mean tangential velocity of disc stars as a function of (the
absolute value of) height over the disc plane (filled circles) for the
three stellar age bins. The dotted lines are at $\pm
\sigma_\phi/2$ indicating the tangential velocity dispersion.  
For the young stars, the
tangential velocity is more or less constant or decreases steadily, as
a function of height over the disc plane.  Older
stars show a more pronounced decrease of $V_\phi$ with height over the
disc plane, with changes of the order or lower than 30 km s$^{-1}$.
 Aq-C-5, Aq-E-5 and Aq-G-5 (three of the four galaxies with more
prominent discs) show the largest gradients in tangential velocity, both for
old and for young stars. In contrast, Aq-A-5, Aq-B-5, Aq-D-5 and Aq-H-5
(except for the youngest stars) tend to have constant tangential velocity as a
function of height over the disc plane. Note that in Aq-D-5 the tangential
velocity shows a slight increase with height over the disc plane for the young
stars. In this galaxy, even the youngest stars define a relatively thick
disc (Fig.~\ref{disk_structure}).

We find that the radial and vertical components of the velocity
dispersion in discs are larger than the tangential ones.  Radial
velocity dispersions are $\sigma_r=80-100$ km s$^{-1}$ (with the
exception of Aq-A-5 which has $220$ km s$^{-1}$) for the oldest stars
and $20-70$ km s$^{-1}$ for the youngest ones; and vertical velocity
dispersions vary from $\sigma_{z}=50-90$ km s$^{-1}$ to $15-70$ km
s$^{-1}$ from the older to the younger populations.  The total
velocity dispersions for simulated discs are between $100$ and $250$ km
s$^{-1}$ for the oldest stars and in the range $30-70$ km s$^{-1}$ for
stars in the youngest age bin.

The results of this section show that the oldest stars define thick
discs which rotate $\sim 2$ times slower and have $2-3$ times larger
(total) velocity dispersions than the young, thinner disc components.
These results are, in broad terms, consistent with results for the thin
and thick discs of the Milky Way (assuming that our separation of thin
and thick discs in terms of stellar age is approximately correct).  In
our Galaxy, thick disc stars are lagging with respect to the thin disc
by $30-70$ km s$^{-1}$ (Gilmore, Wyse \& Norris 2002), with $V_{\rm
  lag}=50$ km s$^{-1}$ usually taken as a canonical value (see also
Vallenari et al. 2006).  On the other hand, in external galaxies the
difference in rotation velocity of thin and thick discs appears to be
diverse and strongly dependent on galaxy mass, with lower mass
galaxies having larger differences between thin and thick velocities
(Yoachim \& Dalcanton 2008).

\begin{figure*}
\begin{center}
\hspace{-0.6cm}
\includegraphics[width=140mm]{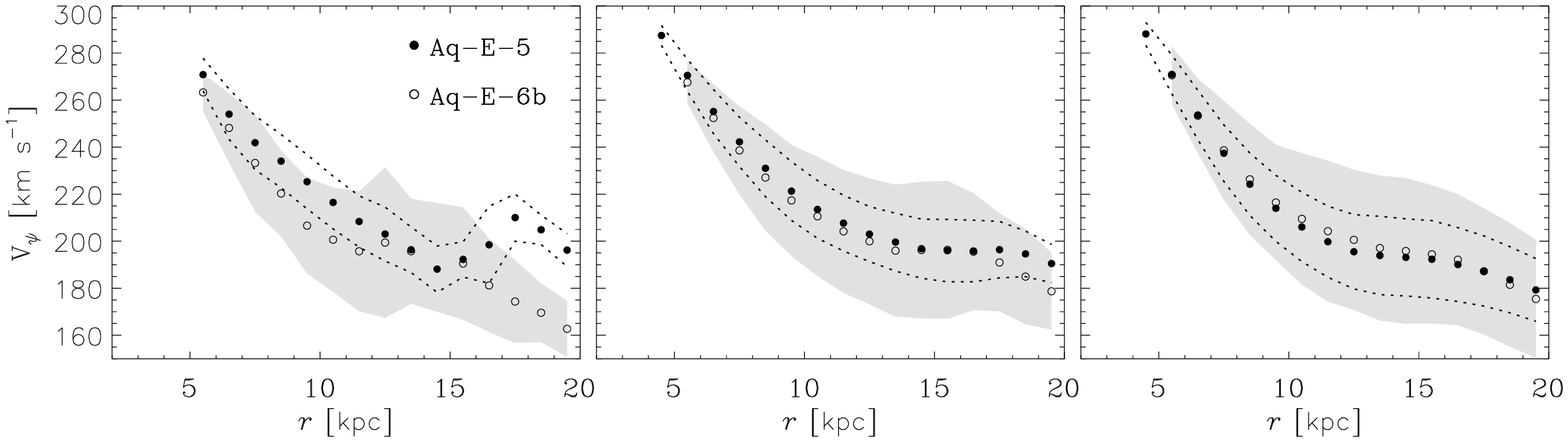}

\end{center}
\caption{  Comparison of two simulations with different resolutions.
The symbols show the tangential velocity as a function of radius for Aq-E-5 (filled circles)
and Aq-E-6b (open circles). 
The width between the dotted lines is the tangential velocity dispersion for Aq-E-5, and
the shaded area corresponds to the tangential velocity dispersion for Aq-E-6b.}  
\label{disk_dynamics_resolution}
\end{figure*}

In terms of velocity dispersions, our results are in relatively good
agreement with the observations of the Milky Way.  Vallenari et
al. (2006) estimated the velocity dispersion ellipsoid of the thin and
thick discs of our Galaxy.  Their results for the thick disc are:
$(\sigma_r, \sigma_\phi, \sigma_z) \sim (74\pm 11, 50\pm 7, 38\pm 7)$
km s$^{-1}$ at the solar radius. For the thin disc, they estimate the
velocity ellipsoid dividing stars into four stellar age bins, finding
$(\sigma_r, \sigma_\phi, \sigma_z) \sim (25-34, 20-32, 10-18)\,{\rm
  km\, s}^{-1}$ (with errors $\lesssim 15\%$). As described above, the
eight simulated galaxies are diverse and show a wide range in velocity
dispersions that agree relatively well with these results (although we
note that simulated galaxies are not expected to resemble the Milky
Way in detail).

Finally, we note that the velocity structure of the simulated stellar
components is complex, in general having important
asymmetries. Fig.~\ref{vtita_maps_stars} shows 2D face-on maps of
tangential velocity for the eight simulations (including both disc and
spheroid stars) within the inner $30$ kpc. From these plots we can
read off the velocity structure of simulated galaxies, the sizes of
bulges and discs, and we can also observe  bar patterns,
particularly in Aq-C-5, Aq-E-5 and Aq-G-5 (see also the next section).
Aq-A-5, Aq-C-5, Aq-D-5 and Aq-E-5 have the largest tangential
velocities, as expected since these are the most massive galaxies and
therefore have higher circular velocities. In the case of Aq-B-5, the
disc starts to dominate at a relatively large radius, showing  a
ring-like structure. As we discuss below, the absence of a disc in
Aq-F-5 is evident from this figure, and also the small disc component
of Aq-H-5.

We find that the dynamical structure of discs found for our simulations
is also present in the lower resolution runs. In all cases, the older
stars define thicker discs compared to the younger populations. 
In particular, Aq-E-5 and Aq-E-6b show very good agreement, as shown in
Fig.~\ref{disk_dynamics_resolution}. In this case, differences
in tangential velocities and velocity dispersions are always lower than $5\%$.
For the youngest stars (left-hand panel),  tangential velocities  are lower  for
Aq-E-6b than for Aq-E-5. Velocity dispersion are, regardless of stellar age,
larger for the lower resolution run.
We detect more significant  differences  for Aq-C-6 and Aq-E-6 with respect to
Aq-C-5 and Aq-E-5 respectively. In these cases, the tangential velocities
are typically $15\%$ lower in the low resolution runs, while velocity dispersions are
$\sim 50\%$ larger. These results show that low resolution runs can 
artificially boost the degree of disc heating.

\section{The spheroids}\label{sec_spheroids}

\subsection{Structure}\label{structure_spheroids}

\begin{figure*}
\begin{center}
{\includegraphics[width=45mm]{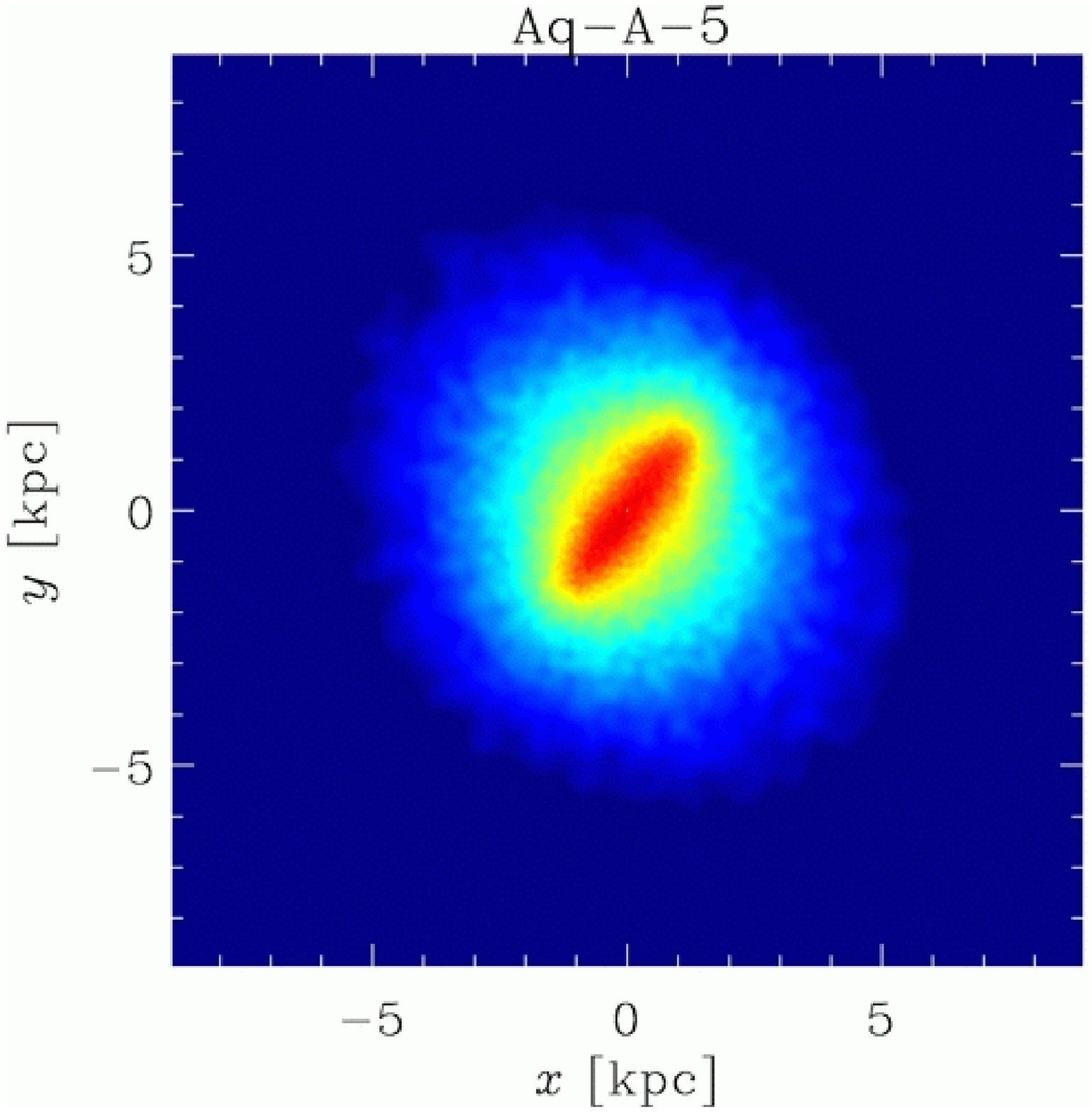}\includegraphics[width=45mm]{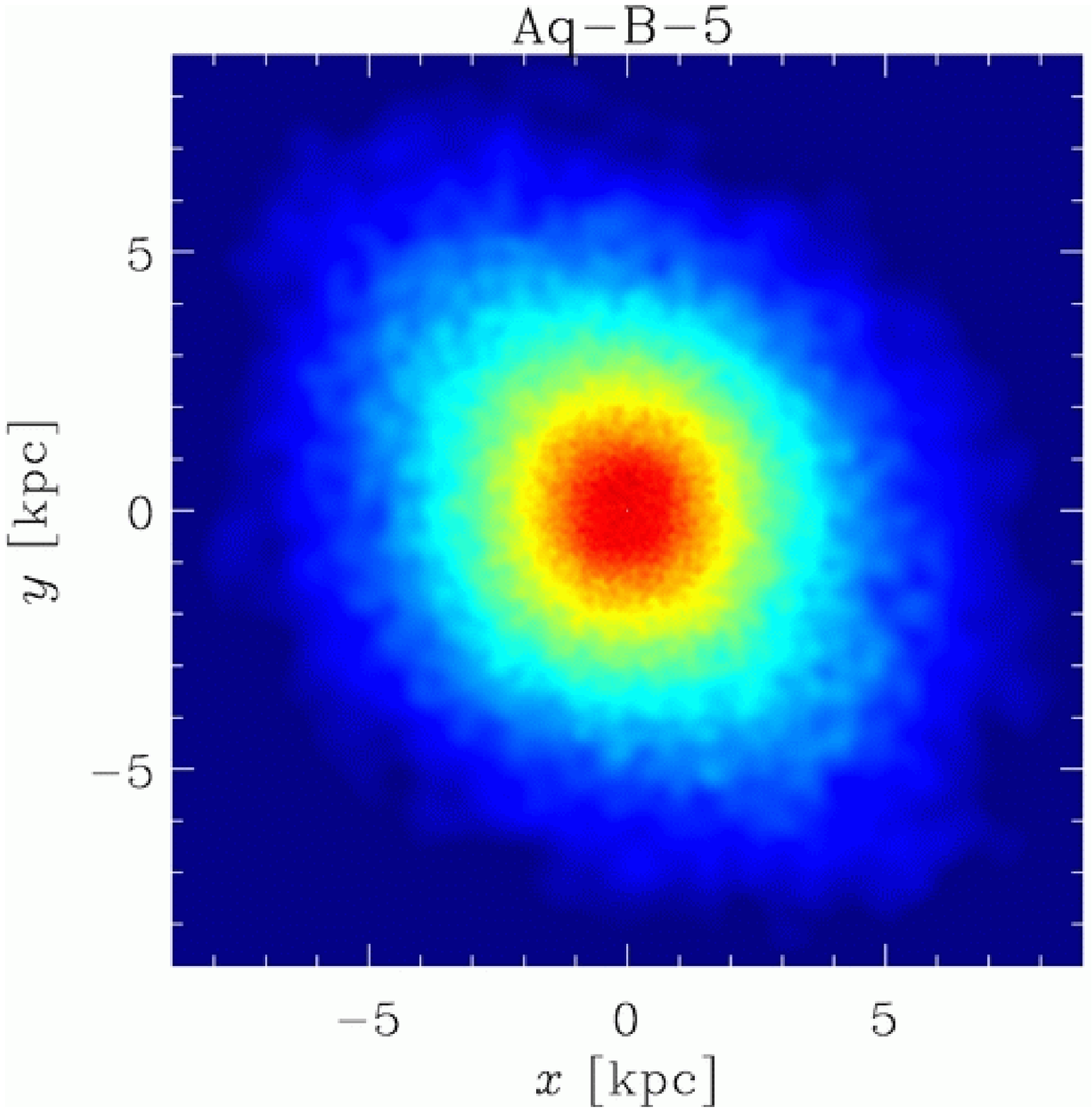}\includegraphics[width=45mm]{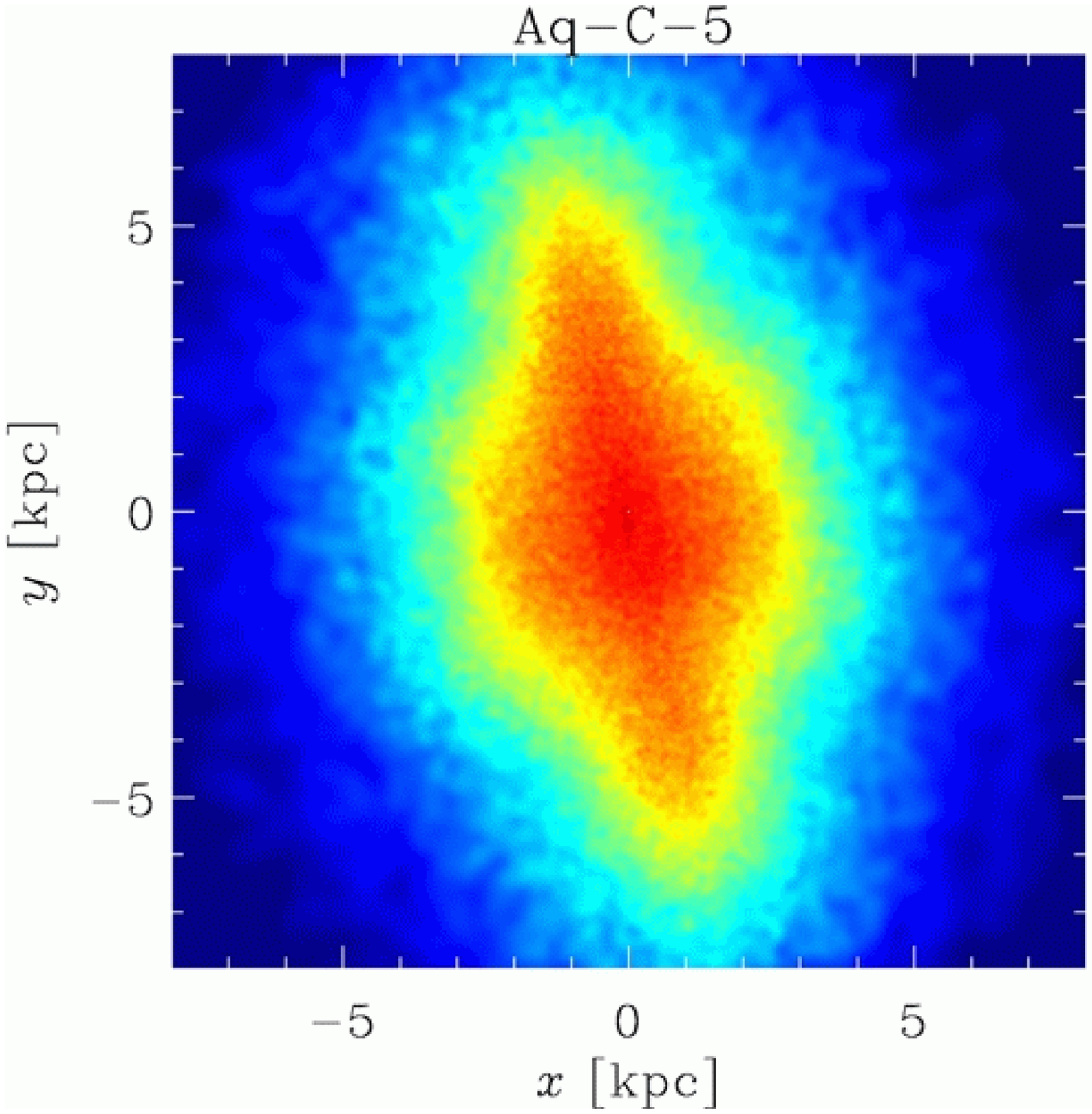}\includegraphics[width=45mm]{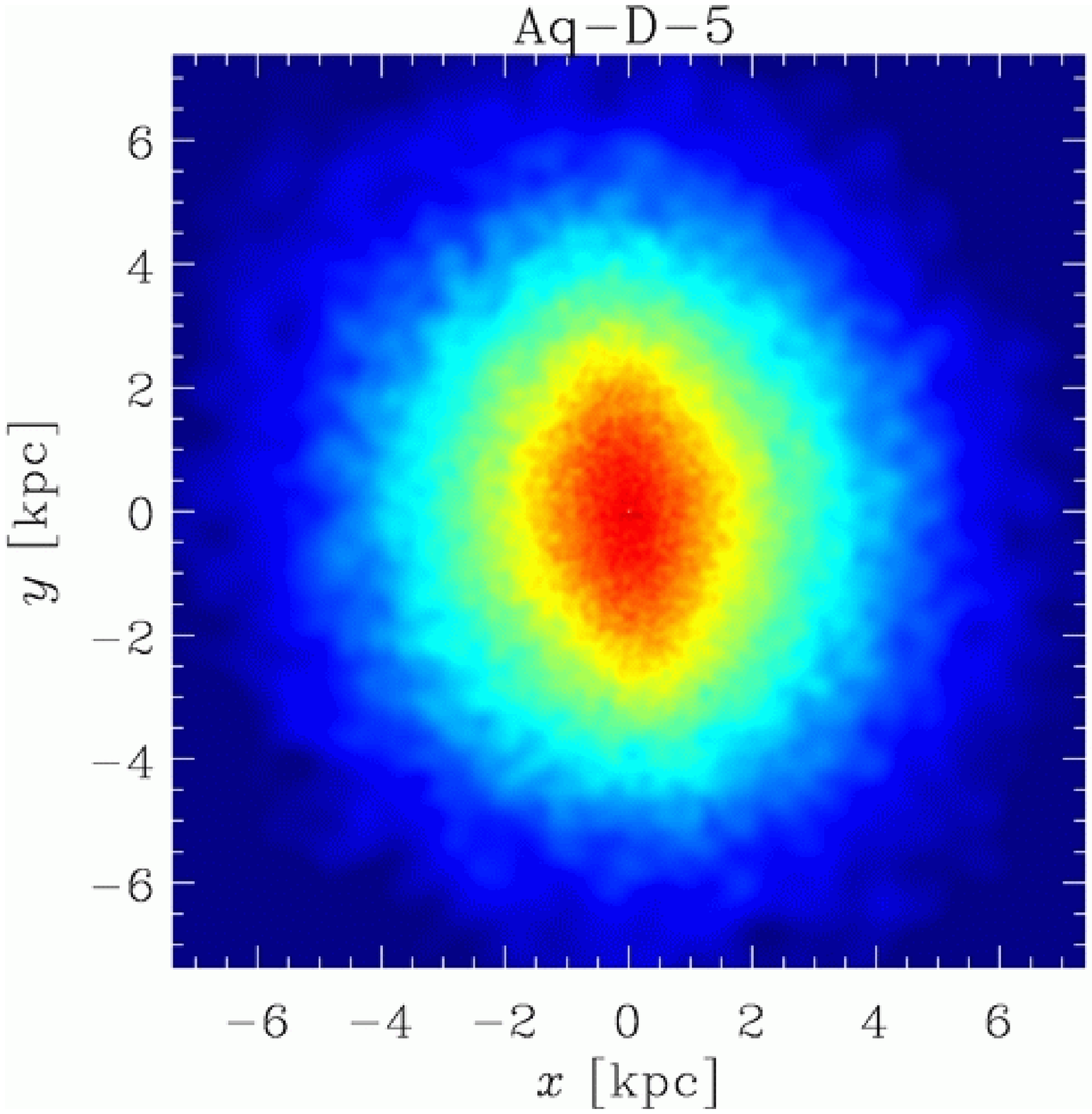}}
\hspace{-0.5cm}\
{\includegraphics[width=45mm]{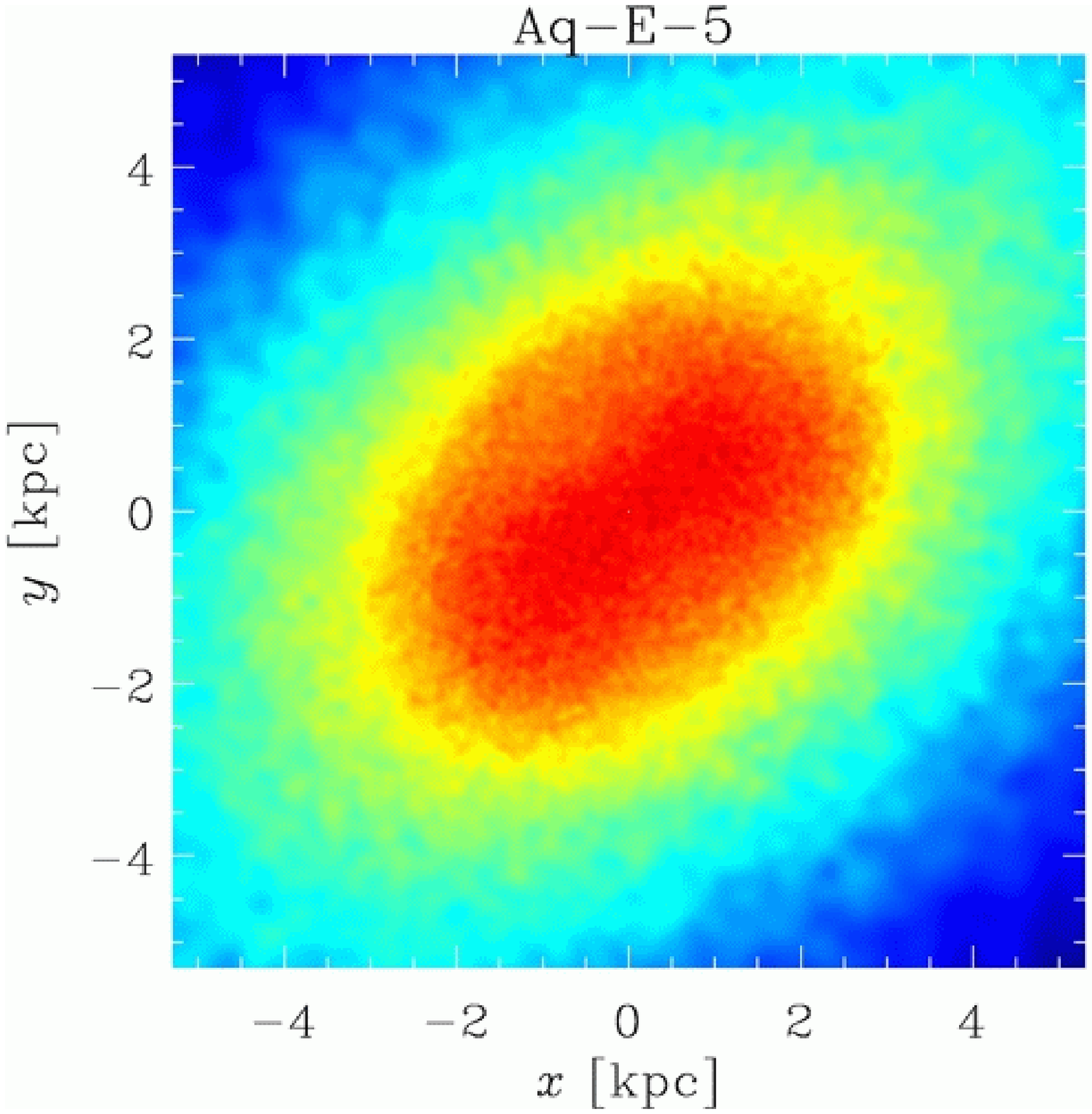}\includegraphics[width=45mm]{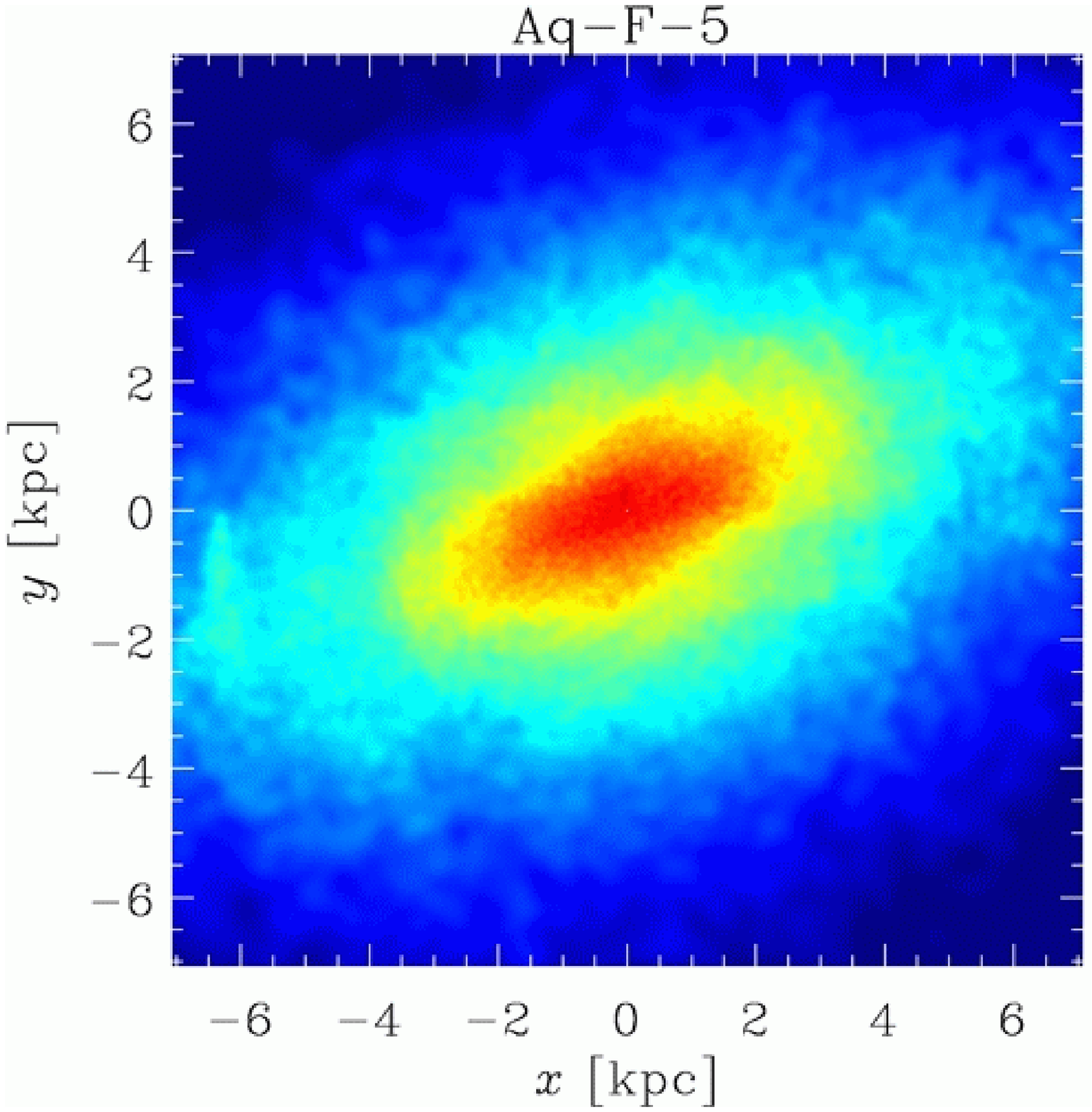}\includegraphics[width=45mm]{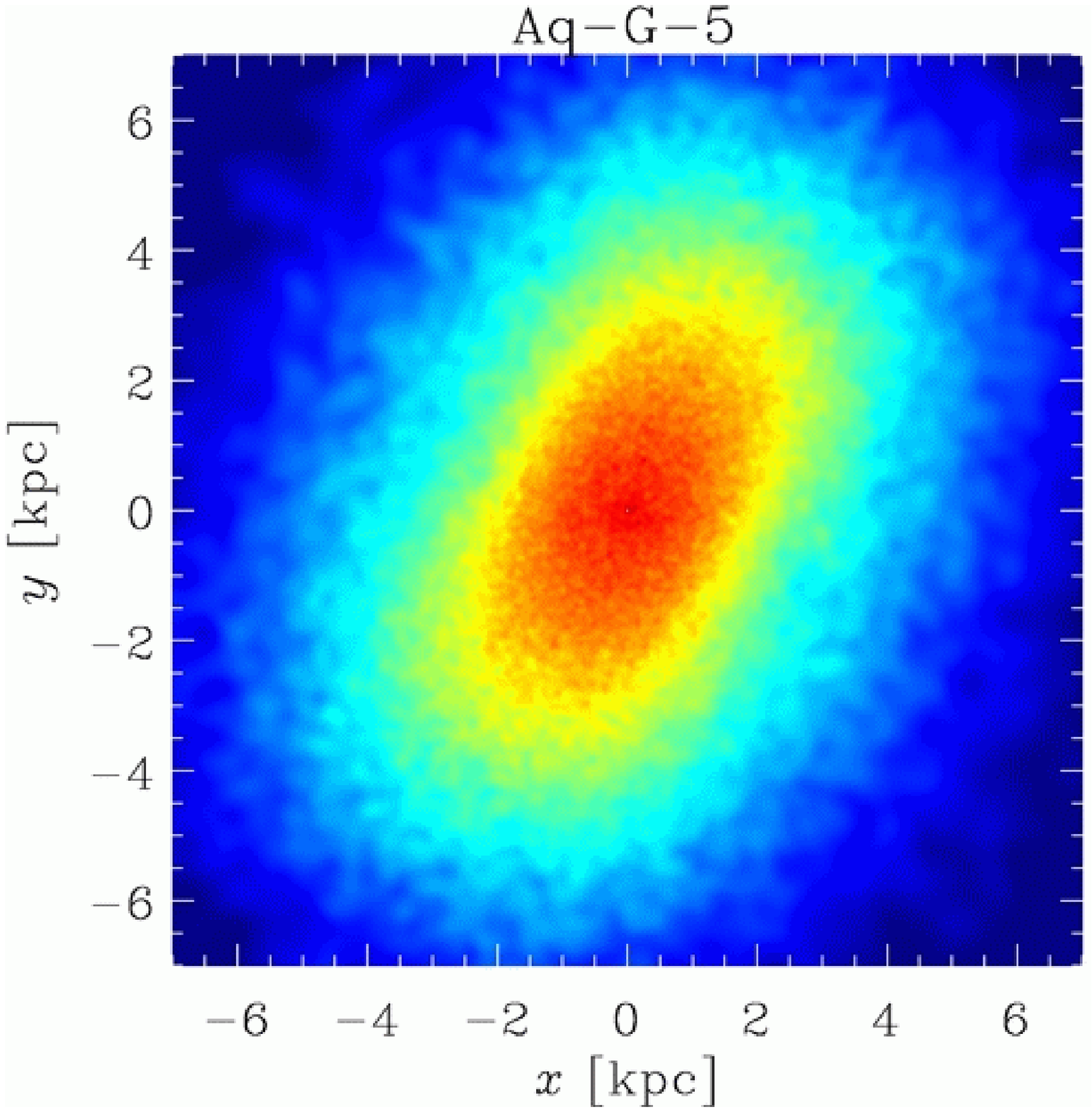}\includegraphics[width=45mm]{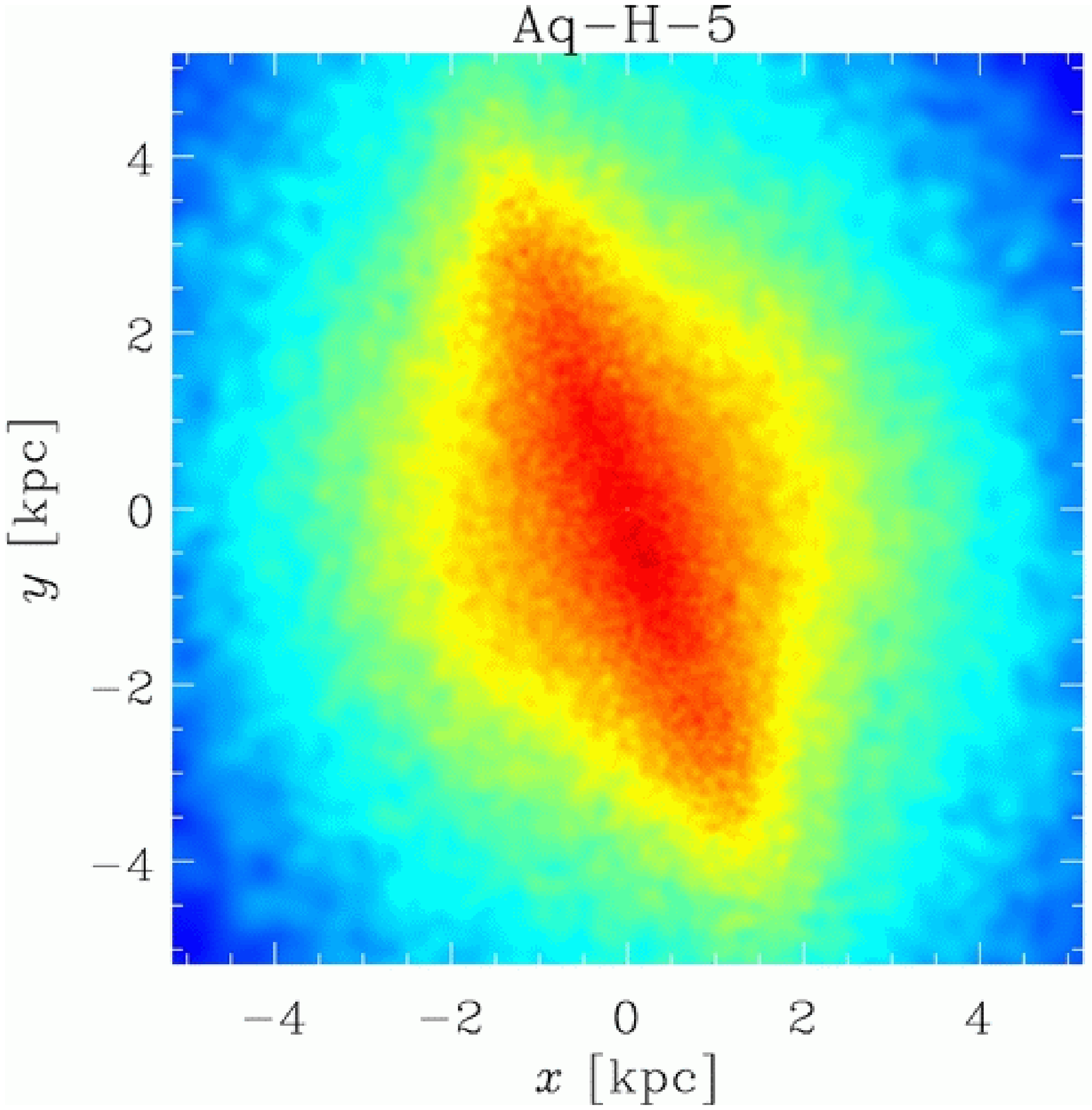}}
\end{center}
\caption{Spatial structure of simulated inner spheroids, up to
  $0.5\, r_{\rm opt}$. The projection is face-on, such that the discs are
  contained in this plane.  Colors represent projected surface mass
  density, on a logarithmic scale that covers $2.5$ orders of
  magnitude.}
\label{maps_innerspheroid}
\end{figure*}

\begin{figure*}
\begin{center}
{\includegraphics[width=45mm]{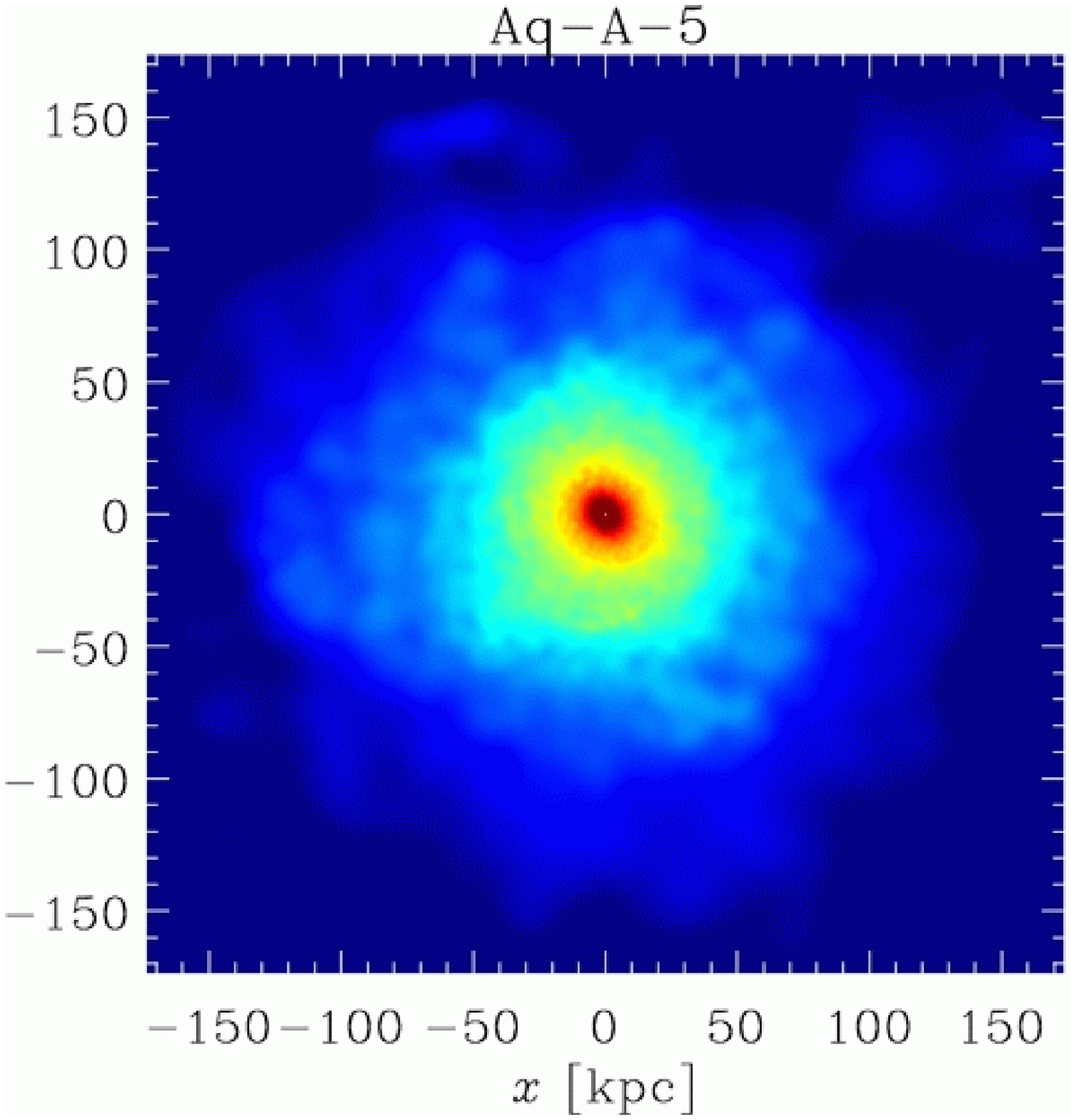}\includegraphics[width=45mm]{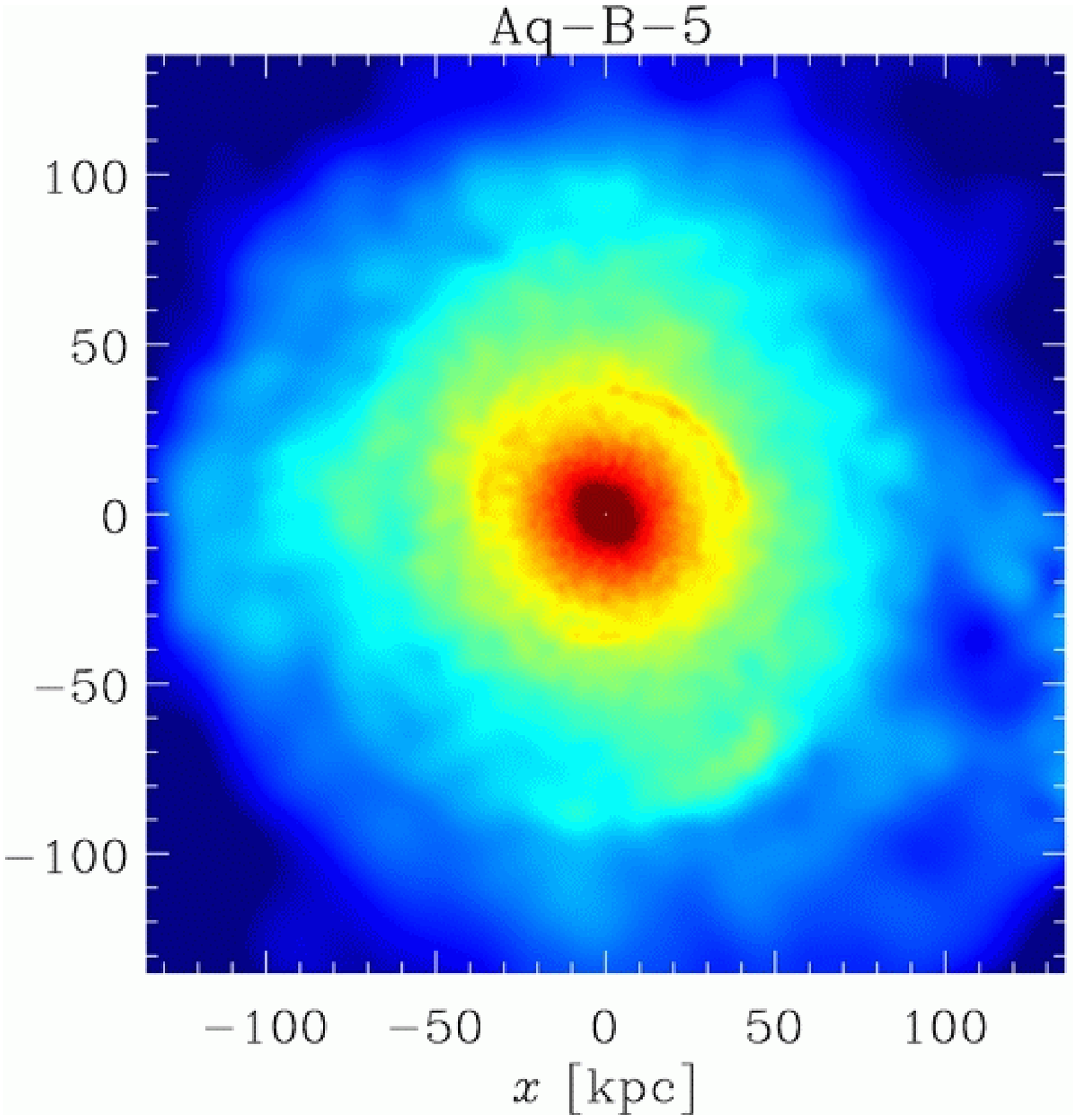}\includegraphics[width=45mm]{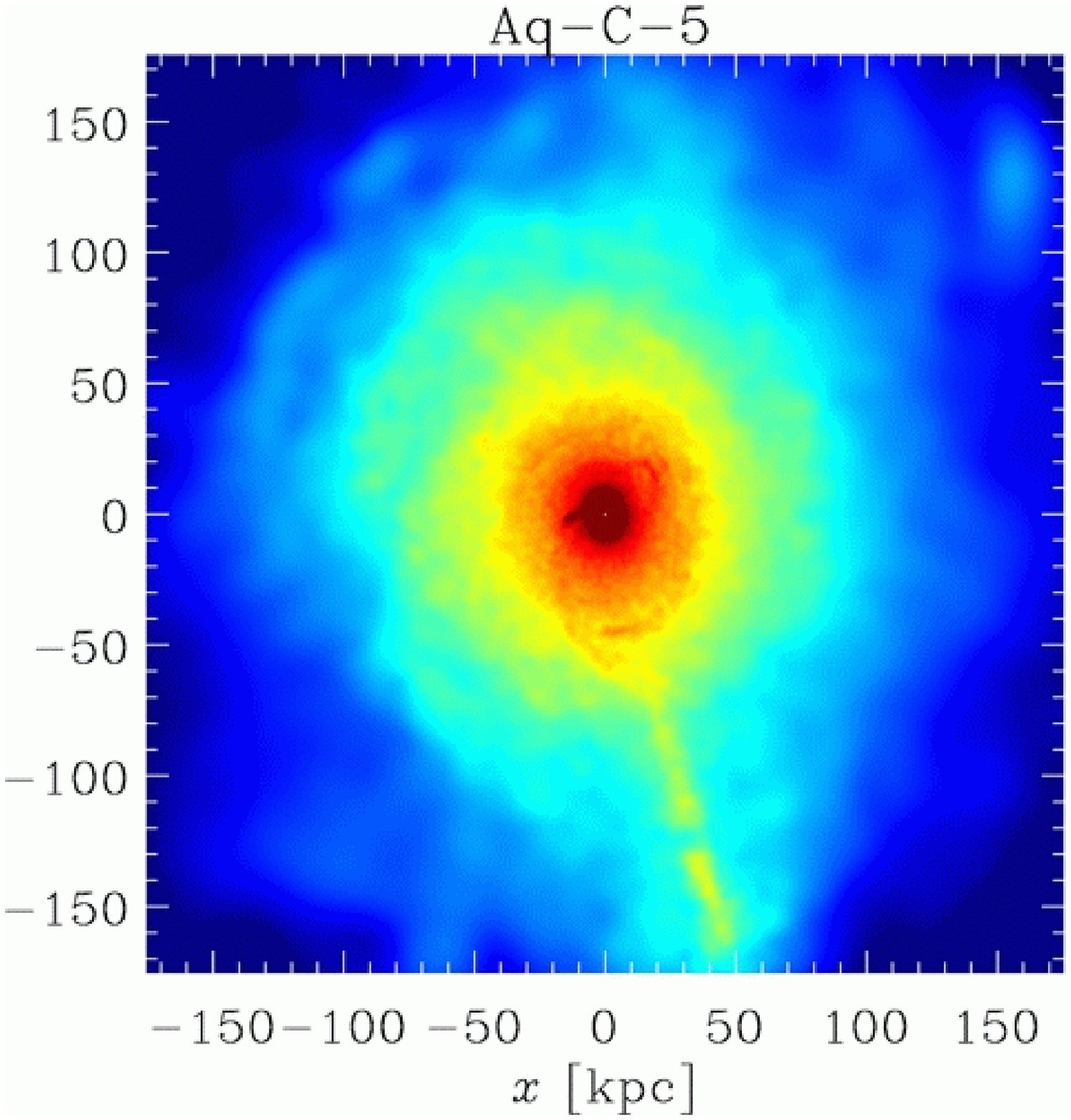}\includegraphics[width=45mm]{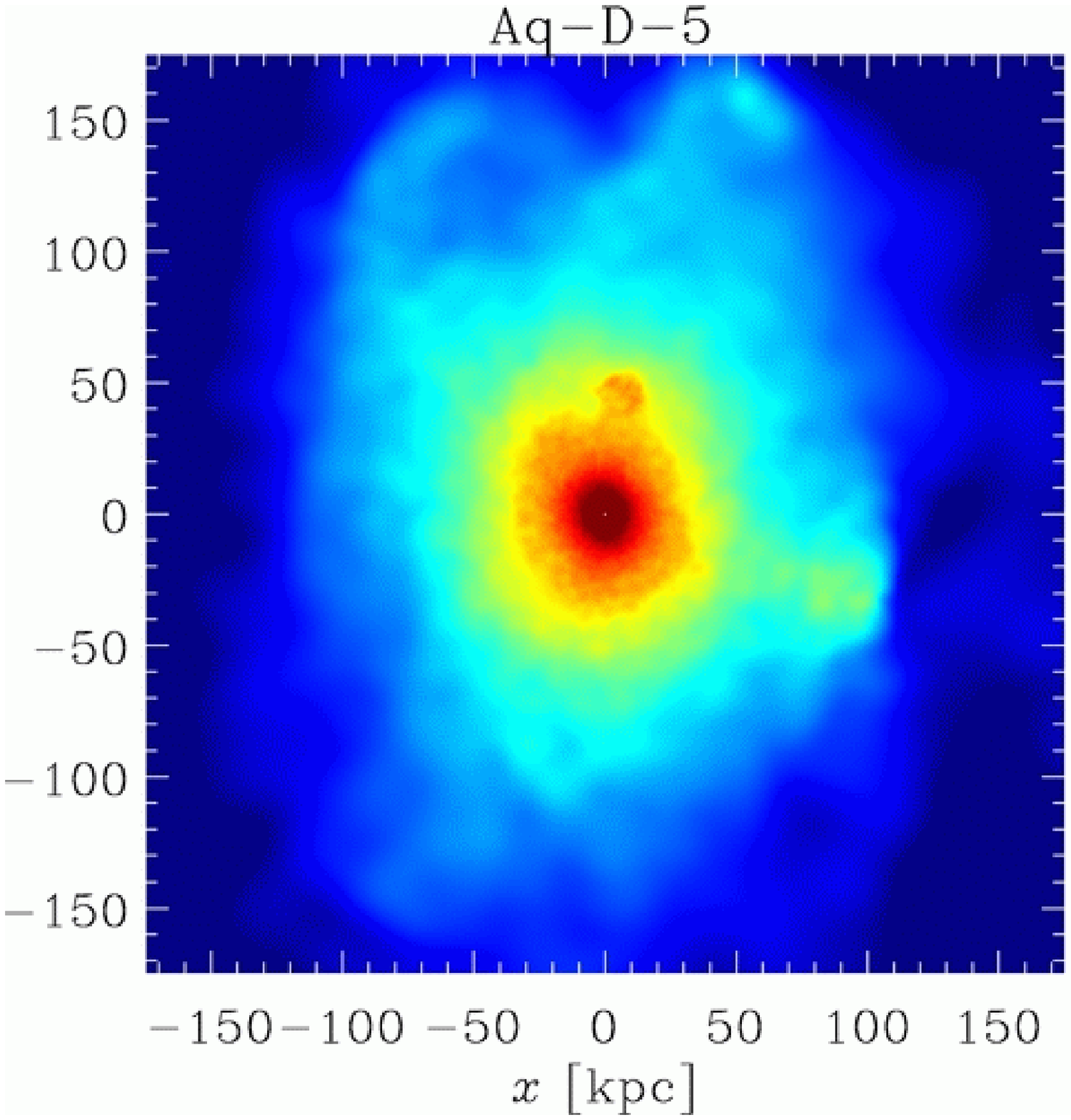}}
\hspace{-0.5cm}\
{\includegraphics[width=45mm]{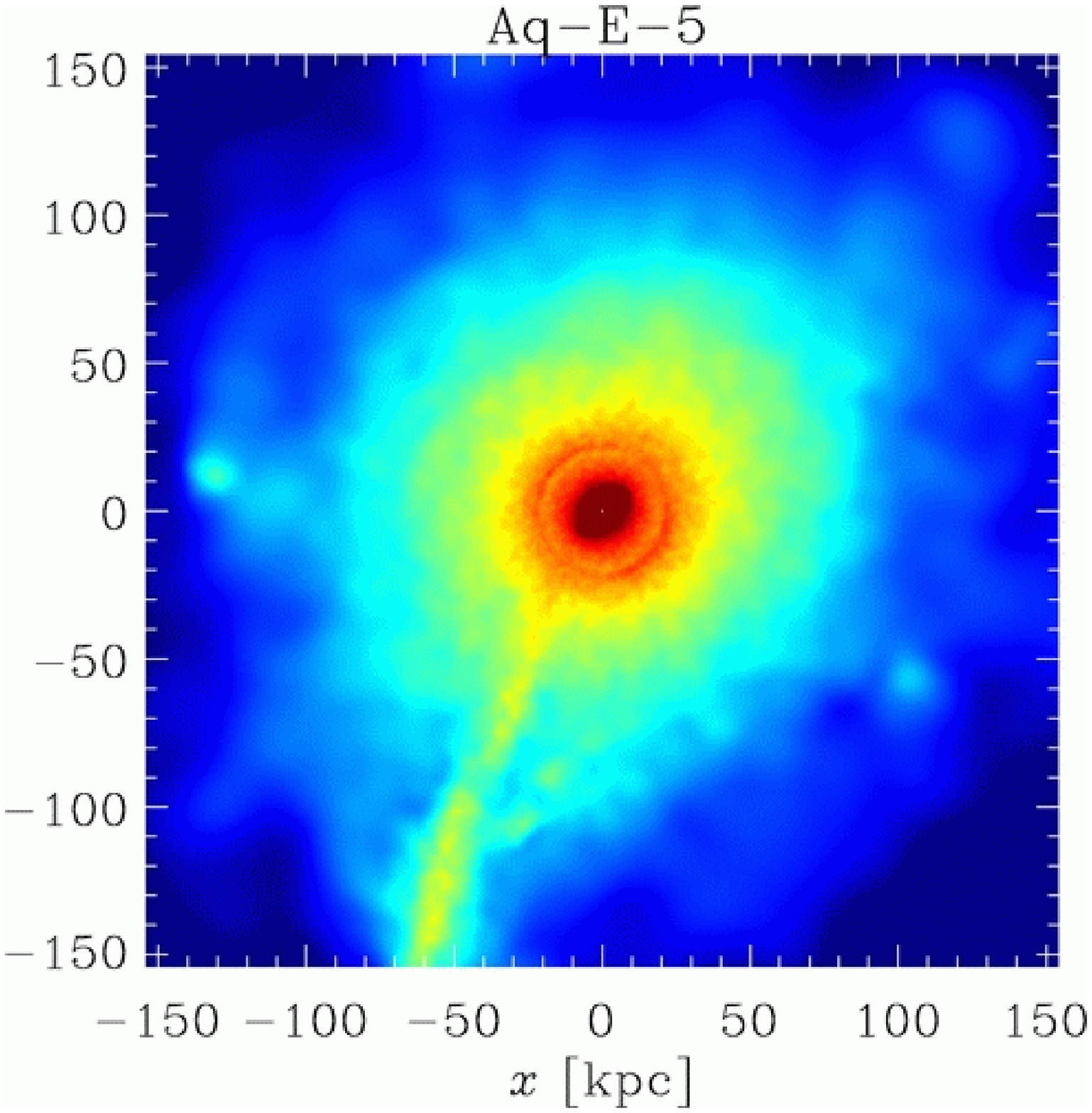}\includegraphics[width=45mm]{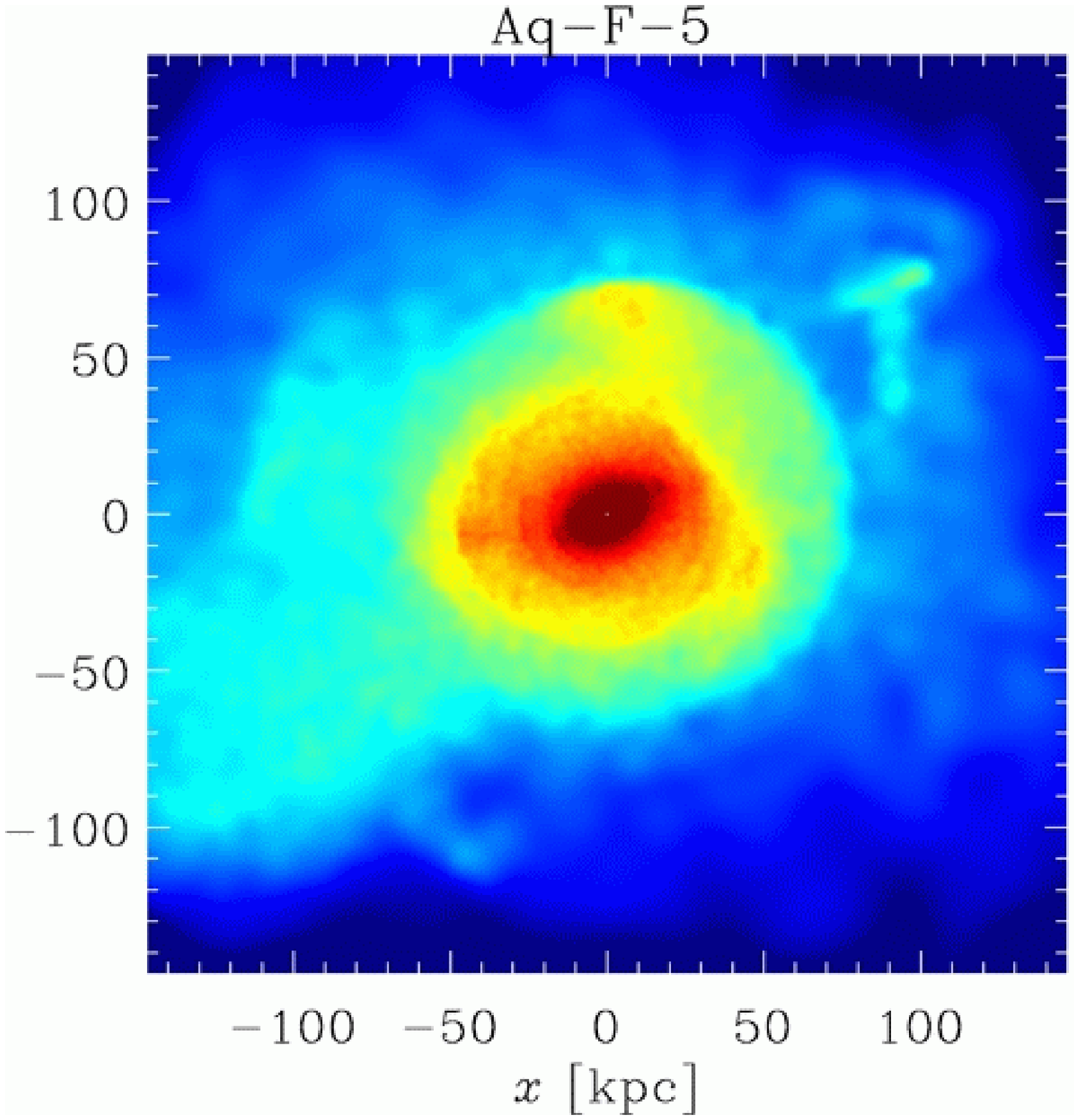}\includegraphics[width=45mm]{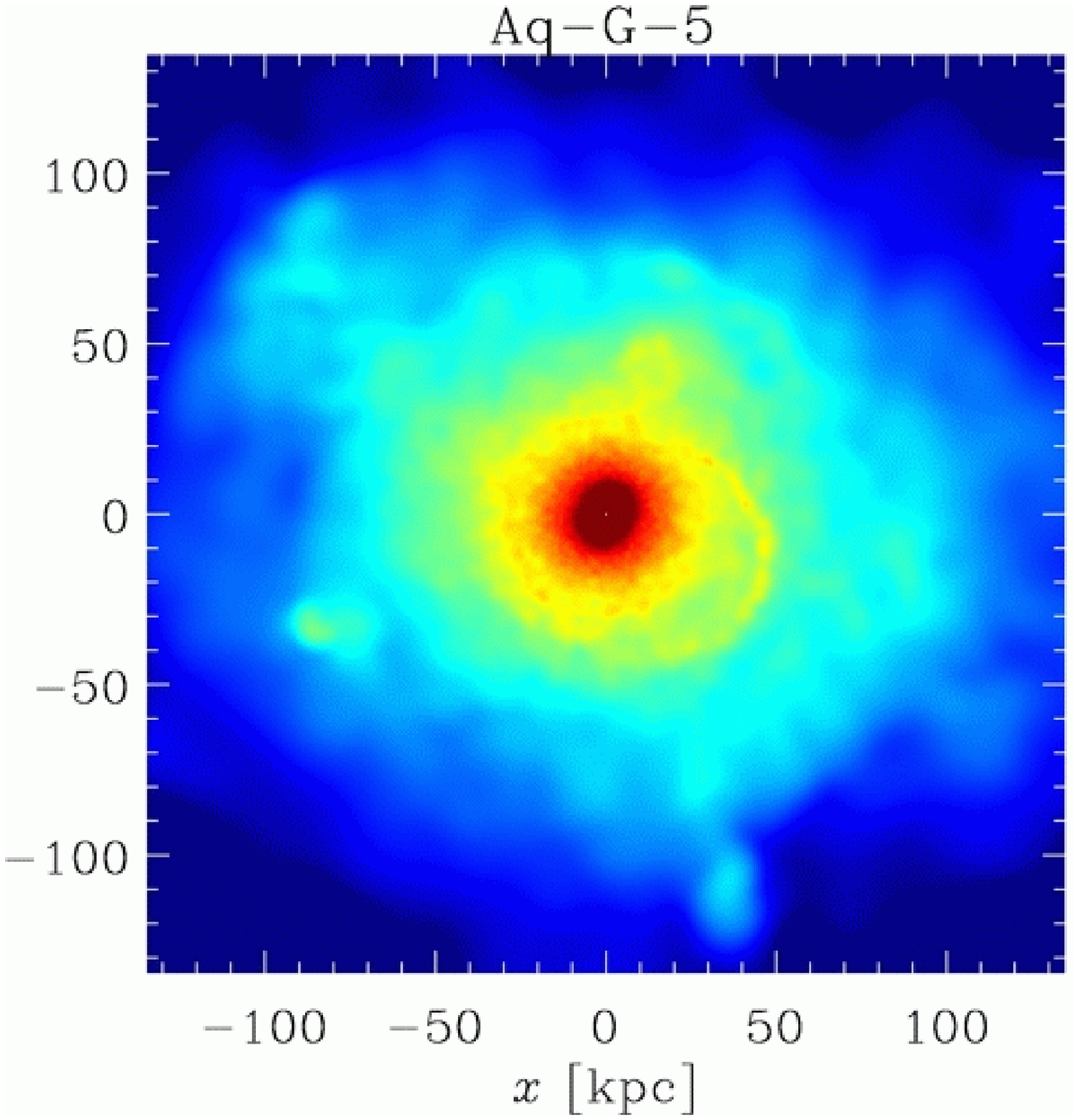}\includegraphics[width=45mm]{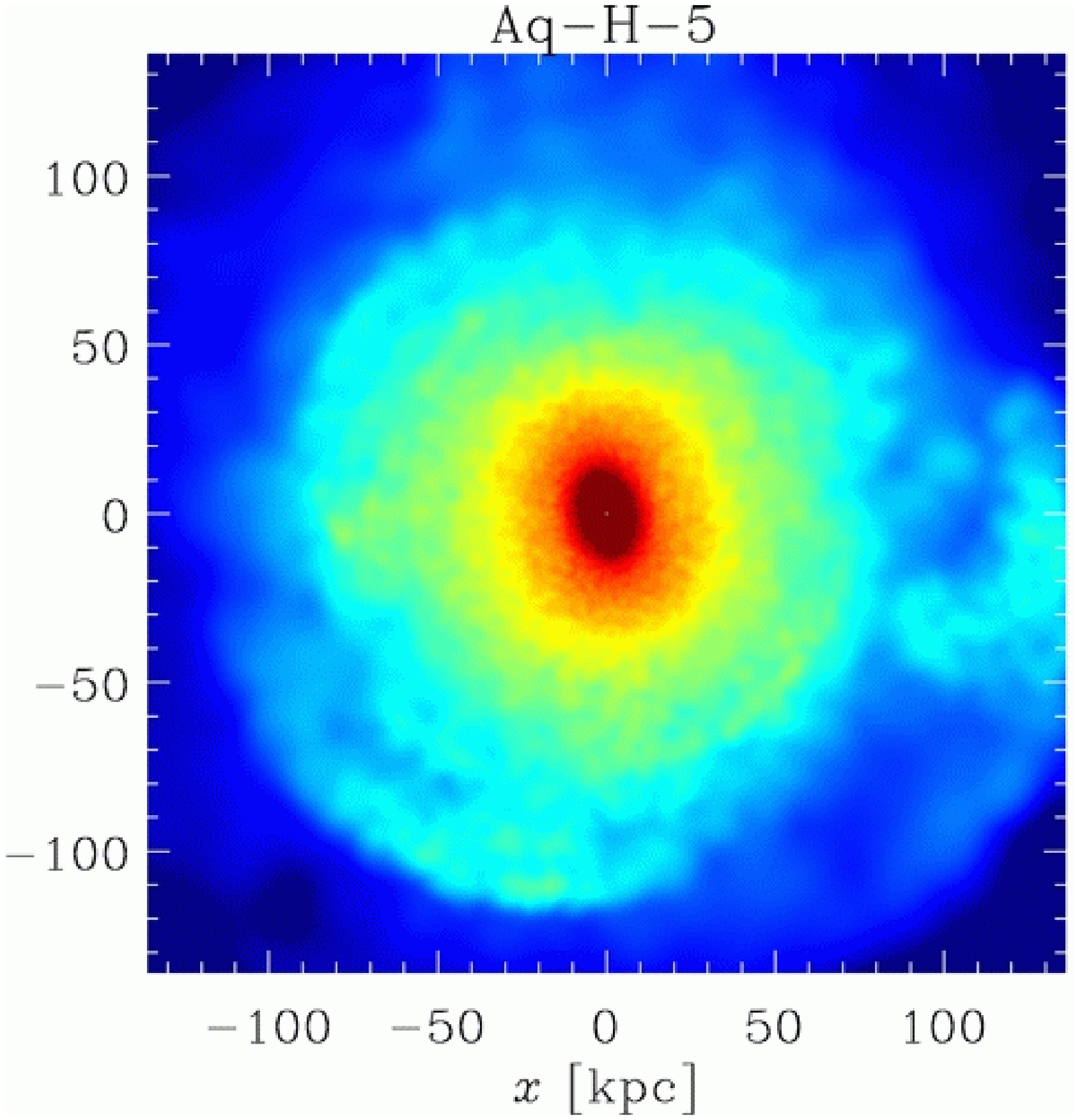}}
\end{center}
\caption{Spatial structure of simulated outer spheroids, up to
  $0.75\,r_{\rm 200}$.  Colors represent projected surface mass density,
  on a logarithmic scale, covering $4$ orders of magnitude. }
\label{maps_outerspheroid}
\end{figure*}

In this section, we discuss the structure of the inner and outer
spheroids of our simulated galaxies.  In
Figs.~\ref{maps_innerspheroid} and \ref{maps_outerspheroid}, we show
maps of surface mass density for inner spheroids (up to $0.5\times
r_{\rm opt}$) and outer spheroids (up to $0.75\times r_{\rm 200}$),
respectively.  The plots are color-coded according to the logarithm of
the projected surface mass density, covering $4$ orders of magnitude
and starting at the maximum density corresponding to each simulation.
The most striking features of the inner spheroids are the presence of
bars (also evident in the velocity distributions, as we discuss below),
and the diversity in shape: in some cases they are almost
axisymmetric, in others strongly ellipsoidal.  Our outer spheroids are
highly asymmetric and rich in structure, in particular, in the
outermost regions, where there are stellar ``streams" and clumps. Note
that these are not gravitationally self-bound systems, and they have
no dark matter halo associated with them, since satellites with dark
matter haloes have already been removed from the analysis.  Aq-F-5
shows the most complex and perturbed stellar distribution, due to its
recent major merger.

\begin{figure*}
\includegraphics[width=175mm]{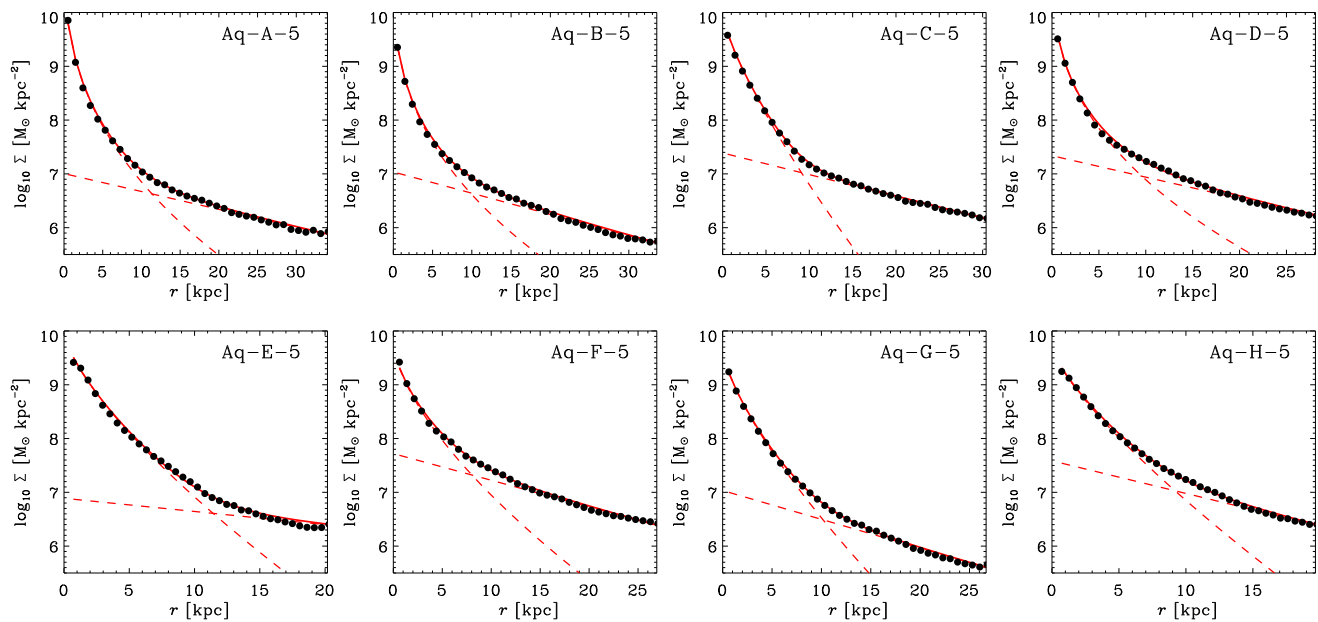}
\caption{Surface mass density profiles projected onto the disc plane
  (filled circles) for the spheroidal components of simulated galaxies
  (up to $2$ times the corresponding optical radius). The solid lines
  show the best fit model assuming a S\'ersic profile in the inner
  parts and a declining exponential law for the outer regions. The
  dashed lines show their relative contributions. }
\label{profiles}
\end{figure*}

In order to investigate the distribution of stars in the spheroidal
components in more detail, we constructed profiles of surface mass
density  (using equally-spaced
bins in $r$) for all simulations, as shown in Fig.~\ref{profiles}. We used
stellar mass and not luminosity, but since these are old populations,
very similar results are found in the two cases.  The profiles were
fitted using a combination of S\'ersic and exponential profiles for
the inner and outer regions, respectively, such that their sum gives
the total surface mass density, namely:
\begin{equation}
\Sigma(r) = \Sigma_{\rm eff}\,{\rm exp}\left[-b_n\left ( \left ({r\over{r_{\rm eff}}}\right )^{1\over{n}}-1\right)\right] + \Sigma_{\rm o}{\rm exp}\left({r\over{r_{\rm o}}}\right ).
\end{equation}
The first term of the right-hand side in this equation is
equivalent to the standard S\'ersic
law, i.e.,
\begin{equation}
\centerline{$\Sigma(r) = \Sigma_{\rm b} {\rm exp}\left [-\left ({r\over{r_{\rm b}}}\right )^{1\over{n}}\right]$} ,
\end{equation}
but has the advantage that it can be expressed in terms of effective
quantities usually quoted in observational studies (e.g. MacArthur et
al. 2003).  We did additional tests replacing the exponential law used
for the outer regions by either a power law or a S\'ersic profile, but
we found that, in general, the fits were poorer.

In order to properly model the break at which the transition between
the exponential and the S\'ersic law takes place, which is clearly
different for the eight simulations, we repeated the fits changing the
break radius and estimated the goodness of the fit using the parameter
$Q$, defined by:
\begin{equation}
Q \equiv {1\over{n}} \ {\sum_{\rm i=1}^{n} ({\rm log_{10}\Sigma}_{\rm i}-{\rm log_{10}\Sigma} _{\rm i,fit})^2}
\end{equation}
where $n$ is the number of data points.
By minimizing $Q$ we selected the best-fit parameters,
which we show in Table~\ref{disk_bulge}.  Furthermore, for each
simulation we have repeated the fits twice, considering the profiles
either to $1.5$ or $2$ times the corresponding optical radii, and
found almost no difference in the best-fit parameters (in all cases
lower than a few percent)\footnote{The fits have also been  repeated
  for the whole radial extent of the outer spheroids, but were then
  poorer. Moreover, since the fraction of stellar mass decreases
  rapidly with increasing radius, the best-fit model does not differ
  strongly with that obtained using the distributions only up to
  $2\times r_{\rm opt}$. Finally, we note that the density profiles in
  the outer regions for the eight galaxies are diverse, and
  show signs of streams and clumps, making a fit almost meaningless.}.  
In Fig.~\ref{profiles} we show the best-fit model
(solid line) as well as the relative contributions of the S\'ersic and
exponential laws (dashed lines).

The inner spheroids (``bulges'' if bars were absent) are characterized
by central surface mass densities between $10^{8.5}$ and $10^9$
M$_\odot$ kpc$^{-2}$, and effective radii in the range $1.5-2.5$
kpc. In terms of their shapes, the S\'ersic parameters $n$ are
typically around $1$, i.e., similar to exponential profiles. Aq-A-5
and Aq-B-5 have the highest $n$ parameters, a reflection of their very
concentrated profiles; on the contrary, Aq-H-5 has the lowest $n$
value indicating a shallower profile in the central regions. The outer
regions of the spheroids have much lower central surface densities,
with $\Sigma_{\rm o}\sim 10^{7}-10^{8}$ M$_\odot$ kpc$^{-2}$, and
scale-lengths of the order of $4-10$ kpc. We find a wide range of
relative contributions of the inner and outer components to the total
spheroid mass, as can be clearly seen in Fig.~\ref{profiles}.   
We note
that, in the cases where bars are present, the profiles should perhaps
include an additional component. This might explain the (small)
differences between the results here and those presented in Scannapieco
et al. (2010)\footnote{Another difference between the technique used
  here and that used in Scannapieco et al. (2010) is that the former
  uses the 1D (mass) density profiles while the latter considers the 2D
  (luminosity-weighted) images.}. However, we do not find larger $Q$ values
for galaxies with bars, suggesting that the number of fitting parameters
is large enough to  fit the 1D profiles of
Fig.~\ref{profiles}, even if bars are not included as an extra component.

\begin{table*} 
\begin{small}
\caption{Disc half-mass radii ($r_{\rm d}$) and best fit values for
the inner and outer spheroid parameters: log$_{10}$($\Sigma_{\rm eff}$), $r_{\rm eff}$, $n$ (S\'ersic law), and log$_{10}$($\Sigma_{\rm o}$) and $r_{\rm o}$ (exponential). Units are kpc
for $r_{\rm d}$, $r_{\rm eff}$ and $r_{\rm o}$  and M$_\odot$ kpc$^{-2}$ for  $\Sigma_{\rm eff}$
and  $\Sigma_{\rm o}$. We also show the   $Q$ parameter used to test the goodness of the fits, as explained in the text.}
\vspace{0.1cm}
\label{disk_bulge}
\begin{center}
\begin{tabular}{lccccccc}
\hline
Galaxy  &   $r_{\rm d}$ &  log$_{10}$($\Sigma_{\rm eff}$) &  $r_{\rm eff}$ &  $n$ &  log$_{10}$($\Sigma_{\rm o}$) & $r_{\rm o}$ & $Q$\\\hline

Aq-A-5    &  21.2 &  9.1&      1.5&      2.5&      7.1&      12.1&
    0.008\\

Aq-B-5    & 24.0 &  8.6&      1.8&      2.7&      7.1&      10.6&
    0.012\\

Aq-C-5   & 12.2  &  8.9&      2.3&      1.2&      7.4&      10.3&
    0.003\\

Aq-D-5   &  11.0 &     8.8&      1.9&      2.5&      7.4&      10.3&
    0.009\\

Aq-E-5   &  12.9 &8.8&      2.5&      1.2&      7.1&      12.3&
    0.016\\

Aq-F-5   &  -  & 8.5&      2.8&      1.7&      7.5&      10.5&
    0.006\\

Aq-G-5  & 10.8 &  8.6&      2.2&      1.5&      7.0&      8.1&
    0.008\\

Aq-H-5   &  10.6& 8.6&      2.6&      1.3&      7.6&      6.9&
    0.005\\
 
\\

Aq-C-6 & 8.2  & 8.6 & 2.6 & 1.0 & 7.3 & 8.9 & 0.012 \\

Aq-E-6b & 12.0 & 8.7 & 2.7 & 1.2 & 7.0 & 11.2 & 0.010 \\

Aq-E-6 & 12.9 & 8.2 & 3.8 & 0.9 & 7.3 & 7.4 & 0.016 \\

\hline
\end{tabular}
\end{center}
\end{small}
\end{table*}

The structural properties of the inner and outer spheroids show very good
agreement with the results for simulations with lower resolution (Table~\ref{disk_bulge}).
We find that  all the fitting parameters change by less  than $10\%$,
with the exception of Aq-E-6 where larger differences ($25-40\%$) are found
for the effective radius, shape parameter and outer spheroid scale-lengths.

\subsection{In-situ fractions}

\begin{figure*}
\includegraphics[width=175mm]{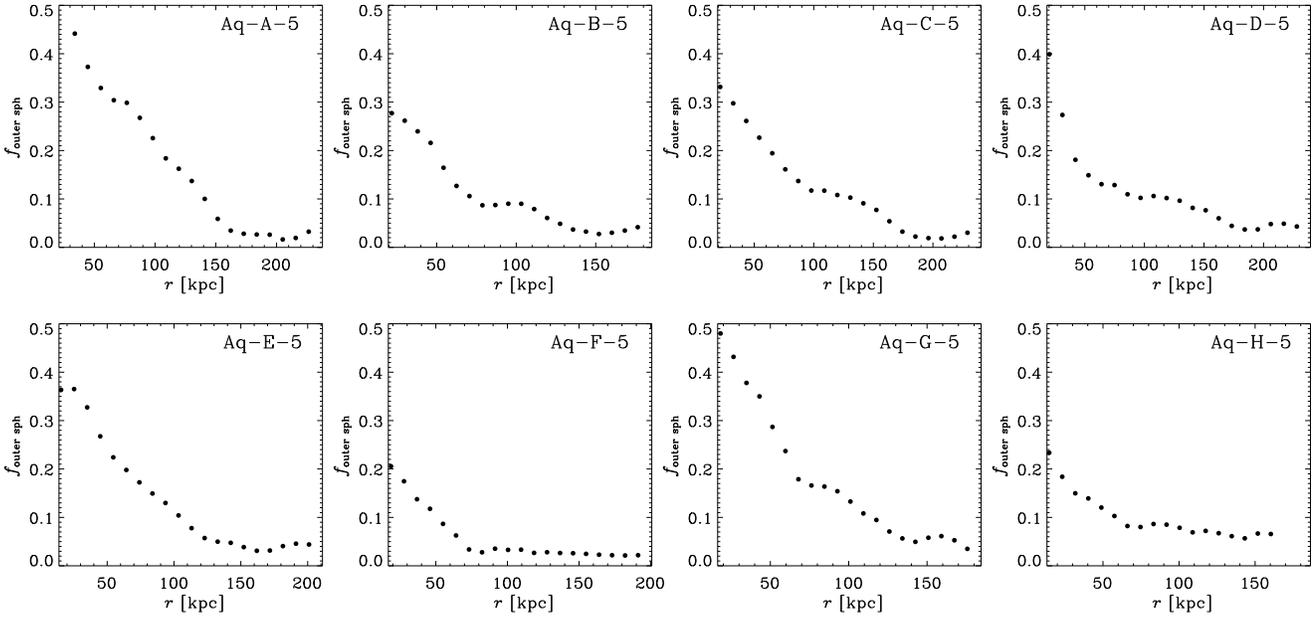}
\caption{{\it In-situ} fraction as a function of radius for stars in the outer spheroidal
components of our simulated galaxies. The overall {\it in-situ} fractions for outer
spheroids are listed in Table~\ref{table_insitu}.}
\label{in_situ_vs_r_outer}
\end{figure*}

The {\it in-situ} fractions for inner and outer spheroids are shown in
Table~\ref{table_insitu}.  For inner spheroids we find, as we did for
discs, that the majority of stars which end up in this component
formed in the galaxy's main progenitor.  {\it In-situ} fractions for
inner spheroids are, in all cases, larger than $0.75$, with the
exception of Aq-F-5 --- due to its recent major merger which brings a
significant amount of new stars into the system.  Conversely, for outer
spheroids we find lower {\it in-situ} fractions, in the range
$0.15-0.35$.  This translates into a high contribution of accreted
stars; in general, $\gtrsim 65\%$ of outer spheroid stars formed in
satellites that later fall into the potential well of the main
progenitor and  disrupt.  Our results clearly show that
``stellar haloes" are more likely formed by a combination of both {\it
  in-situ} and accreted stars.  Furthermore, the {\it in-situ}
fractions of outer spheroids show a clear anti-correlation with
radius, as shown in Fig.~\ref{in_situ_vs_r_outer}.  Our results are in
good agreement with those of Zolotov et al. (2009), who detect an
anticorrelation between {\it in-situ} fraction and radius for stellar
haloes, and also a wide range of stellar halo {\it in-situ} fractions,
which arise due to the variety of merger histories of their simulated
galaxies.

The {\it in-situ} fractions for the inner spheroidal components in
the three lower resolution simulations are in excellent agreement  with
their higher resolution counterparts, with differences smaller than $4\%$ 
(Table~\ref{table_insitu}).
For outer spheroids, the differences are larger, in particular for Aq-C-6 which
has a  $40\%$ lower {\it in-situ fraction} compared to Aq-C-5.
For Aq-E-6b and Aq-E-6, differences are smaller ($11\%$ and $4\%$ respectively)
with respect to Aq-E-5. Note that outer spheroids are expected to
be more strongly affected than the inner components, 
due to poorer resolution in the low mass satellites which contribute
many of their stars.

\subsection{Dynamical properties}

The dynamical structure of simulated inner spheroids can be read from
Fig.~\ref{vtita_maps_stars}, where we show the face-on 2D distribution
of mean tangential velocities (note that all stars, including disc and
spheroidal components, are shown in these maps).  The relative
contribution and size of bulges and bars can be clearly seen in these
plots: Aq-A-5 has an asymmetric velocity distribution even in the very
central regions; Aq-B-5 has an extended bulge with a smooth transition
in velocity increasing towards the disc region; Aq-C-5 has a more
extended disc and the bulge (the region where no significant tangential
velocity is detected) is relatively small; Aq-D-5 has a spherical
bulge with a rapid transition to the disc region; Aq-E-5 shows signs
of rotation from the very inner regions and its velocity structure is
also rather asymmetric; Aq-F-5, the only galaxy with no disc
component, has very low tangential velocities regardless of distance
to the centre; Aq-G-5 has no clear indication of a significant bulge
(but it has a strong bar, see below); and Aq-H-5 has an asymmetric
velocity structure with low rotation (this is also one of the less
massive galaxies).  Furthermore, the presence of bars is clear in
Aq-C-5, Aq-E-5 and Aq-G-5.  In this latter case, the bar seems to
dominate the inner regions, and there is no clear indication of a
significant true bulge.  A more detailed study in terms of the
dynamical properties of stars, and also through the use of
bulge/disc/bar decompositions (Scannapieco et al.  2010) shows that
Aq-A-5 also has an important bar component (see also
Fig.~\ref{maps_innerspheroid}).

\begin{figure*}
\includegraphics[width=175mm]{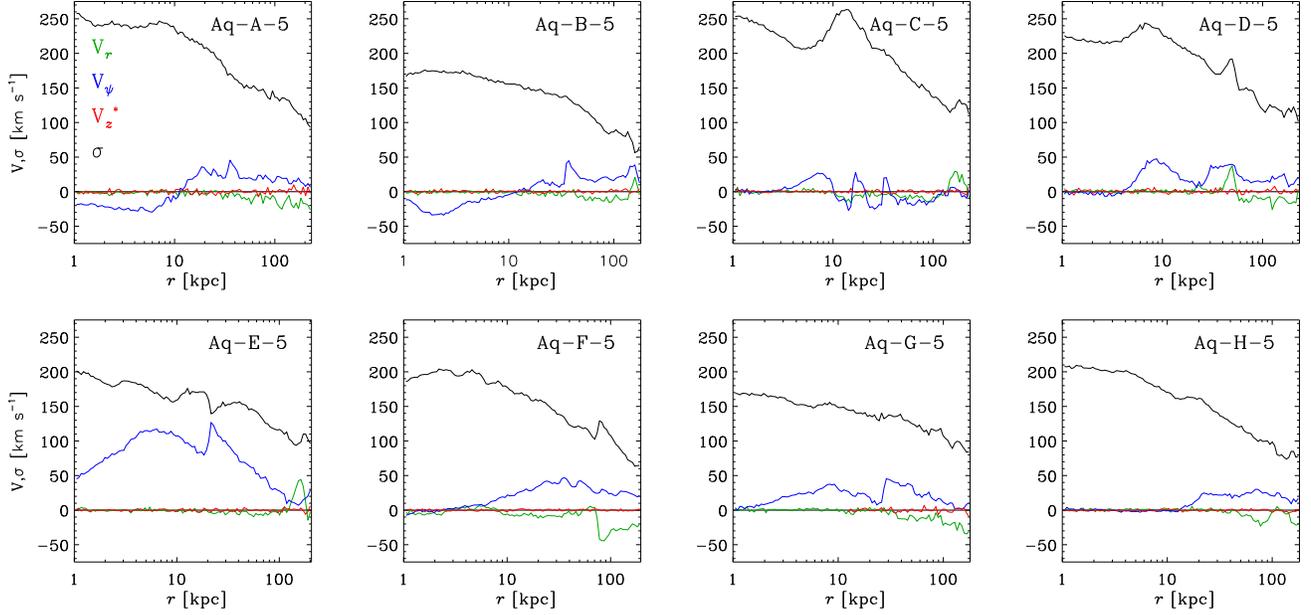}
\caption{Radial (green), tangential (blue) and vertical (red)
  velocities as a function of projected radius for the spheroidal
  components of our simulated galaxies (up to the corresponding virial
  radii). In the case of the vertical velocities, we use $V_z^*\equiv
  V_z\ {\rm sign}(z)$, in order to distinguish between inflows
  ($V_z^*<0$) and outflows ($V_z^*>0$).  We also show the profiles of
  total stellar velocity dispersion (black lines).}
\label{vel_and_sigma_vs_r_spheroids}
\end{figure*}

In Fig.~\ref{vel_and_sigma_vs_r_spheroids} we show the profiles of
radial, tangential and vertical velocities for spheroid stars, as well
as the corresponding total velocity dispersions.  For the vertical
velocities, we use $V_z^*\equiv V_z\ sign(z)$, in order to distinguish
between inflows ($V_z^*<0$) and outflows ($V_z^*>0$).  The radial and
vertical velocity components are very small in all simulations.  In
some cases, there are signs of non-zero but small radial velocities in
the outermost regions, but these are not significant, specially taking
into account that the amount of stellar mass in these regions is very
small.  We do find non-zero tangential velocities, with a great
variety of patterns: Aq-A-5 and Aq-B-5 are counter-rotating in the
inner parts and co-rotating in the outer regions (always in relation
to the overall rotation of the disc); the spheroidal component of
Aq-E-5 has a net rotation (S09); the other galaxies show signs of
co-rotation, mainly outside the inner regions. The total velocity
dispersions decline with radius, varying typically from $150-250$ km
s$^{-1}$ in the inner regions to $50-100$ km s$^{-1}$ near the virial
radius.

\begin{figure*}
\includegraphics[width=175mm]{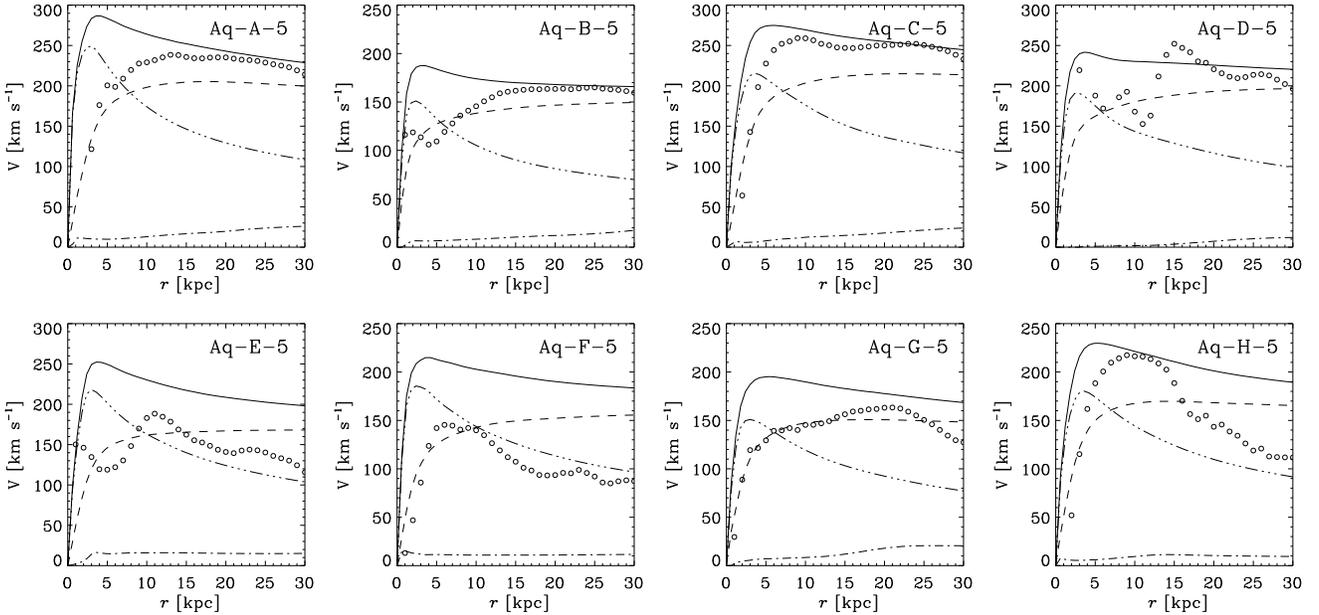}
\caption{Circular velocity curves for our eight simulated systems. 
  Solid lines are total circular velocities, while dashed,
   triple dotted-dashed  and dotted-dashed  lines are for the dark
  matter, stellar and gaseous components, respectively.  Open circles
  show the mean tangential velocities for the gas components -- in
  this case, the tangential velocities are calculated after projecting
  galaxies according to the angular momentum of the gas
  component. This is not always aligned with the angular momentum of
  stars.}
\label{rot_curves}
\end{figure*}

The dynamical properties of the inner and outer spheroids in our
lower resolution
simulations are in good agreement with their higher resolution counterparts,
and span the same ranges in the three components of the velocity
and of the velocity dispersion.

\section{Circular velocity curves and gas dynamics}\label{sec_vcirc}

Fig.~\ref{rot_curves} shows the total circular velocities, $V_{\rm
  circ}= \sqrt{GM(r) / r}$ as a function of radius for our eight
simulations (solid lines).  All simulations have similar circular
velocity curves, reaching maximum $V_{\rm circ}$ at small radii, and
declining smoothly at larger distances.  The characteristics of the
circular velocity profiles are determined by the mass distributions
and relative mass contributions of the stellar, gaseous and dark
matter components.  In this way, the shape of the circular velocity
curves in the inner regions is dominated by the stellar spheroids,
which are massive and centrally concentrated (Fig.~\ref{profiles}).
The circular velocities for the stellar components are, in all cases,
strongly peaked at $r\sim 5$ kpc, as indicated by the triple
dotted-dashed lines.  At larger radii, the stellar mass grows only
modestly, and the stellar circular velocities decline.

\begin{figure*}

\hspace{-1cm}\includegraphics[width=160mm]{sim_names_A-D.ps}\vspace{-0.1cm}

\vspace{0.2cm}

{\includegraphics[width=40mm]{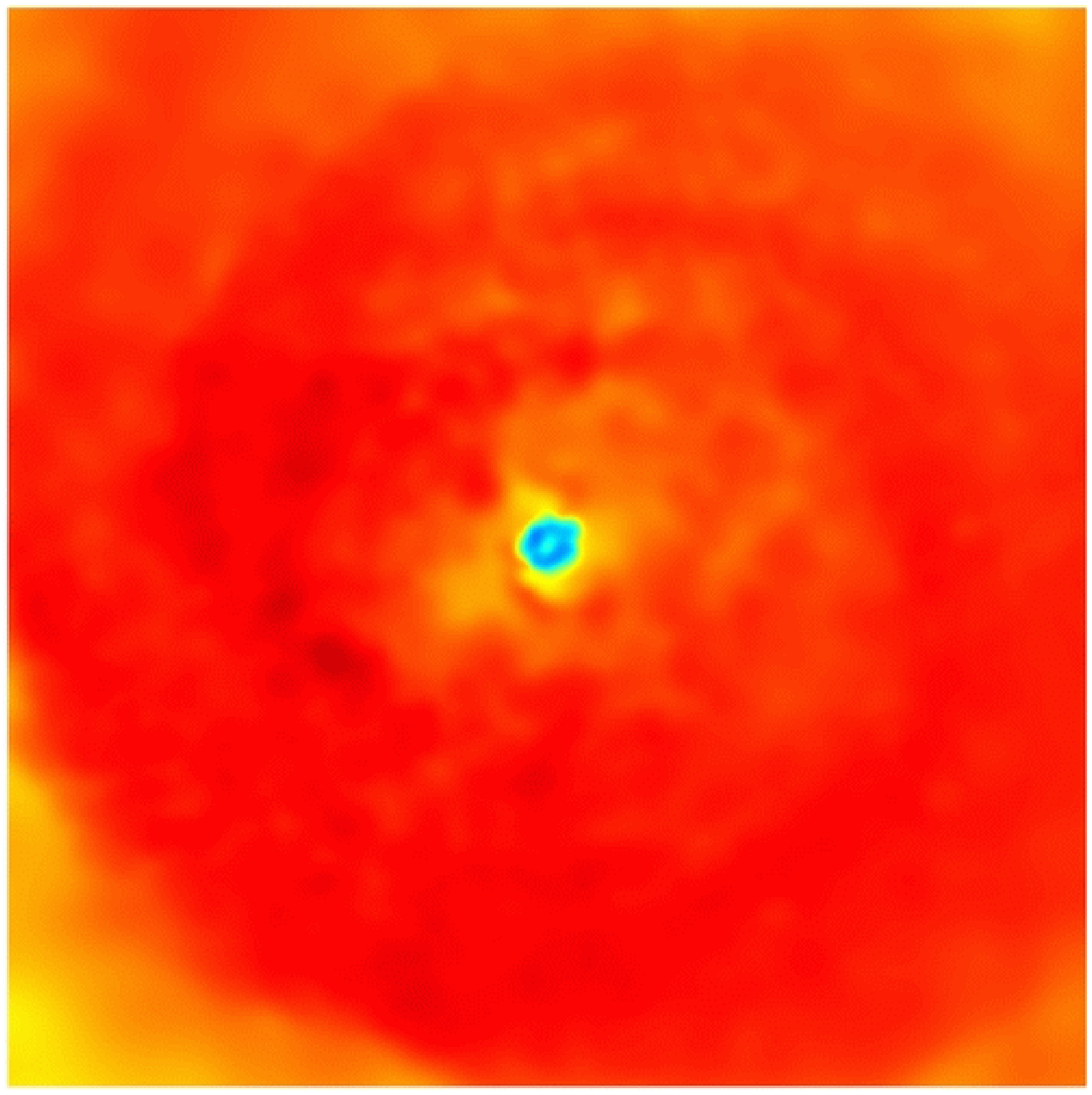}\includegraphics[width=40mm]{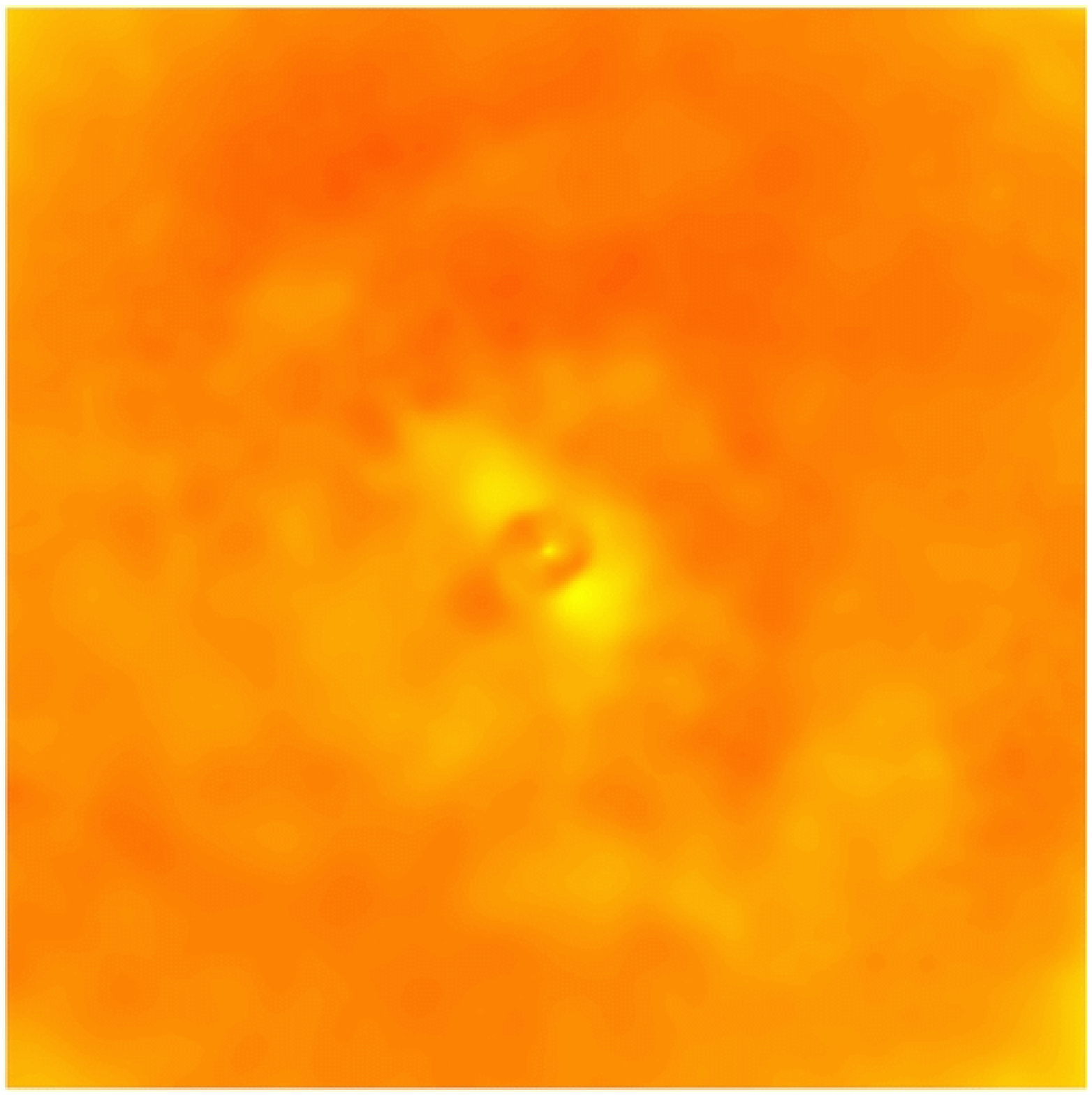}\includegraphics[width=40mm]{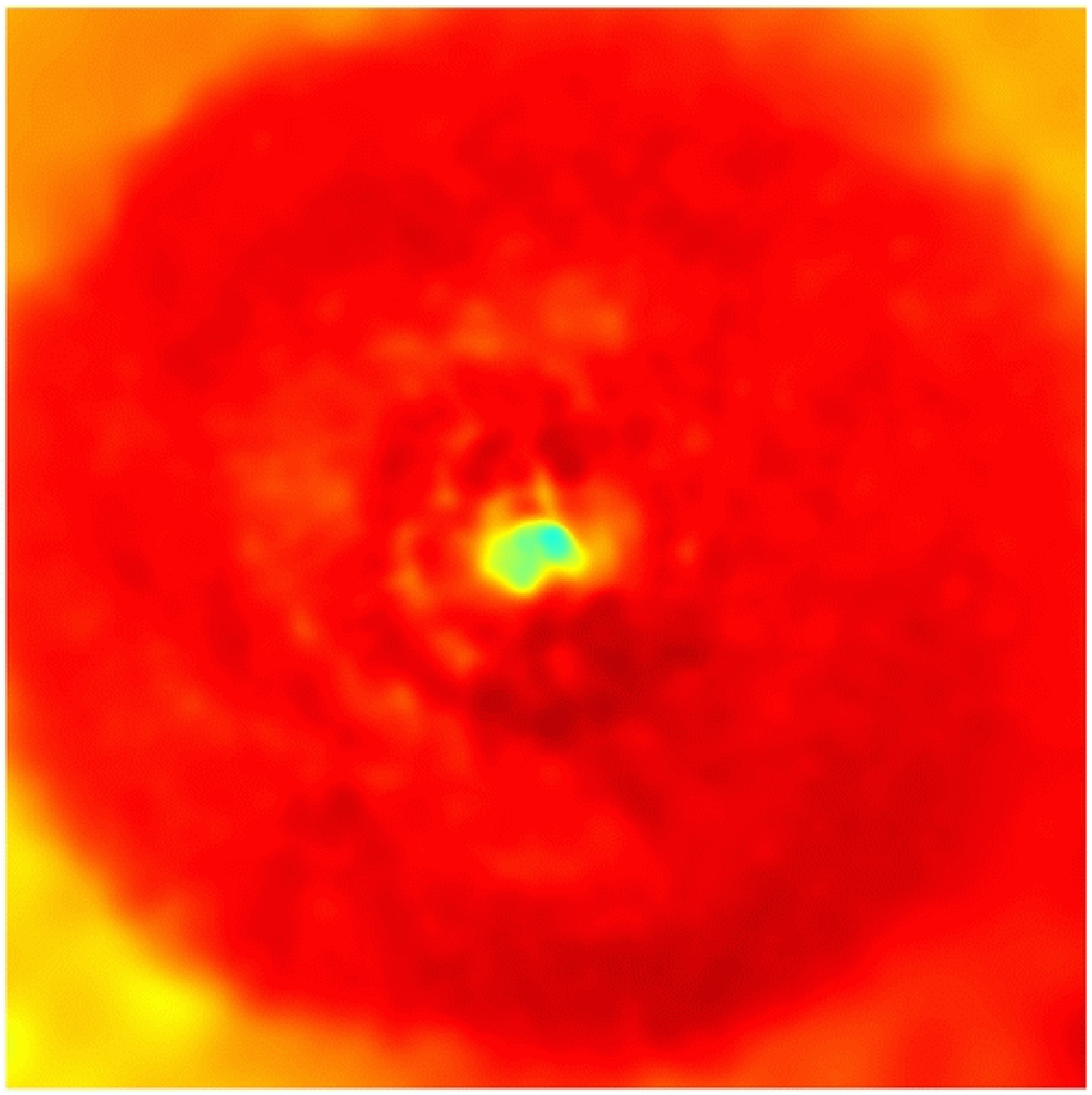}\includegraphics[width=40mm]{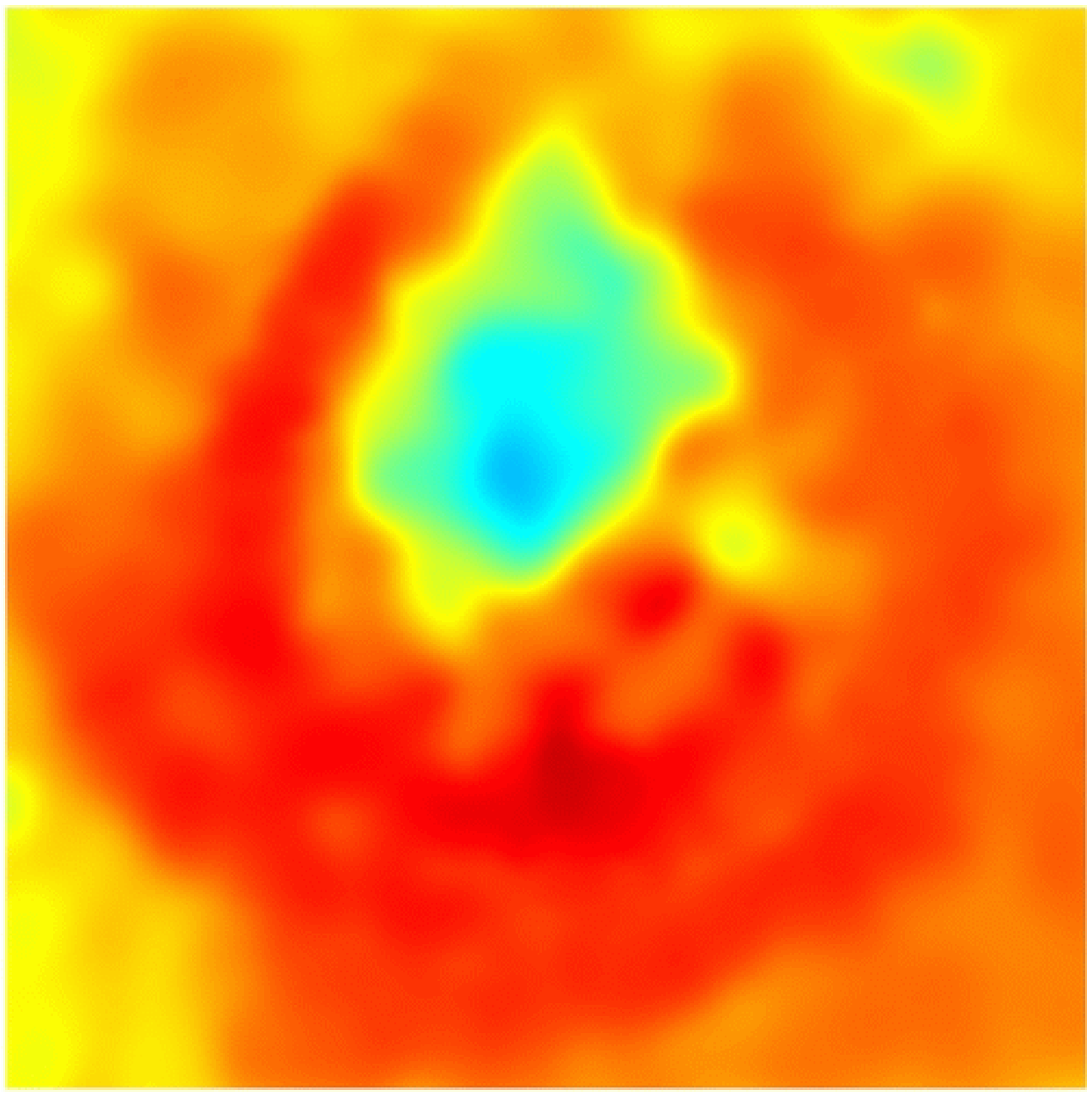}\includegraphics[width=14mm]{legend_vel_vert.eps}}

\hspace{-1cm}\includegraphics[width=160mm]{sim_names_E-H.ps}\vspace{-0.1cm}

\vspace{0.2cm}

{\includegraphics[width=40mm]{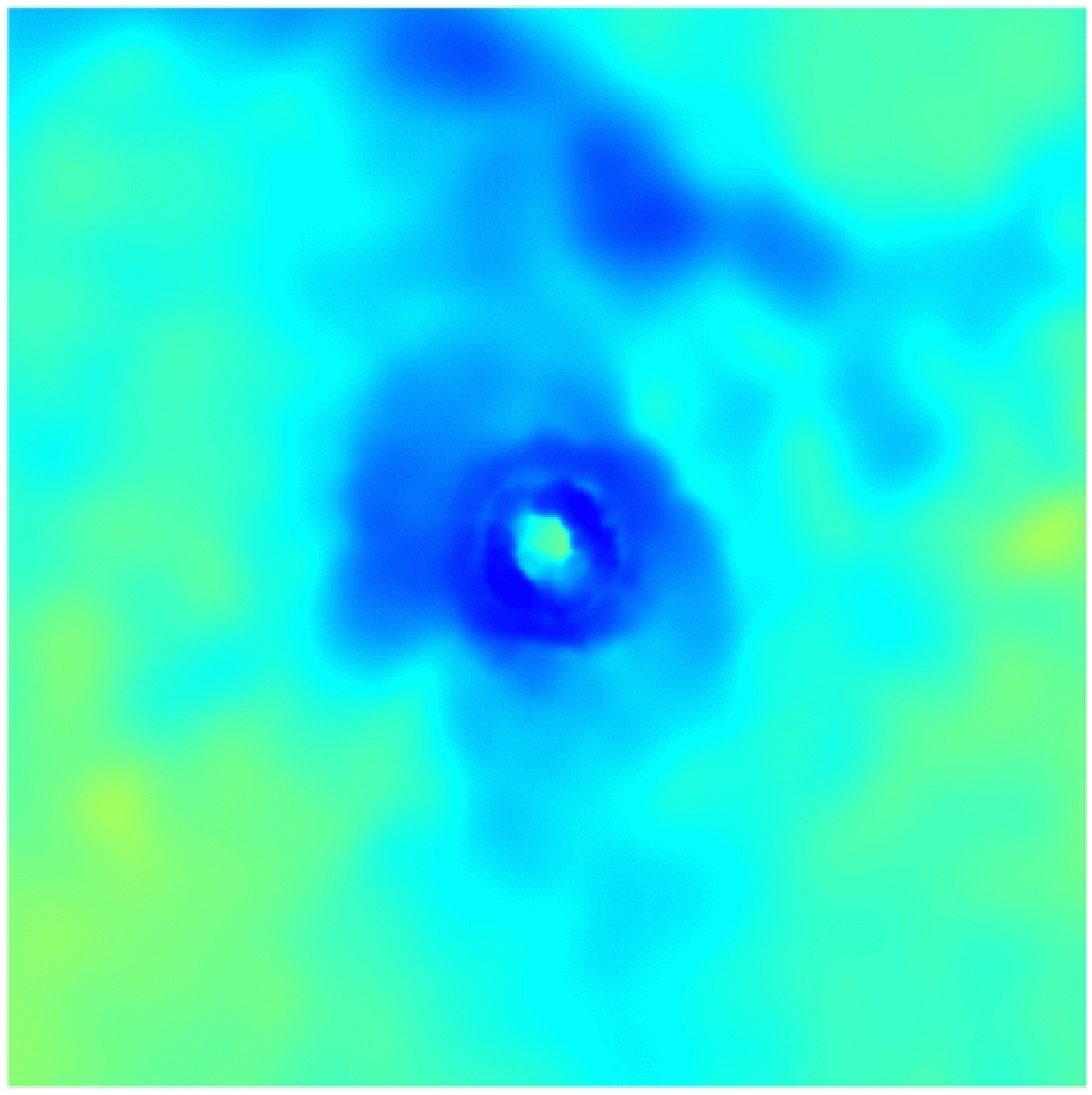}\includegraphics[width=40mm]{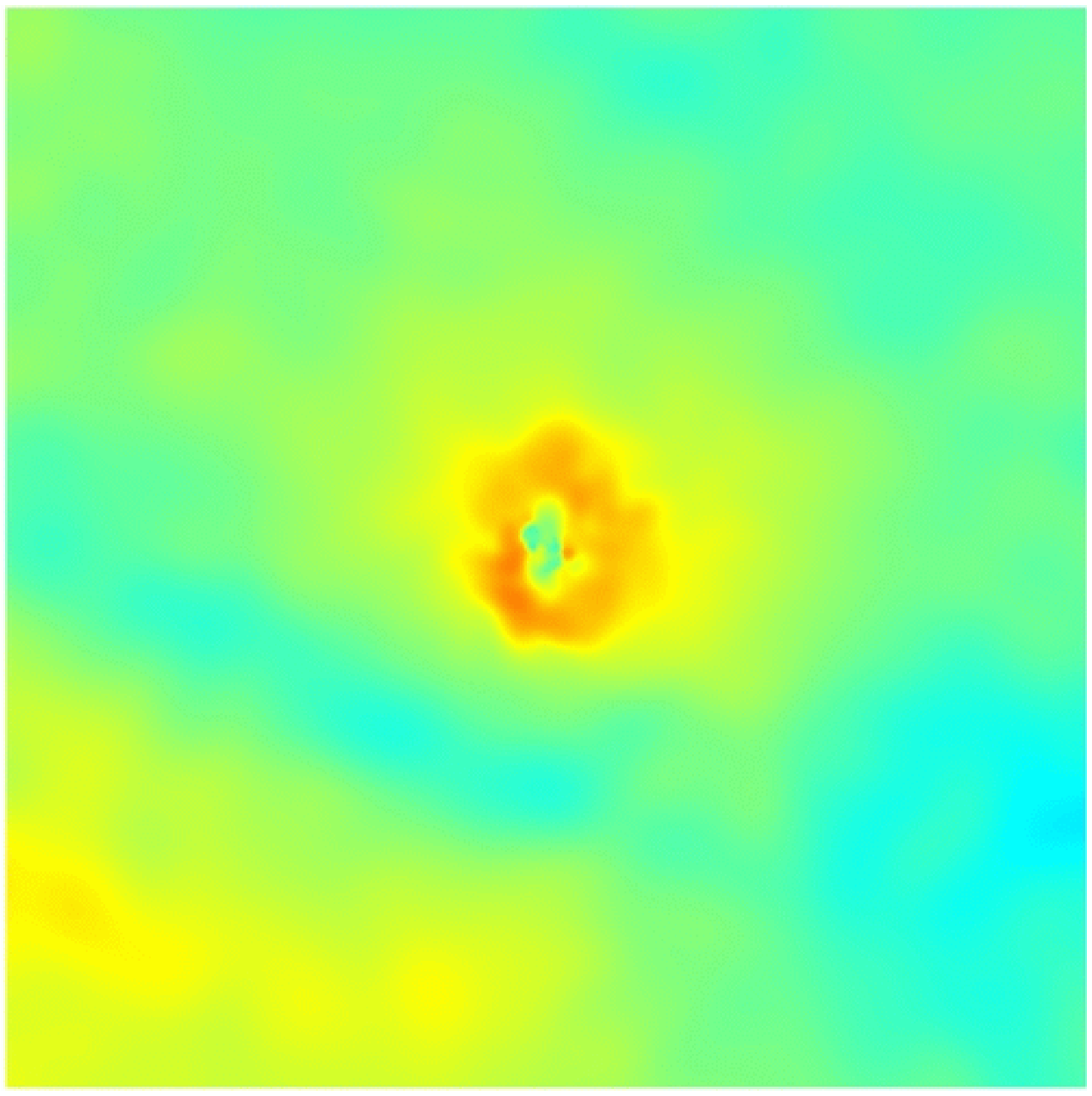}\includegraphics[width=40mm]{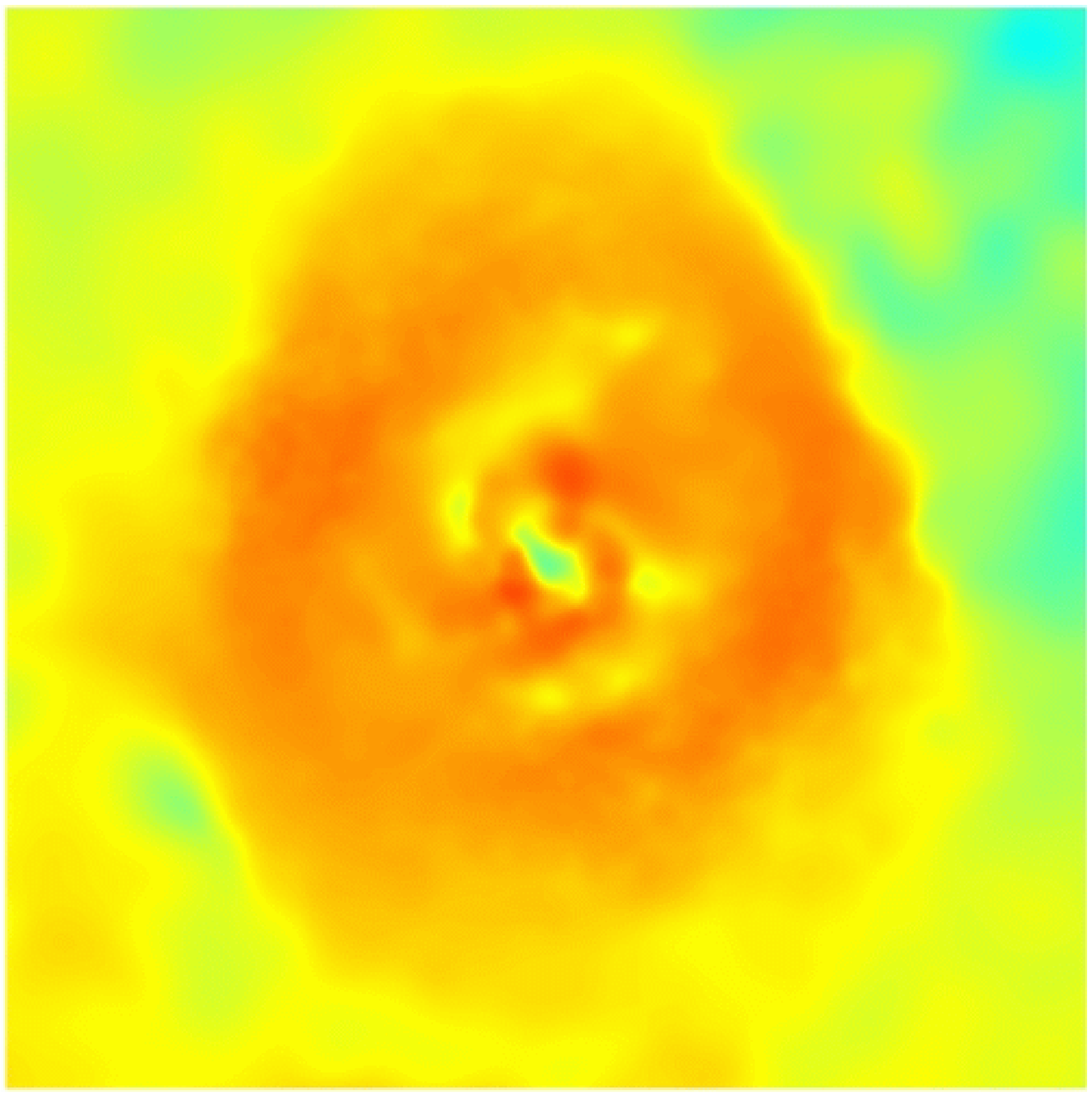}\includegraphics[width=40mm]{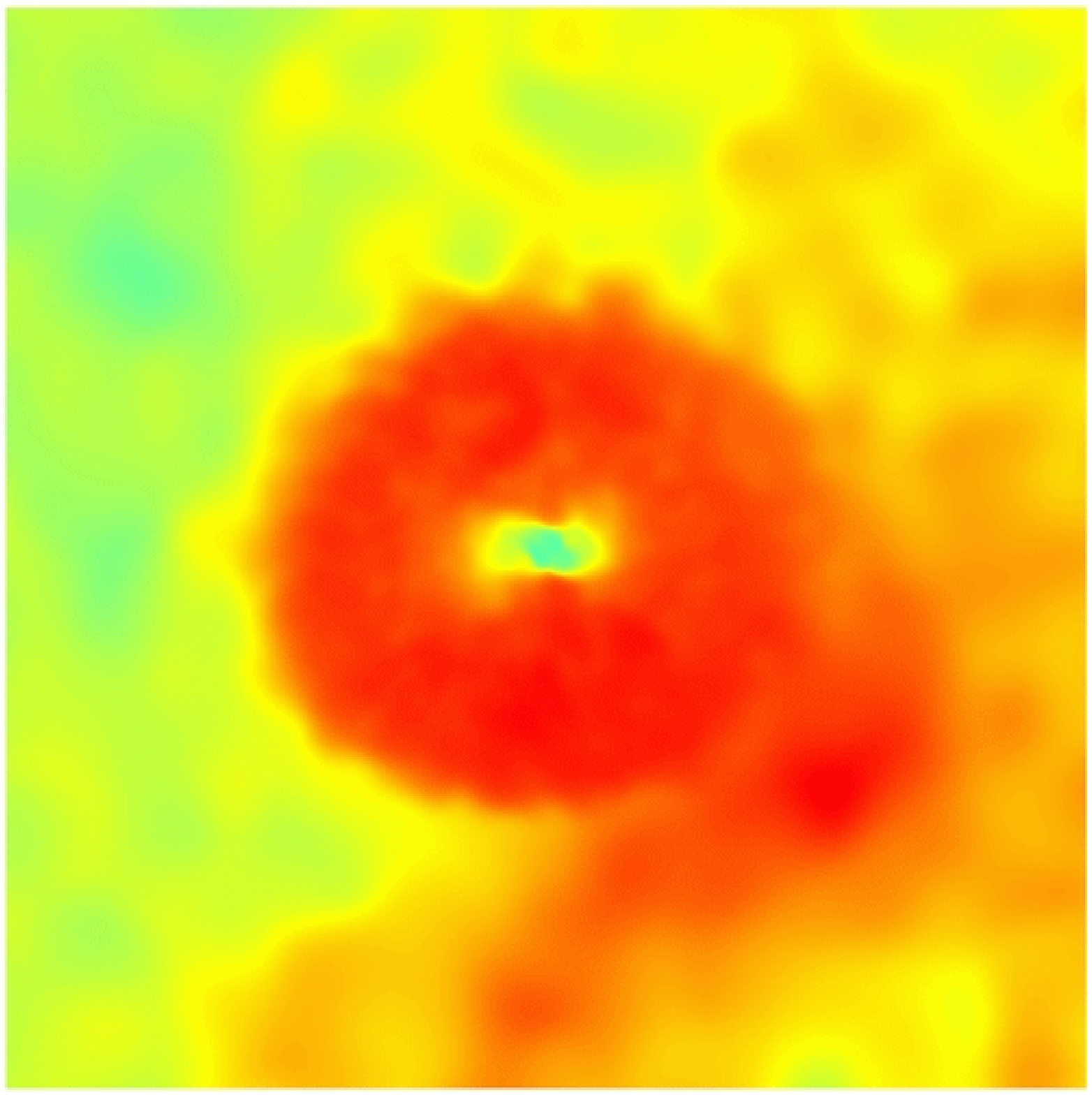}\includegraphics[width=14mm]{legend_vel_vert.eps}}

\caption{Maps of tangential velocity for the gas components of
our  simulated galaxies. In this projection, the stellar discs are
  face-on, and colors represent the mean tangential velocity in the
  corresponding bin, as indicated in the velocity scale. The plots are
  $30$ kpc across. }
\label{vtita_maps_gas}
\end{figure*}

At $r\gtrsim 5-10$ kpc, the total circular velocities start being
dominated by the dark matter. The dark matter components exhibit flat
circular velocity curves, as indicated by the dashed lines in
Fig.~\ref{rot_curves}.  However, due to the non-negligible
contribution of the stellar component, the total circular velocity curves are
not completely flat in the outer regions. The differences between the
peak circular velocity and the velocity at $r= 30$ kpc are of the
order of $10$ to $20\%$ and are larger than in most observed galaxies. 
This is a common problem in cosmological
galaxy formation simulations, but in our case it is not very severe.

In the case of the gas components (dotted-dashed lines), we find very
low circular velocities, indicative of the small amount of left over
cold gas in the central regions 
(Fig.~\ref{baryonic_mass_evol}, see also
Table~\ref{simulations_table}).  The circular velocities are a proxy
for the mass, rather than real velocities. From
Figs.~\ref{baryonic_mass_evol} and~\ref{rot_curves}, it is clear that
the gas contributes very little to the total circular velocity;
however, the gas is, in most cases, on nearly circular orbits 
in the disc plane.  This can be seen from the open circles in
Fig.~\ref{rot_curves}, which represent the mean gas velocity as a
function of radius for the different simulations (we have done this
calculation after doing an extra projection, in order to get the
rotation plane of the gas which is not always the same as that of
the stellar discs).  The mean gas velocities are relatively high, but
 their structure is complex, due to the non trivial interplay
between cooling and heating by SN feedback. This complexity can also
be observed in Fig.~\ref{vtita_maps_gas}, where we show the mean
tangential velocities for the gas, in a 2D view (with the galaxies
face-on). The velocity distributions usually present strong
asymmetries and, in some cases such as Aq-E-5, the mean tangential velocities are
negative,  indicating that the gas disc is
counter-rotating with respect to the stellar disc.  This results from
significant late accretion of gas with misaligned angular momentum (see S09).
As for the stars, we detect little rotation in
the gas component of Aq-F-5, the galaxy with a recent major merger.

The circular velocity curves are similar for varying resolution, although 
the inner parts show differences that reflect those found
for the total stellar masses 
 in each simulation (Table~\ref{resolution_global}).
In particular, Aq-C-6 has a $12\%$ lower peak velocity and a $8\%$ lower
velocity at $30$ kpc compared to Aq-C-5; and in Aq-E-6b (Aq-E-6) the
peak velocities and velocities at $30$ kpc are $4\%$ ($26\%$) and
$0.1\%$ ($9\%$) lower than those found for Aq-E-5.

\section{Conclusions}
\label{conclu}

We have continued our study of the properties of discs and spheroids 
in eight  simulations of galaxy
formation in a $\Lambda$CDM cosmology. 
The simulations
correspond to  haloes with present day virial masses in the range
$7-16\times 10^{11}$ M$_\odot$ and  spin parameters between $0.01$
and $0.05$. We use a kinematic
decomposition  to separate discs from spheroids. 
Four of the eight galaxies have significant discs,
three have  small discs, and one is a pure spheroid. None could
represent a late-type spiral. We compared
the formation histories, {\it in-situ} fractions,
structure and dynamical properties within and between galaxies, subdividing
also by age and by radius. Our main results are
summarized below.

{\it Formation time-scales.}  We found significant differences between the
formation histories and time-scales of discs and spheroids.  Spheroids
are formed early  and on short time-scales, while discs are
younger and have broad age distributions, often with a number of different
bursts.  Typical (mass-weighted) ages for spheroids and discs are
$\gtrsim 10$ Gyr and [$4-9$] Gyr, respectively.  Because spheroids
are old, we detect no difference in their
 mass- and luminosity-weighted ages. On the contrary,
luminosity-weighted ages for discs are [$2-8$] Gyr, significantly
smaller than the mass-weighted estimates. The mean 
ages of our simulated discs and spheroids are  in good agreement with observational results.

{\it Thin and thick discs.}  We find clear signs of the presence of
more than one component in our simulated discs, reminiscent of observed
thin and
thick discs. The youngest stars define thin structures, with high
tangential velocities and low velocity dispersions, whereas the oldest
disc stars define thicker discs, with lower rotational velocities
and higher velocity dispersions.  Assuming that thin and thick disc
components can be distinguished  by the age of their stars (we
adopted $9$ Gyr as the boundary), we determined typical rotation velocities
and velocity dispersions for thick/thin simulated discs. These agree
reasonably well with observations of the Milky Way. We note, however,
that our simulations cover a range of galaxy masses and halo
properties, and therefore of typical rotation velocities and velocity
dispersions.

{\it Structure of discs, bulges and bars.} As expected in the context
of the $\Lambda$CDM cosmology assumed in this work, the stellar
components of our eight simulations show great variety in their
structure. Discs have a wide range of thicknesses and scale-lengths,
and are usually complex, with misaligned components or boxy shape.
The inner regions of spheroids also show great diversity: half of
the simulated galaxies have bar components, which can dominate
over the bulges. S\'ersic fits to the mass density profiles generally
give shape parameters $n\sim 1$, i.e. they are similar to
exponential. As found in other studies, outer spheroids or ``stellar
haloes'' are very rich in structure, with streams, clumps and shells,
similar to observational results (Mart\'{\i}nez-Delgado et al. 2010).  In broad terms, the
sizes and structure of our bulges and discs agree well with observational
results, although our bulges are too massive with respect to discs.

{\it In-situ fractions.}  These are useful to
investigate the dominant formation channels for the different stellar
components.  We find that discs have the highest {\it in-situ}
fractions, typically $\gtrsim 0.9$, and all disc stars younger than
$9$ Gyr formed in the disc itself.  In one of the simulated galaxies,
however, $15\%$ of the final disc mass is contributed by a satellite
that came in on a nearly coplanar orbit. These stars are generally old
but are able to stay in the disc.  Inner spheroids (i.e. bulges and
bars) also have relatively high {\it in-situ} fractions, generally
larger than $0.8$, and consequently a low contribution from accreted
stars formed in systems other than the main progenitor.
Conversely, outer spheroids have a large contribution from
accreted stars and low {\it in-situ} fractions, in the range
$0.15-0.35$. As expected, the lowest {\it in-situ} values are
detected for systems with recent massive mergers. Our
results suggest that the outer regions of galaxies are populated by a
combination of {\it in-situ} and accreted stars, with the relative
fraction reflecting the particular formation and accretion history of
the host. We also find a negative dependence of {\it
  in-situ} fraction on radius, the outermost
regions of simulated galaxies are almost entirely populated by
accreted stars.  These results are in good agreement with other
studies (e.g. Zolotov et al. 2009; Cooper et al. 2010).

{\it Dynamical properties.}  We find complex and highly asymmetric
dynamical structures for the stellar and gaseous components of
our simulated galaxies. They reflect not only the diversity of our discs,
bulges and bars, but also the non-trivial interplay between cooling
and heating mechanisms.  Circular velocity curves are relatively flat for
$r\gtrsim 5$ kpc, and have peak velocities which are $10$ to $20\%$
higher than the value at larger radii.

We tested the effects of resolution by comparing the results of our
simulations to several  additional simulations of $4$ and $8$ times
lower mass resolution. We find relatively good agreement for the formation
time-scales, structure and  {\it in-situ} fractions of discs and spheroids.
The dynamical properties show more significant  variation with
 resolution, in particular for the discs. 
We find that in simulations of  higher resolution the discs tend to  have
higher tangential velocities and lower velocity dispersions. These
results indicate that low resolution runs  suffer from
artificial disc heating. However, the trends found for the velocities
and velocity dispersions with age are still captured in the low resolution runs.

In broad terms, our simulations agree with previous 
studies, and also show similar deficiencies, the most
important one being the presence of overly dominant bulges.  It is still
under debate whether these deficiencies are due to insufficient resolution,
or are a consequence of a poor description of the gas hydrodynamics or
the star formation and feedback processes.  On the other hand, the persistent failure
to reproduce late-type galaxies in a cosmological context in typical
galactic haloes may indicate that additional physical processes, not yet
 considered in simulations,  play a role in the regulation of star formation in
galaxies.  Our understanding of important processes
related to galaxy formation -- from the formation of stars to the
effects of supernova and black hole feedback -- is still quite rudimentary, 
so it is remarkable that cosmological simulations nevertheless produce
galaxies with structural and dynamical properties in reasonable
agreement with observational results.

\section*{Acknowledgments}

We thank the referee  for a thorough reading of this work and for his/her
helpful
comments and suggestions.
The simulations were carried out at the Computing Centre of the
Max-Planck-Society in Garching.  This research was supported by the
DFG cluster of excellence `Origin and Structure of the Universe'.  CS
thanks M. Boylan-Kolchin for providing the data for Fig.~\ref{MAH},
Lauren MacArthur for her help in the interpretation of observational
results, and M. Steinmetz, M. Vlajic, J. Wang and M. Williams for
useful and stimulating discussions.

\begin{appendix}

\section{Mass assembly histories}\label{Ap_MAH}

\begin{figure}
\includegraphics[width=80mm]{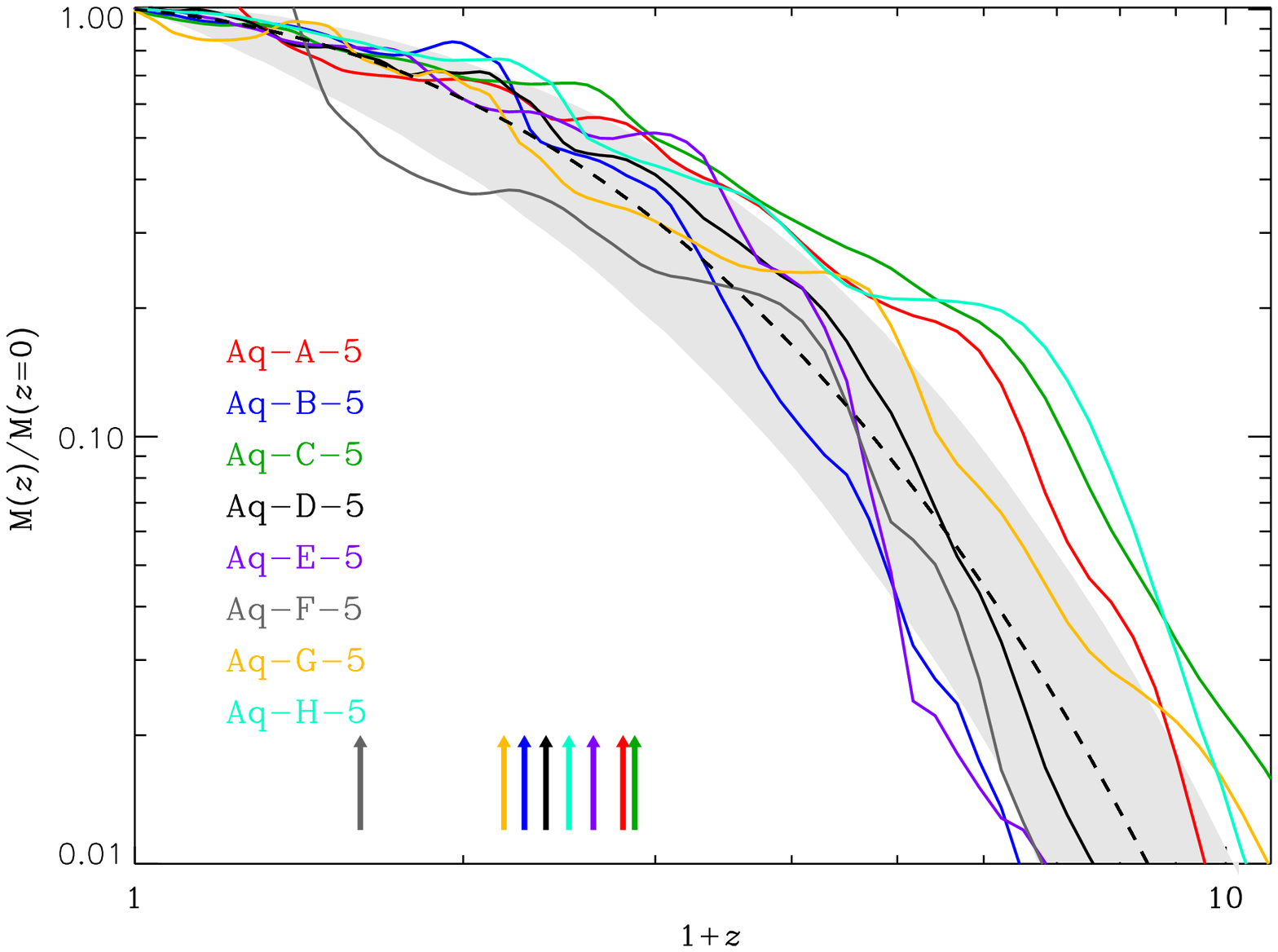}
\caption{Mass assembly histories of the eight simulated haloes. $M(z)$
  is the mass within the virial radius, and it is normalized to its
  present-day value.  The dashed line shows the median mass assembly
  history of Milky Way-mass haloes from the MS-II simulation, the grey
  shaded area containing $68\%$ of the distribution (Boylan-Kolchin et
  al. 2010). The arrows indicate the formation redshift of each halo,
  defined as the time where $M(z)$ reaches half its final value. }
\label{MAH}
\end{figure}

Fig.~\ref{MAH} shows the mass assembly history of our eight simulated
haloes, where $M(z)$ accounts for all mass within the virial radius,
and is normalized to its present-day value.  The median mass assembly
history of $\sim 7500$ Milky Way-mass haloes from the MS-II Simulation
is also shown, as given by Boylan-Kolchin et al. (2010).  Clearly, the
mass assembly histories of our eight haloes are a
representative sample of Milky Way-mass haloes formed in a
$\Lambda$CDM universe, although most of them form earlier than the
median relation.  Our plot is similar to that shown by
Boylan-Kolchin et al. (2010), but we include the two additional haloes
analysed in this paper.

\end{appendix}


\begin{thebibliography}{}

\bibitem{ab}
Abadi M.G., Navarro J.F., Steinmetz M., Eke V.R., 2003a, ApJ, 591, 499

\bibitem{ab2}
Abadi M.G., Navarro J.F., Steinmetz M., Eke V.R., 2003b, ApJ, 597, 21

\bibitem{agertz}
Agertz O., Teyssier R., Moore B., 2010, MNRAS, 410, 1391

\bibitem{barker}
Barker M.K., Sarajedini A., Geisler D., Harding P., Schommer R., 2007, ApJ, 133, 1125


\bibitem{bensby}
Bensby T., Zenn A.R., Oey M.S., Feltzing S., 2007, ApJ, 663, 13

\bibitem{mike}
Boylan-Kolchin  M., Springel V., White S.D.M., Jenkins A., Lemson G., 2009, MNRAS,
398, 1150

\bibitem{mike2}
Boylan-Kolchin  M., Springel V., White S.D.M., Jenkins A.,  2010, MNRAS,
406, 896

\bibitem{brook}	
Brook C.B., Kawata D., Gibson B.K., Flynn C., 2004, MNRAS, 349, 52


\bibitem{brooks}
Brooks A.M., Governato F., Quinn T., Brook C.B., Wadsley J., 2009, ApJ,
694, 396


\bibitem{BC03}
Bruzual G., Charlot S., 2003, MNRAS, 344, 1000

\bibitem{bullock}	
Bullock J.S., Dekel A., Kolatt T.S., Kravtsov A.V., Klypin A.A., Porciani C., Primack J.R.,
2001, ApJ, 555, 240


\bibitem{ceverino}
Ceverino D., Klypin A., 2009, ApJ, 695, 292

\bibitem{coop}
Cooper A.P., et al., 2010, MNRAS, 406, 744

\bibitem{gilmore}
Gilmore G., Wyse R.F.G., Norris J.E., 2002, ApJ, 574, 39


\bibitem{governato}
Governato F., et al., 2004, ApJ, 607, 688

\bibitem{governato2}
Governato F., Willman B., Mayer L., Brooks A., Stinson G.,
Valenzuela O., Wadsley J., Quinn T., 2007,  MNRAS, 374, 1479


\bibitem{governato3}
Governato F., et al. 2009, MNRAS, 398, 312


\bibitem{guo}
Guo Q., White S.D.M., Li C., Boylan-Kolchin M., 2010, MNRAS, 404, 1111


\bibitem{hm}
Haardt F., Madau P., 1996, ApJ, 461, 20

\bibitem{jung}
 Jungwiert B., Combes F., Palou\v{s} J., 2001, A\&A, 376, 85

\bibitem{juric}
Juri\'c M., et al., 2008, ApJ, 673, 864

\bibitem{lauren}
MacArthur L.A., Courteau S., Holtzman J.A.,
2003, ApJ, 582, 689


\bibitem{lauren}
MacArthur L.A., Courteau S., Bell E., Holtzman J.A.,
2004, ApJS, 152, 175


\bibitem{lauren2}
MacArthur L.A., González J.J., Courteau S., 2009, MNRAS, 395, 28



\bibitem{martig}
Martig M., Bournaud F., 2010, ApJ, 714, 275


\bibitem{md}
Mart\'{\i}nez-Delgado D., et al., 2010, AJ, 140, 962

\bibitem{mayer}
Mayer L., Governato F., Kaufmann T., 2008, Advanced Science Letters, 1, 7

\bibitem{NB91}
Navarro J.F., Benz W., 1991, ApJ, 380, 320

\bibitem{nord}
Norstr\"om B.,  et al., 2004, A\&A, 418, 989

\bibitem{okamoto}
Okamoto T., Eke V.R., Frenk C.S., Jenkins A., 2005, MNRAS, 363, 1299


\bibitem{piontek}
Piontek F., Steinmetz M., 2010, MNRAS, 410, 2625

\bibitem{robertson}
Robertson B., Yoshida N., Springel V., Hernquist L., 2004, ApJ, 606, 32

\bibitem{sanchez}
S\'anchez-Bl\'azquez P., Courty S., Gibson .K, Brook C.B., 2009, MNRAS, 398, 591


\bibitem{till1}
Sawala T., Scannapieco C., Maio U., White S.D.M., 2010, MNRAS, 402, 1599


\bibitem{sunrise}
Scannapieco C., Gadotti D.A., Jonsson P., White S.D.M., MNRAS, 407, 41

\bibitem{paperI}
Scannapieco C., Tissera P.B., White S.D.M., Springel V., 2005, MNRAS, 364, 552 

\bibitem{paperII}
Scannapieco C., Tissera P.B., White S.D.M., Springel V., 2006, MNRAS,
371, 1125 

\bibitem{paperII}
Scannapieco C., Tissera P.B., White S.D.M., Springel V., 2008, MNRAS,
389, 1137 

\bibitem{disks}
Scannapieco C., White S.D.M., Springel V., Tissera P.B., 2009, MNRAS, 396, 696 (S09)

\bibitem{Sellwood}
Sellwood J.A., Binney J.J., 200, MNRAS, 336, 785


\bibitem{sofue}
Sofue Y., Honma M., Omodaka T., 2009, PASJ, 61, 227


\bibitem{sm}
Sommer-Larsen J., Dolgov A., 2001, ApJ, 551, 608

\bibitem[Springel \& Hernquist 2003]{SH03}
Springel V., Hernquist L., 2003, MNRAS, 339, 289 

\bibitem{gadget}
Springel V., 2005, MNRAS, 364, 1105

\bibitem{aquarius}
Springel V., et al., 2008, MNRAS, 391, 1685


\bibitem{stinson}
Stinson G, Bailin J., Couchman H., Wadsley J., Shen S., Brook C.B., Quinn T., 2010, MNRAS, 408, 812

\bibitem{SD93}
Sutherland R.S., Dopita M.A., 1993, ApJS, 88, 253 

\bibitem{tinsley}
Tinsley B. M., 1974, ApJ, 192, 629

\bibitem{pat}
Tissera P.B., White S.D.M., Pedrosa S., Scannapieco C., 2010, MNRAS, 406, 922


\bibitem{val}
Vallenari A., Pasetto S., Bertelli G., Chiosi E., Spagna A., Lattanzi M., 2006, A\&A, 451, 125

\bibitem{wang}
Wang J., et al., 2010, MNRAS, 412, 1081


\bibitem{will}
Williams B.F., Dalcanton J.J., Dolphin A.E., Holtzman J., Sarajedini A., 2009, ApJ, 695, 15

\bibitem{yoachim}
Yoachim P., Dalcanton J.J., 2006, AJ, 131, 226


\bibitem{zolotov}
Zolotov A., Willman B., Brooks A.M., Governato F., Brook C.B., Hogg D.W.,
Quinn T., Stinson G., 2009, ApJ, 702, 1058



\end{thebibliography}
\end{document}